\documentclass[aps,prc,twocolumn,amsmath]{revtex4}
\usepackage[normalem]{ulem}

\usepackage{graphicx}
\usepackage{epsfig}
\usepackage{color}
\usepackage{multirow,ulem}
\usepackage{booktabs}
\usepackage{adjustbox,pdflscape,longtable,rotating}

\usepackage{color,graphicx}
\usepackage{mathptmx}                
\usepackage{dcolumn}                 
\usepackage{bm}                      
\usepackage[utf8]{inputenc}
\def\roughly#1{\mathrel{\raise.3ex\hbox{$#1$\kern-.75em%
\lower1ex\hbox{$\sim$}}}}

\begin{document}

\title{Relativistic hypernuclear compact stars with calibrated equations of state}

\author{Morgane Fortin}\email{fortin@camk.edu.pl}
\affiliation{N. Copernicus Astronomical Center, Polish Academy of Sciences,
  Bartycka 18, 00-716 Warszawa, Poland}

\author{Adriana R. Raduta}\email{araduta@nipne.ro}
\affiliation{National Institute for Physics and Nuclear Engineering (IFIN-HH),
  RO-077125 Bucharest, Romania}

\author{Sidney Avancini}\email{sidney.avancini@ufsc.br}
\affiliation{Departamento de Fisica, Universidade Federal de Santa Catarina,
  88040-900 Florianopolis, Santa Catarina, Brazil}

\author{Constança Provid\^encia}\email{cp@uc.pt}
\affiliation{$^4$CFisUC, Department of Physics, University of Coimbra, 3004-516 Coimbra, Portugal}

\begin{abstract}
  Within the covariant density functional theory of hypernuclear matter we build
  a series of equations of state for hypernuclear compact stars, by calibrating
  the coupling constants of the $\Xi$-hyperon to the experimental binding energy
  of the single-$\Xi$ hypernuclei $^{15}_{\Xi^-}$C and $^{12}_{\Xi^-}$Be.
Coupling constants of the $\Lambda$-hyperon to nucleons have been calibrated on a vast collection of experimental data on single $\Lambda$-hypernuclei and we employ those values.
  Uncertainties on the couplings of the $\Sigma$-hyperon to nuclear matter,
  due to lack of experimental data, are accounted for by allowing for a wide variation
  of the well depth of $\Sigma$ at rest in symmetric saturated nuclear matter.
  To account for uncertainties in the nucleonic sector at densities much larger than $n_0$,
  a rich collection of parametrizations is employed, some of them in agreement with
  existing constraints from nuclear physics and astrophysics.
  Neutron star properties are investigated with all these calibrated equations of
  state. The effects of the presence of hyperons on the radius, on the tidal deformability,
  on the moment of inertia, and on the nucleonic direct Urca process are
  discussed. The sensitivity of the hyperonic direct Urca processes to uncertainties
  in the nucleonic and hyperonic sectors is also addressed.
  It is shown that the relative variations of the radius, tidal deformability and
  moment of inertia from the values that characterize purely nucleonic stars
  are linearly correlated with the strangeness fraction.
  The maximum  radius deviation, obtained for most massive neutron stars, is $\approx 10\%$.
  The reduction of the maximum mass, triggered by nucleation of strangeness, is estimated
  at $\approx 15 - 20\%$, out of which 5\% comes from insufficient information
  on the $\Sigma$-hyperon interactions.
  A total of 44 calibrated hyperonic equations of state are published as Supplemental Material.
\end{abstract}

\date{\today}


\maketitle

\section{Introduction}

The recent detection of gravitational waves emitted during
the inspiral phase of a neutron star-neutron star merger  GW170817 \cite{gw170817}  together with
the following up electromagnetic signal opened a new door to the study of neutron stars (NS)
\cite{GBM2017}.
NSs have been acknowledged since long ago to be perfect test grounds of cold and
dense baryonic matter, with thermodynamic conditions complementary to those
produced in terrestrial laboratories.
In the innermost shells non-nucleonic degrees of freedom such as hyperons and
kaon or pion condensates or a quark gluon plasma have been predicted to exist \cite{Glend2000}
in addition to or instead of the nucleonic ones.
Understanding the way in which these "exotic" degrees of freedom affect the structure and
evolution of NS ultimately allows one to confirm or, on the contrary, rule out their presence.
Information thus implicitly gained on the so far insufficiently constrained
interaction potentials makes NSs a promising research field for nuclear physics.

The major source of uncertainties that affect NSs comes from the nucleonic sector,
which dominates in all but pure quark stars and whose
behavior at densities much larger than the saturation density of symmetric nuclear matter ($n_0$)
and high isospin asymmetry remains poorly known, despite intense theoretical and experimental
effort. To account for this state of facts NS studies typically allow for the widest
collection of equations of state (EoS) compatible with constraints from nuclear physics
experiments, {\it ab initio} calculations and astrophysical observations.
In the present work we adopt the same strategy.
Additional sources of uncertainties are related to the above-mentioned "exotic" species.
They are less serious than the ones in the first category, as only NSs with masses exceeding
the threshold value for nucleation of those species are affected.
Some of the first new degrees of freedom that are expected to be populated are the hyperons,
which make the object of our present study. Another possibility is the
$\Delta$ resonance \cite{Glend2000} which has recently been investigated
by several authors \cite{Drago2014,Cai2015,Ribes2019,Li:2019}, but which  is not considered in the
present study.

With the aim of building EoS as realistic as possible, we continue the work started in
Refs. \cite{Fortin17,Fortin18,Providencia19}, where the $\Lambda$-nucleon and
$\Lambda-\Lambda$ interaction potentials have been constrained based on a vast collection
of experimental data on single- and double-$\Lambda$ nuclei, by constraining the
$\Xi$-nucleon potentials on experimental data on single-$\Xi$ nuclei.
As in Refs. \cite{Fortin17,Fortin18,Providencia19},
we calculate the binding energies of nuclei with variable number of nucleons
and one hyperon by solving the Dirac equations of the nucleons
and the hyperon obtained from the assumed Lagrangian density.
The coupling constants between the $\Xi$ and the scalar $\sigma$ meson are
be tuned on the binding energy of $^{15}_{\Xi^-}$C, measured by the KEK-E373 experiment
\cite{kiso}.
The thus obtained EoS are called calibrated, as they comply with
the maximum available experimental information.
The third hyperonic species expected to nucleate in NS cores is $\Sigma$.
Experimental data on strong-interaction level shifts, widths and yields
collected from $\Sigma^-$ atoms and
inclusive $(\pi^-, K^+)$ spectra on medium to heavy targets
indicate a repulsive $\Sigma$N potential.
According to Refs. \cite{Gal2010,Gal2016,Sugimura2014,Honda2017,Harada2018},
these data are compatible with a wide range of the
well depth of $\Sigma$s at rest in
saturated symmetric nuclear matter,
$10 \lesssim U_{\Sigma}^{(N)} \lesssim 50$ MeV.
Theoretical studies performed within the chiral effective field theory
support the repulsive character of $U_{\Sigma}^{(N)}$, though the magnitude
is estimated to lesser values $\approx 15$ MeV \cite{Haidenbauer2015}.
Not being able to constrain the couplings of the $\Sigma$-hyperon,
we investigate how the uncertainties that affect $U_{\Sigma}^{(N)}$
impact the NS properties. Special attention is given to the
chemical composition of NS, particularly sensitive to negatively
charged particles and susceptible to being indirectly determined from
NSs' thermal evolution.
Notable effects are expected to occur for the less repulsive potentials,
which favor earlier onset of $\Sigma$s.
This expectation relies on the fact that, as soon as they appear,
any negatively charged particles
 partially replace the electrons in the net charge neutrality equation and,
consequently, alter the $\beta$-equilibrium conditions which determine the
relative abundances. In extreme scenarios, also the threshold of nucleonic
dUrca may be affected.

The first high-precision measurement of a massive pulsar mass,
corresponding to PSR J$1614-2230$ \cite{Demorest10,j1614a}
with $M=1.908\pm 0.016M_\odot$, (in the following masses are given with a precision at the 1-$\sigma$ level that a 68\%.3 confidence level) challenged the nuclear physics community
on whether two solar mass NSs can accommodate non-nucleonic degrees of freedom
\cite{Vidana10,Demorest10}.
The particular case of the onset of hyperons, commonly known as
the hyperon(ization) puzzle, was addressed at length in Ref. \citep{Chatterjee15},
where several scenarios that reconcile large masses and hyperonic degrees
of freedom have been identified.
They include i) a sufficiently hard nucleonic EoS and
ii) going beyond the simple $SU(6)$ symmetry ansatz to fix the vector meson couplings
\cite{Bednarek2012,Weissenborn12,Weissenborn13,colucci_13,vandalen_14,Fortin16,Sun19}.
Other massive pulsars have been detected in the meanwhile,
in particular the pulsars PSR J$0348+0432$ \cite{Antoniadis13},
with a mass $2.01 \pm 0.04M_\odot$, 
and the millisecond pulsar MSP J0740+6620 \cite{Cromartie2019},
with a mass $2.14^{+0.10}_{-0.09} M_\odot$.
It is worthwhile to note that massive NSs can be theoretically obtained also
by assuming a deconfinement phase transition to quark matter
\cite{Alford06,Weissenborn11,Bonanno2012,Masuda2013,Alford2013,Klahn2013,Zdunik2013,logoteta2013,Drago2016,Pereira2016,Fukushima2016,Alford2017}.
In the present paper we add some more information to the
issue and study, within a relativistic mean-field (RMF) approach with
model parameters fitted to experimental data,
under which conditions NS cores do accommodate hyperons and how
these extra particle degrees of freedom modify NSs' geometric and
chemical properties. 
The extent to which one may learn information on chemical
composition from thermal evolution is addressed elsewhere.

The paper is organized as follows.
Section \ref{sec:eos} presents the nucleonic EoSs on which our hyperonic EoSs
are built and the way in which $\Xi$-meson coupling constants are determined
from experimental $\Xi$-hypernuclei data.
Physical (maximum mass and radius, tidal deformability and moment of inertia
versus gravitational mass) and chemical properties of hypernuclear compact stars
built upon our set of calibrated EoSs are discussed in Sec. \ref{sec:prop}. 
Special attention is given to the uncertainties related to the $\Sigma$ potential.
The conclusions are drawn in Sec. \ref{sec:concl}.

\section{Equations of state}
\label{sec:eos}

In the following we define the set of EoSs that we choose to perform our study.
The parametrizations employed to describe the nucleonic sector are introduced
in Sec. \ref{ssec:model}.
The way in which experimental data on hypernuclei are used to calibrate the
hyperon-meson coupling constants is detailed in Sec. \ref{ssec:hyperon}.

\subsection{Nucleonic EoS}
\label{ssec:model}

The phenomenological EoS considered in our study have been obtained
in the framework of RMF. In this category of models
the nucleons interact among each others by exchanging scalar-isoscalar ($\sigma$),
vector-isoscalar ($\omega$), vector-isovector ($\rho$) and, in more sophisticated
cases (not considered here) also vector-isoscalar ($\delta$) mesons.
For a general review of these types of models, see Ref. \cite{Dutra2014}.
The chosen models fall into two classes:
a) models with constant couplings and nonlinear meson terms, generically called
NL models and
b) models with density dependent couplings, generically called DD models. 
We recall that the nonlinear meson terms have been introduced
in order to correctly describe the properties of symmetric saturated nuclear matter,
when the coupling constants do not depend on density. 

We presently consider the set of models discussed in Ref. \cite{Providencia19}.
It consists of the NL models:
FSU2 \cite{Chen2014}, FSU2H and FSU2R \cite{Tolos17,Negreiros18},
NL3 \cite{nl3}, NL3$\omega\rho$ \cite{Pais16,Horowitz01},
TM1 \cite{tm1}, TM1$\omega\rho$ \cite{Providencia13,Bao2014,Pais16},
TM1-2 and TM1-2$\omega\rho$ \cite{Providencia13},
and the DD models DD2 \cite{typel10} and DDME2 \cite{ddme2}.

From the astrophysical point of view, the common feature of all these models 
is the ability to provide NS masses in excess to 2$M_\odot$ stars.
  In regards to the incompressibility parameter of symmetric nuclear matter
  at saturation, $K_{\infty}$, they span values between
  238 MeV (FSU, FSU2R, FSU2H) and 281.2 (TM1, TM1$\omega\rho$, TM1-2, TM1-2$\omega\rho$),
  with DD2, DDME2 and NL3 and NL3$\omega\rho$ having intermediate values as
  follows: 242.7, 250.9 and, respectively, 271.5 MeV.
  As is frequent for RMF models, these values are larger than the
  generally accepted constraints coming from isoscalar giant monopole and dipole resonances
  in nuclei, $240 \pm 20$ MeV \cite{Shlomo_EPJA_2006}, or, the more recently,
  $211.9 \pm 24.5$ MeV \cite{De_PRC_2015}.

In regards to the isovector channel, they fall into two classes.
i) models with moderate values of the slope of the symmetry energy at saturation,
which thus comply with a compilation of constraints coming
from experimental nuclear physics, {\em ab initio} calculations of
pure neutron matter and astrophysical observations and are
customarily expressed as $40 \lesssim L \lesssim 62$ MeV \cite{Lattimer13} or
$30 \lesssim L \lesssim 86$ MeV \cite{Oertel17}. They are:
FSU2H, FSU2R, NL3$\omega\rho$, TM1$\omega\rho$, TM1-2$\omega\rho$,
DD2 and DDME2.  These models have $L$ in the range 44-55 MeV;
ii) models with high values of the slope of the symmetry energy at saturation,
$L\gtrsim 100$ MeV. They are: FSU2, NL3, TM1, TM1-2
with values of $L$ in the range  108-119 MeV.
For all these models a table with the corresponding slope values and also other
nuclear matter saturation properties  can be found in Ref. \cite{Providencia19}.
At this point we recall that
large values of $L$ are still compatible with the Lead Radius Experiment (“PREX”) \cite{prex},
and recent analysis of elliptic flow in heavy ion collisions  \cite{Cozma2018}.
The models in the second class are mainly kept for the sake of completeness
and because some of the models in the first class have been derived from them
by introducing additional meson couplings.
  It should be
  pointed out that although most of the models that we consider predict
  well-accepted nuclear matter properties at saturation density they
  are not constrained at high densities, except for the fact that they have to
  also predict two solar mass stars.
The same is true for the hyperonic models which are
  constrained by properties defined at saturation density. Besides,
  another source of uncertainty is the
  density dependence of the hyperonic couplings in  the density
  dependent hyperonic models, which has  been considered to behave as
  the respective nucleonic couplings.
\subsection{Hyperonic EoS with calibrated meson couplings}
\label{ssec:hyperon}

The starting point of the meson-exchange RMF model for hypernuclei
is the covariant Lagrangian density,
\begin{equation}
  {\cal L}={\cal L}_N+{\cal L}_Y+{\cal L}_M,
  \label{eq:L}
  \nonumber
  \end{equation}
where ${\cal L}_N$,  ${\cal L}_Y$ and ${\cal L}_M$ respectively stand
for the nucleonic,
hyperonic and purely mesonic components.
In addition to the couplings accounted for when describing
infinite nuclear matter - and listed in Sec \ref{ssec:model} -,
in the case of (hyper)nuclei ${\cal L}_N$ additionally contains
couplings with the photon field $A^{\mu}$.
For details, see Ref. \cite{Dutra2014}.
The hyperonic term of Lagrangian density is given by
\begin{equation}
  {\cal L}_Y=\bar{\Psi}_Y \left[
    \gamma_{\mu} D^{\mu}_Y - m^{*}_Y + \frac{f_{\omega Y}}{2m_Y}
    \sigma^{\mu\nu} \partial_\nu \omega_\mu
    \right] \Psi_Y,
  \label{eq:LY}
\end{equation}
with
\begin{equation}
D^{\mu}_Y=i\partial^{\mu}-g_{\omega  Y} \omega^{\mu}
-g_{\rho Y} \boldsymbol{\tau}_Y \cdot \boldsymbol{\rho}^{\mu}
-e \frac{\tau_{Y,3}-1}{2} A^{\mu},
\label{eq:DY}
\end{equation}
where $m^{*}_Y=m_Y-g_{\sigma Y} \sigma$ stands for the Dirac effective mass,
$e$ is the elementary electric charge,
$\boldsymbol{\tau}_Y$ is the isospin operator,
$\omega^{\mu}$ and $\boldsymbol{\rho}^{\mu}$ are the fields associated to the
$\omega$ and $\rho$ mesons and ${\Psi}_Y$ is the $Y$-hyperon field.
$g_{\sigma Y}$, $g_{\omega  Y}$ and $g_{\rho Y}$ correspond to the coupling
constants of the various meson fields with the $Y$-hyperon.
The third term in eq. (\ref{eq:LY}), proportional to $f_{\omega Y}/2M_Y$,
represents the tensor coupling between the $Y$ hyperon with the $\omega$ field.
It impacts the spin-orbit splitting and, in principle, can be
determined from spectroscopic data.
The last term in Eq. (\ref{eq:DY}), proportional to $e$, describes
the interaction of the $Y$-hyperon with the Coulomb field and is meaningful
for charged hyperons only.

Note that the $\sigma^*$ and $\phi$-mesons were not included in Eq.~(\ref{eq:DY})
since, for the case of $\Xi$-hyperon of interest here,
there are no experimental data that could be exploited in order to fix them.
We do, however, include the $\phi$-meson in the next section, devoted
to the EoS of stellar matter. In that case the coupling is determined based on
SU(6) flavor symmetry arguments: $g_{\phi \Lambda}=-\sqrt{2}/3 g_{\omega N}$, $g_{\phi \Sigma}=-\sqrt{2}/3 g_{\omega N}$, $g_{\phi \Xi}=-2\sqrt{2}/3 g_{\omega N}$ and $g_{\phi N}=0$.
This contribution brings repulsion to the interaction and,
given that we do not include the $\sigma^*$-meson,
this interaction might be overestimated.

Once the coupling constants are given, the single particle Dirac equations for
baryons and the Klein-Gordon equations for mesons and photon  are
obtained in the mean-field approximation by the variational method
\cite{Shen06,Avancini07,Sun2016}.
The numerical procedure in order to solve the Dirac and Klein-Gordon
variational equations consists in expanding both the meson fields and
the baryon single-particle Dirac wave functions in terms of a spherical harmonic oscillator basis.
Thus, one has to solve a self-consistent system  of nonlinear matrix equations.
In the case of the electromagnetic field the Poisson equation is solved directly
by using the associated Green's function since the basis expansion method  is very slowly convergent.
As the translational symmetry is broken in the mean-field approximation,
the result has to be corrected for the center of mass motion.
The correction is more important for light systems.
As in Refs. \cite{Fortin17,Fortin18}, we adopt the microscopic expression
$E_{C.M.}=-\langle \hat {\bf P}^2\rangle/2M$, where $M=\sum_B M_B$ is the
total mass of the hypernucleus and $ \hat {\bf P}=\sum_B \hat {\bf P}_B$
is the total momentum operator.

The binding energy of the hyperon $Y$ in a nucleus with $A_n$ neutrons,
$A_p$ protons and $A_Y$ hyperons is given by the difference between
the energies of that hypernucleus and the hypernucleus with $A_Y-1$ hyperons.
Note that removal of charged hyperons, as $\Xi^-$ of interest here,
implies modification of the number of protons.

Experimental data on hypernuclei binding energy may be used in order to calibrate
the interaction between the hyperon and the scalar meson fields \cite{Mares1989,Jennings1990,Sugahara1994,Mares1994,Shen06}.
The issue was most recently addressed by Fortin {\it et al.} \cite{Fortin17,Fortin18},
who accounted for a vast collection of experimental data on single-$\Lambda$ hypernuclei
in $s$ and $p$ shells and effective nucleonic interactions.
These authors have thus confirmed that binding energies are directly related to
the well depth of the hyperon at rest in symmetric nuclear matter,
as customarily assumed in the literature merely based on heuristic arguments,
and that, once the flavor symmetry model is fixed, the value of the hyperon
coupling constant represents a fixed fraction of the coupling $g_{\sigma N}$. 
In the present work we use the $g_{\sigma \Lambda}$ coupling constants obtained
in Refs. \cite{Fortin17,Fortin18}.

So far experimental data exist only for two light $\Xi^-$-hypernuclei:
$^{12}_{\Xi^-}{\rm Be}$ and $^{15}_{\Xi^-}{\rm C}$.
The former was produced in $(K^-,K^+)$ reactions on a
$^{12}{\rm C}$ target \cite{khaustov00}.
The second, known as the Kiso event \cite{kiso}, corresponds to
an intermediate state in the reaction
$\Xi^-+^{14}{\rm N} \to ^{15}_{\Xi^-} {\rm C} \to ^{10}_\Lambda {\rm Be}+^5_\Lambda {\rm He}$.
Both sets of data indicate an attractive $N \Xi$ interaction and
bound $\Xi$-hypernuclei. See also the recent results in Ref. \cite{Yoshida2019}

Indeed, double differential cross section for $^{12}_{\Xi}{\rm Be}$ production
was found to be compatible with a $\Xi$-nucleus potential well depth of about 14 MeV,
within the Woods-Saxon prescription \cite{khaustov00}.
In regards to the binding energy of $\Xi^-$ in $^{12}_{\Xi^-}{\rm Be}$,
estimations performed within the cluster model \cite{Hiyama2008}
and the coupled-channels antisymmetrized molecular dynamics approach
\cite{Matsumya2011} provide values between 3 and 5.5 MeV.
As expected, these values depend on the $N \Xi$ interaction.
When this interaction is adjusted such as to give a value consistent
with the experimental spectrum in Ref. \cite{khaustov00}, as has been done
in Ref. \cite{Hiyama2008}, the binding energy is of $\approx 5$ MeV.

The interpretation of the Kiso event is more problematic,
for the final state of the daughter nucleus $^{10}_{\Lambda}{\rm Be}$
was not unambiguously identified.
Typically two scenarios are assumed for dealing with data:
a) In the first case it is assumed that $^{10}_{\Lambda}{\rm Be}$
is in its ground state. Then, $^{15}_{\Xi^-}{\rm C}$ is considered to be in the state
$^{14}{\rm N(g.s.)}+\Xi^-(1s)$.
b) In the second scenario $^{10}_{\Lambda}{\rm Be}$ is assumed to be
produced in an excited state.
If this is the case, $^{15}_{\Xi^-}{\rm C}$ corresponds to the state 
$^{14}{\rm N(g.s.)}+\Xi^-(1p)$.
The binding energy of $\Xi^-$, $B_{\Xi^-}= E(^{15}_\Xi{\rm C})-E(^{14}{\rm N})$,
has different values in cases a) and b).
The maximum value corresponds to the case a) and amounts to $4.38 \pm 0.25$ MeV.
The lower limit, of $1.11 \pm 0.25$ MeV, corresponds to the second excited state
of $^{10}_{\Lambda}{\rm Be}$ with an excitation energy of 3.2 MeV,
as calculated by different models \cite{Hiyama_2012,Millener_2012}.
Note that the energy spectrum of $^{10}_{\Lambda}{\rm Be}$ were recently investigated
using $(e,e'K^+)$ reactions by Gogami {\it et al.} \cite{Gogami_PRC_2016},
who found that the first excited state lies at $\approx 2.7$ MeV.
The two scenarios together with their compatibility with the information
obtained for the binding energy of $\Xi$ in $^{12}_{\Xi^-}$Be have been
investigated in Ref. \cite{Sun2016}, within RMF and Skyrme-Hartree-Fock approaches.
The conclusion reached by Sun {\it et al.} \cite{Sun2016} is that
the most plausible interpretation is the one corresponding to 
$^{14}{\rm N (g.s.)}+\Xi^- (1p)$.

In order to constrain the couplings of the $\sigma$ meson to $\Xi$-hyperon
we employ the procedure described in Refs. \cite{Fortin17,Fortin18}
and experimental data corresponding to
$^{15}_{\Xi^-}{\rm C}$ \cite{kiso}.
We alternatively assume that $^{15}_{\Xi^-}{\rm C}$ corresponds to
the $^{14}{\rm N(g.s.)}+\Xi^-(1s)$ and, respectively,
$^{14}{\rm N (g.s.)}+\Xi^- (1p)$ states.
In both cases we calculate also
$^{12}_{\Xi^-}{\rm Be}$ and compare with data in Ref. \cite{khaustov00}.

Other coupling constants are fixed as follows.
For the coupling between the $\Xi$ and the $\omega$-meson,
we use the SU(6) value $g_{\omega  \Xi}  =2/3 g_{\omega  N}$.
For the coupling between the $\Xi$ and the $\rho$-meson,
we assume $g_{\rho \Xi}=g_{\rho N}$.
In DD models, we suppose the same density dependence for hyperon-
and nucleon-meson couplings. Let us recall that other
calibration constraints have been employed in older works,
as proposed in Ref. \cite{gm1991}, or more recent works \cite{vandalen_14}.
Finally, in order to get a weaker interaction between 
$\Xi$ and the nuclear spin orbit, the tensor coupling is included as in
Refs. \cite{Mares1994,Sun2016}, with $f_{\omega\Xi}=0.4  g_{\omega  N}$.

Since $^{15}_\Xi$C is constituted by a symmetric nucleus $^{14}_7$N,
with isospin 0, and a $\Xi$-hyperon, which has
nonzero isospin, the $\rho$-meson field is finite due to the
self-interaction of the $\Xi$ with itself.
To remove this spurious contribution, we follow the 
procedure proposed in Refs. \cite{Mares1994,Sun2016}.
It consists in performing two calculations of the hypernucleus energy.
1) In the first one, the coupling of the $\rho$-meson to the nucleons is put to 0
while the coupling of the $\rho$-meson to the $\Xi$-hyperon is kept fixed.
The corresponding energy is $E(g_{\rho N}=0,g_{\rho \Xi})$.
2) In the second case, the coupling constants of the $\rho$-meson to both
nucleons and $\Xi$-hyperon are put to 0.
The corresponding energy is $E(g_{\rho N}=0,g_{\rho \Xi}=0)$.
The spurious energy is given by
$E_{sp}=E(g_{\rho N}=0,g_{\rho \Xi})-E(g_{\rho N}=0,g_{\rho \Xi}=0)$
and can be straightforwardly removed from the full calculation.
We employ this procedure to correct the energies of both
$^{15}_\Xi{\rm C}$ and $^{12}_\Xi {\rm Be}$. 

The values of the coupling constant $g_{\sigma \Xi}$, expressed
as the fraction $x_{s \Xi} = g_{\sigma \Xi}/g_{\sigma N}$,
obtained from the fit of the binding energy of the $\Xi$-hyperon
in the $^{15}_\Xi$C hypernucleus are given in Table \ref{tabBe}.
Different effective nucleon interactions,
introduced in Sec. \ref{ssec:model}, are considered.
The two scenarios which assume that $\Xi^-$ occupies the $1s$ or,
alternatively, the $1p$ state of $^{14}{\rm N}$ are considered separately.
They correspond to the binding energies $BE=4.4$ MeV and, respectively,
$BE=1.1$ MeV.
Also given are the values of $U^{(N)}_\Xi$, the well depth of $\Xi$ at rest
in symmetric saturated nuclear matter, and the binding energy of
the $\Xi$-hyperon in the hypernucleus $^{12}_\Xi$Be.
Finally the last column lists, for comparison,
the binding energy of $\Xi$ in the $1s$ state of $^{15}_{\Xi^-}$C
obtained when the coupling constant given in the fifth column is used.
If we take into account that the experimental data of $^{12}_\Xi {\rm Be}$
\cite{khaustov00} have been interpreted as compatible with $U_\Xi (n_0) \sim -14$ MeV,
our results confirm the conclusion of Ref. \cite{Sun2016},
suggesting as the most plausible scenario the one in which $^{15}_{\Xi}{\rm C}$
is produced in an excited state.
In regards to the binding energy of $\Xi$ in $^{12}_{\Xi}{\rm Be}$, the
situation is less clear as the values provided by Refs. \cite{Hiyama2008,Matsumya2011},
in the range 3-5.5 MeV, sit in between the values we obtained for the two scenarios.
We nevertheless note that three interactions (FSU2R, TM1 and TM1$\omega \rho$)
provide, for the second scenario, values similar to those of Refs.
\cite{Hiyama2008,Matsumya2011}.

\begin{table}[h]
  \caption{Coupling constant fraction $x_{s\Xi}=g_{\sigma\Xi}/g_{\sigma N}$,
    well depth of $\Xi^-$ at rest in symmetric matter at saturation density
    ($U_{\Xi}^{(N)}$) and, respectively,
    binding energy of $\Xi$ in $^{12}_{\Xi 1s}$Be as obtained
    from the fit of the binding energy of $^{15}_{\Xi}$C.
    Results corresponding to the hypothesis according to which
    $^{10}_{\Lambda}$Be is produced in the ground state or, alternatively,
    the first excited state 
    are reported in columns 2-4 and, respectively, 5$-$7.
    The binding energies of $^{15}_{\Xi 1s}$C and $^{15}_{\Xi 1p}$C are
    $BE_\Xi=4.4$ MeV and $BE_\Xi=1.1$ MeV.
    For the second hypothesis, also the energy of $^{15}_{\Xi 1s}$C is provided,
    on the last column.
    Results correspond to different nucleon effective interactions. } 
\bigskip
\label{tabBe}
\begin{tabular}{lrrrrrrrr}
\hline
Model& \multicolumn{4}{c}{$^{15}_{\Xi 1s}$C}(BE=4.4 MeV)
  &$\phantom{m}$&\multicolumn{3}{c}{$^{15}_{\Xi 1p}$C} (BE=1.1 MeV)\\
\cline{2-4} \cline{6-9}
& $x_{s\Xi}$ & $U_\Xi^{(N)} $&  $^{12}_{\Xi 1s}$Be& &
 $x_{s\Xi}$ & $U_\Xi^{(N)}$ & $^{12}_{\Xi 1s}$Be
&$^{15}_{\Xi 1s}$C\\
(MeV) & & & (MeV) & &  & (MeV) & (MeV) & (MeV) \\
\hline
DD2                            & 0.304 & -11.10 & 2.35  & & 0.320 & -17.50 & 6.48 & 9.02\\
DDME2                          & 0.306 & -12.49 & 2.38  & & 0.321 & -18.78 & 6.31 & 8.83\\
FSU2R                          & 0.296 & -11.80 & 2.51  & & 0.316 & -17.51 & 5.87 & 8.12\\
FSU2                           & 0.296 & -10.00 & 2.64  & & 0.311 & -15.69 & 6.15 & 8.19\\
FSU2H                          & 0.296 &  -10.00 & 2.68  & & 0.310 & -15.47 & 6.47 & 7.91\\
TM1                            & 0.295 &  -9.78 & 2.59  & & 0.310 & -14.93 & 5.48 & 7.69\\
TM1$\omega\rho$                & 0.295 &  -9.80 & 2.51  & & 0.310 & -14.94 & 5.34 & 7.68\\ 
TM1-2                          & 0.292 &  -8.71 & 2.58  & & 0.309 & -14.62 & 6.62 & 8.79\\
TM1-2$\omega\rho$              & 0.292 &  -8.74 & 2.59  & & 0.309 & -14.63 & 6.56 & 8.77\\
NL3                            & 0.296 &  -9.88 & 2.84  & & 0.310 & -15.36 & 7.31 & 7.93\\
NL3$\omega\rho$                & 0.296 &  -9.90 & 2.73  & & 0.311 & -15.39 & 7.17 & 7.93\\
\hline
\end{tabular}
\end{table}

\begin{figure*}[th]
\begin{tabular}{cc}
\includegraphics[width=.4\linewidth]{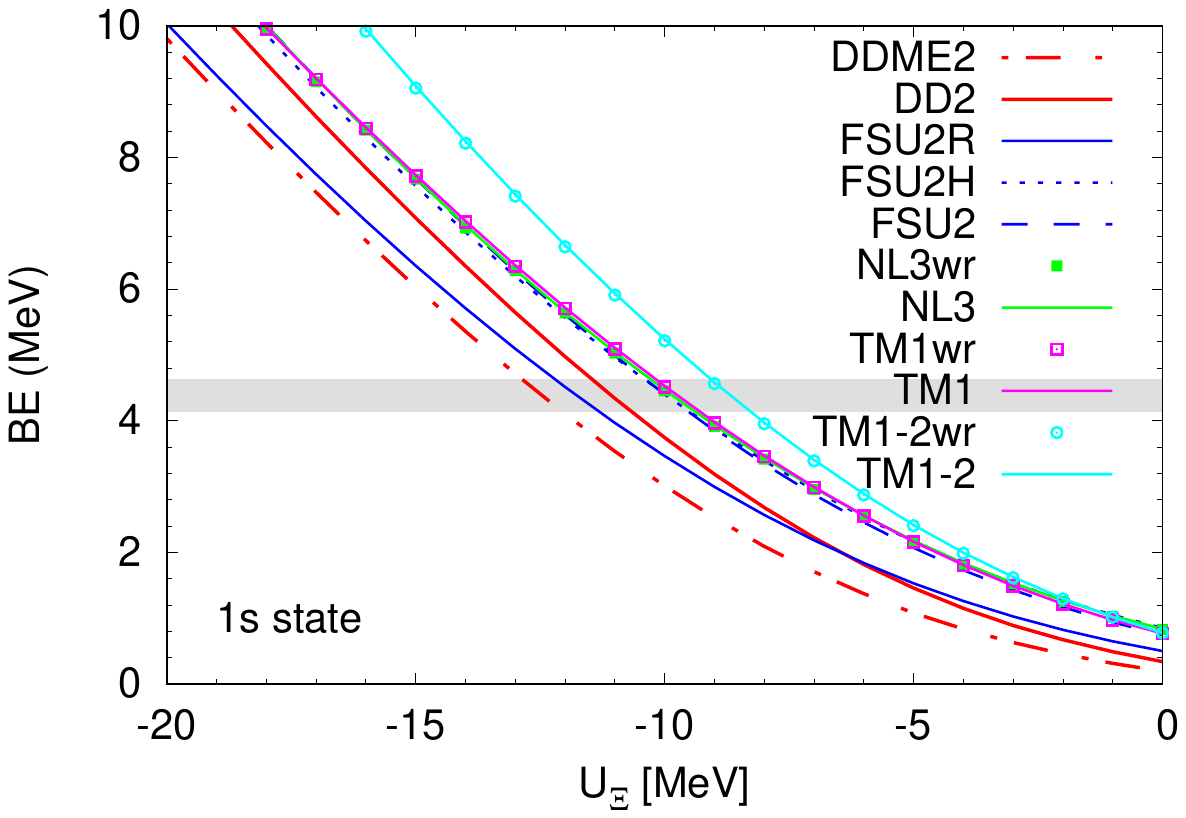}&
\includegraphics[width=.4\linewidth]{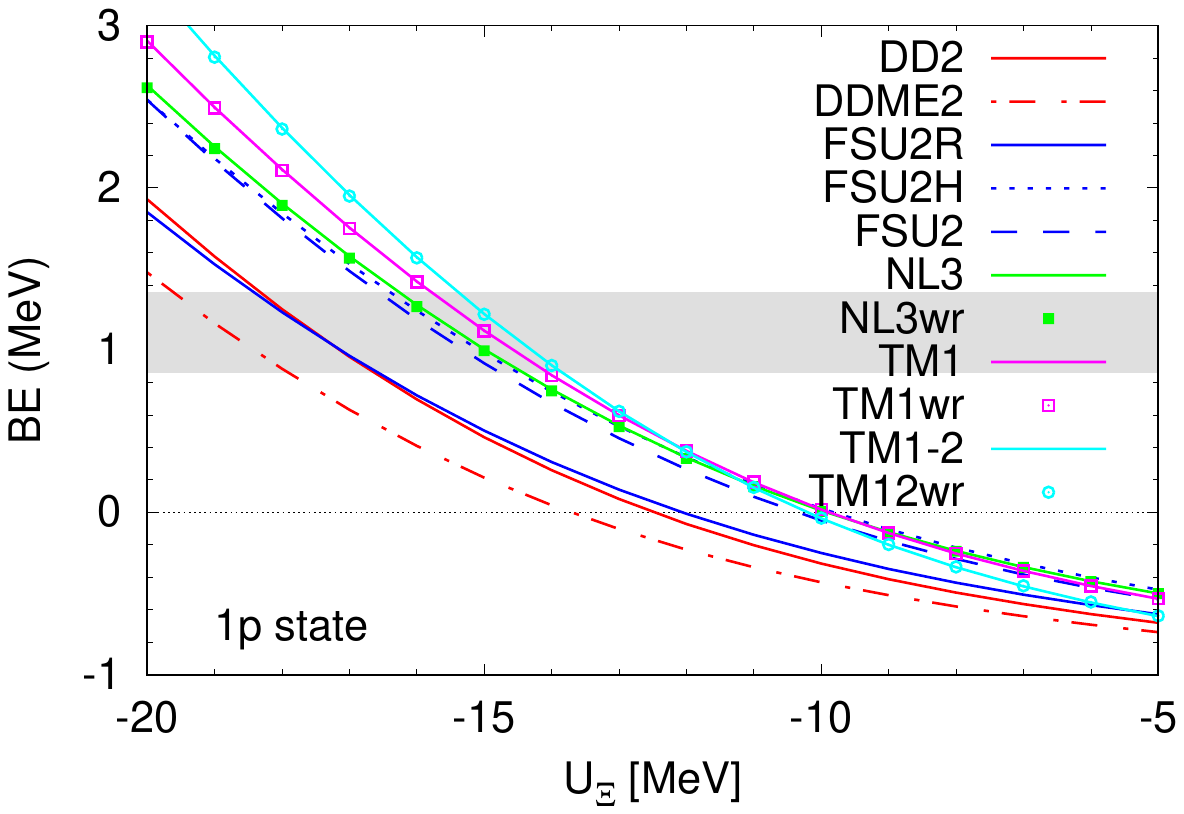}\\
\includegraphics[width=.4\linewidth]{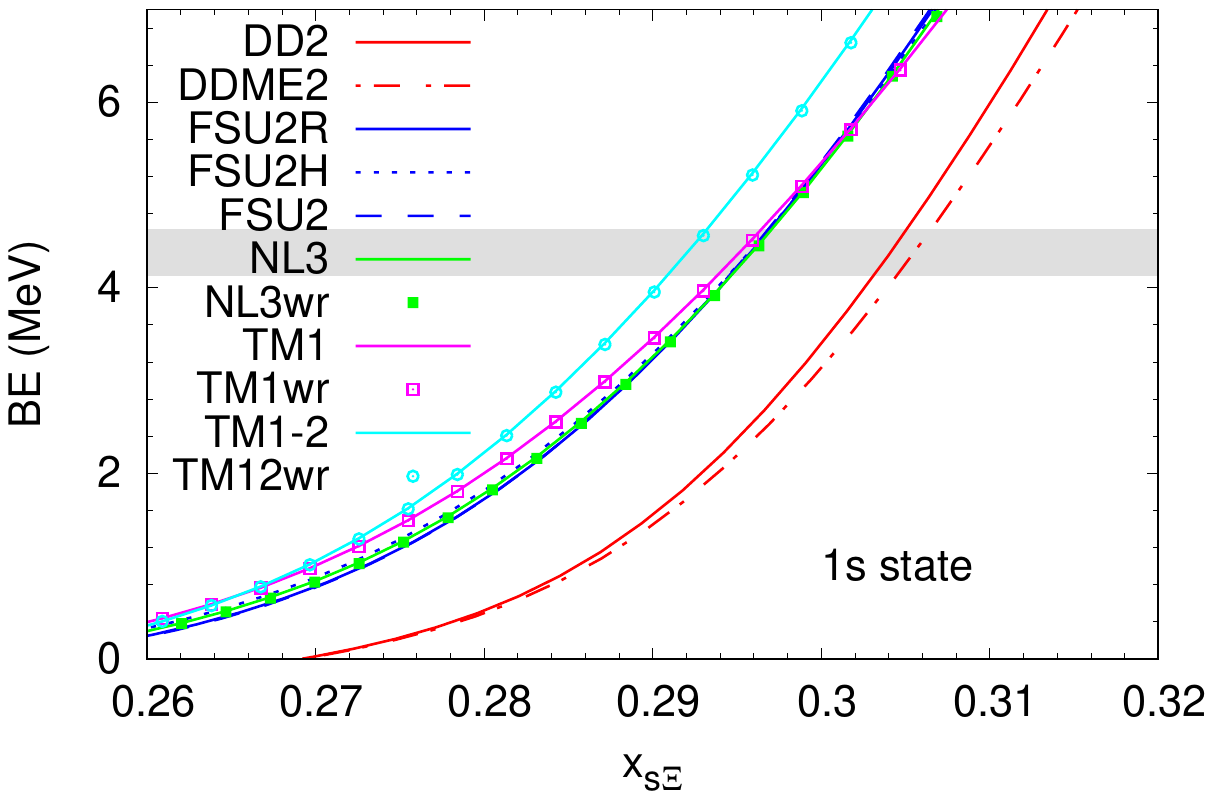}&
\includegraphics[width=.4\linewidth]{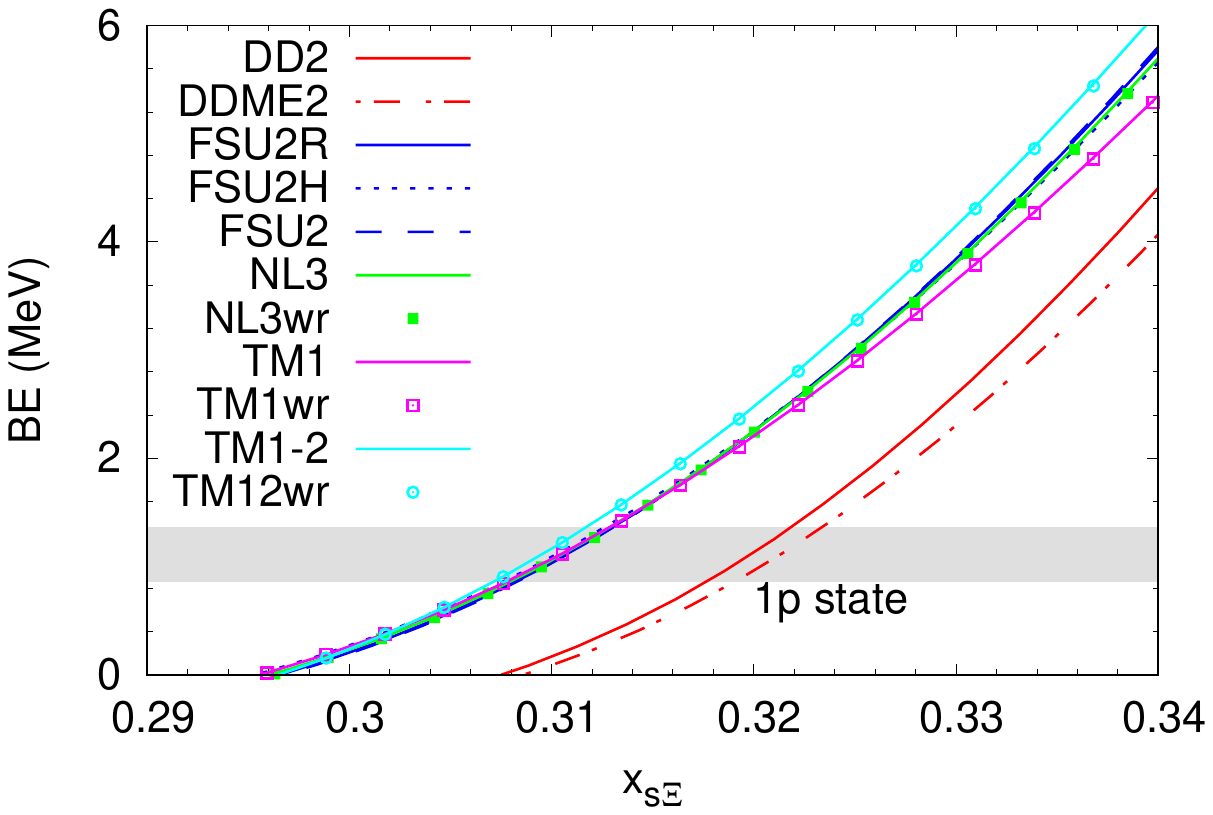}\\
\end{tabular}
\caption{Binding energy of $^{15}_{\Xi^-}$C as a function of
  $U_\Xi^{(N)}$ (top panels) and $x_{s\Xi}$ (bottom panels), under the assumptions
  that $\Xi^-$ occupies a $1s$ state (left panels) or, alternatively,
  a $1p$ state (left panels).
  Note that the scales are not the same in the different panels.
  Gray band: experimental binding energy of the Kiso event.}
\label{xi14A}
\end{figure*}

The binding energy of the $\Xi$-hyperon in the hypernucleus $^{15}_{\Xi^-}$C
is plotted in Fig. \ref{xi14A} as a function of the $U^{(N)}_\Xi$ potential
(top panels) and, respectively, $x_{s \Xi}$ (bottom panels).
Left (right) panels correspond to the assumptions according to which $\Xi$
occupies a $1s$ ($1p$) state. The different nucleonic effective interactions
presented in Sec. \ref{ssec:model} are considered.
A gray band identifies the binding energy obtained in the Kiso event
\cite{kiso}.
Some conclusions are in order:
a) if $^{15}_{\Xi}{\rm C}$ is in the ground state, {\it i.e.} $\Xi$ occupies the $1s$ state,
$-12.5 \lesssim U^{(N)}_\Xi \lesssim -8.7$ MeV, with the most attractive values
corresponding to the two DD models and FSU2R;
this corresponds to $0.295 \leq x_{s \Xi} \leq 0.306$; see Table \ref{tabBe},
b) if $^{15}_{\Xi}{\rm C}$ is in an excited state, {\it i.e.} 
$\Xi$ occupies the $1p$ state, $U^{(N)}_\Xi$ is more attractive,
$-18.8 \lesssim U^{(N)}_\Xi \lesssim -14.6$ MeV; again, the most attractive values
correspond to the two DD models and FSU2R;
the coupling constants have larger values, $0.31 \leq x_{s \Xi} \leq 0.32$.
The relative stability of $x_{s \Xi}$ to the modification of the nucleonic EoS, for
  each considered scenario, reflects the relatively small dispersion among the considered
  EoS, over the subsaturation density domain explored by a hyperon
  bound in a nucleus.
  Note that a similar situation corresponds, according to \cite{Fortin17,Fortin18,Providencia19},
  also to $x_{s \Lambda}$ and the explication is the same.

Contrary to what occurs for $\Lambda$ and $\Xi$, there is no hypernuclear
data on which the $\Sigma$N interaction can be tuned.
As a consequence we treat $g_{\sigma \Sigma}$ as a free parameter and vary its
values such as to explore $-10 \leq U^{(N)}_{\Sigma} \leq 40$ MeV.
We recall that, according to Ref. \cite{Gal2016}, 
$U^{(N)}_\Sigma(n_0) \approx 30\pm 20$ MeV.
As assumed for the couplings between $\Xi$ and $\omega$ and $\rho$ mesons
and for the same reasons,
$g_{\omega \Sigma}=2/3 g_{\omega N}$ and $g_{\rho \Sigma}=g_{\rho N}$.

\section{Properties of hypernuclear compact stars}
\label{sec:prop}

In the following we discuss the properties of hypernuclear compact stars
built upon the calibrated EoS discussed in Sec. \ref{sec:eos}.
In particular, we analyze the effect of the different nucleon effective
interactions and $U^{(N)}_{\Sigma}$-potential values
on the onset and abundances of hyperons as well as
on the maximum mass, radii, tidal deformability and moment of inertia.

\begin{figure}[th]
\begin{tabular}{cc}
\includegraphics[width=.95\linewidth]{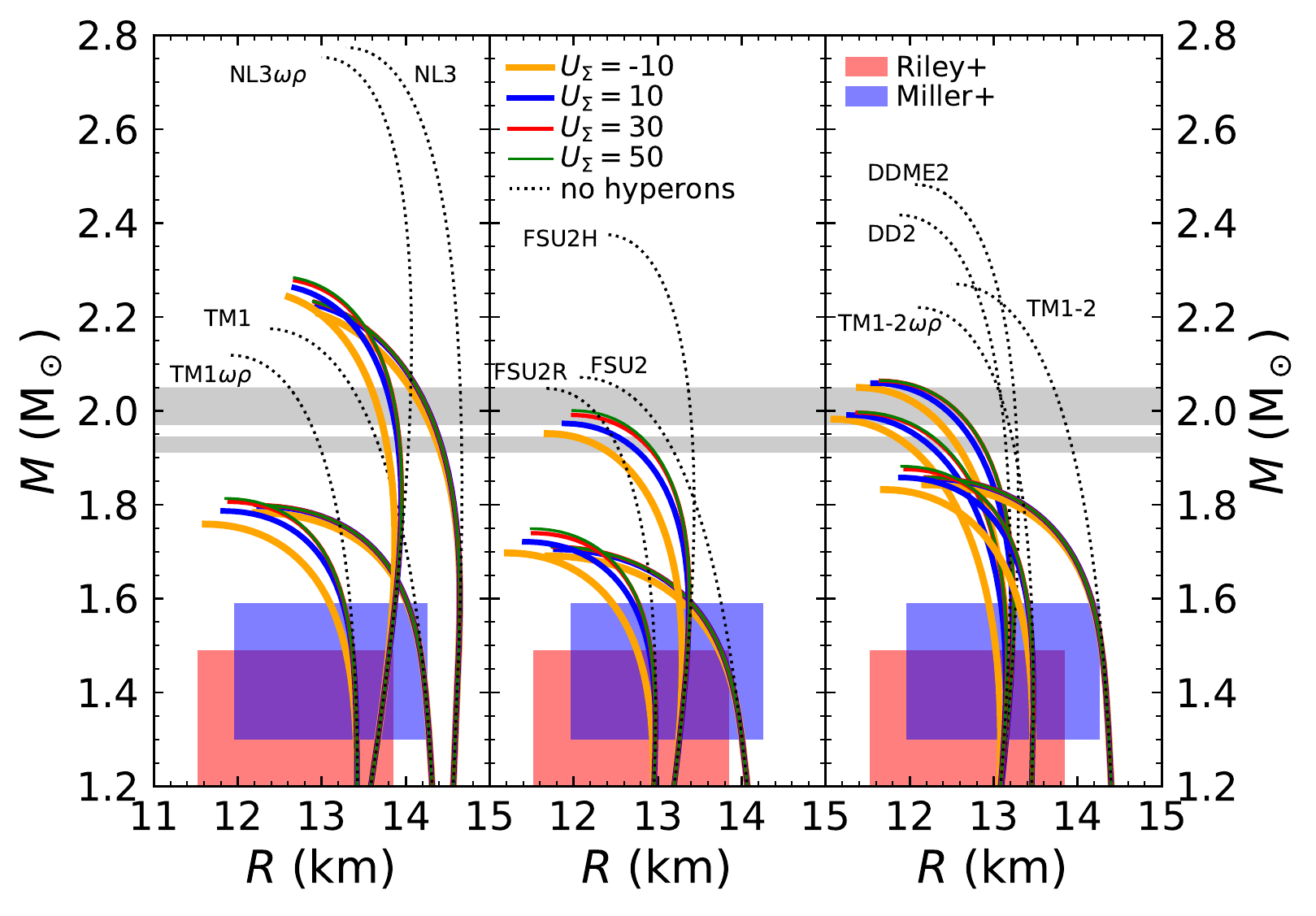}\\
\includegraphics[width=.95\linewidth]{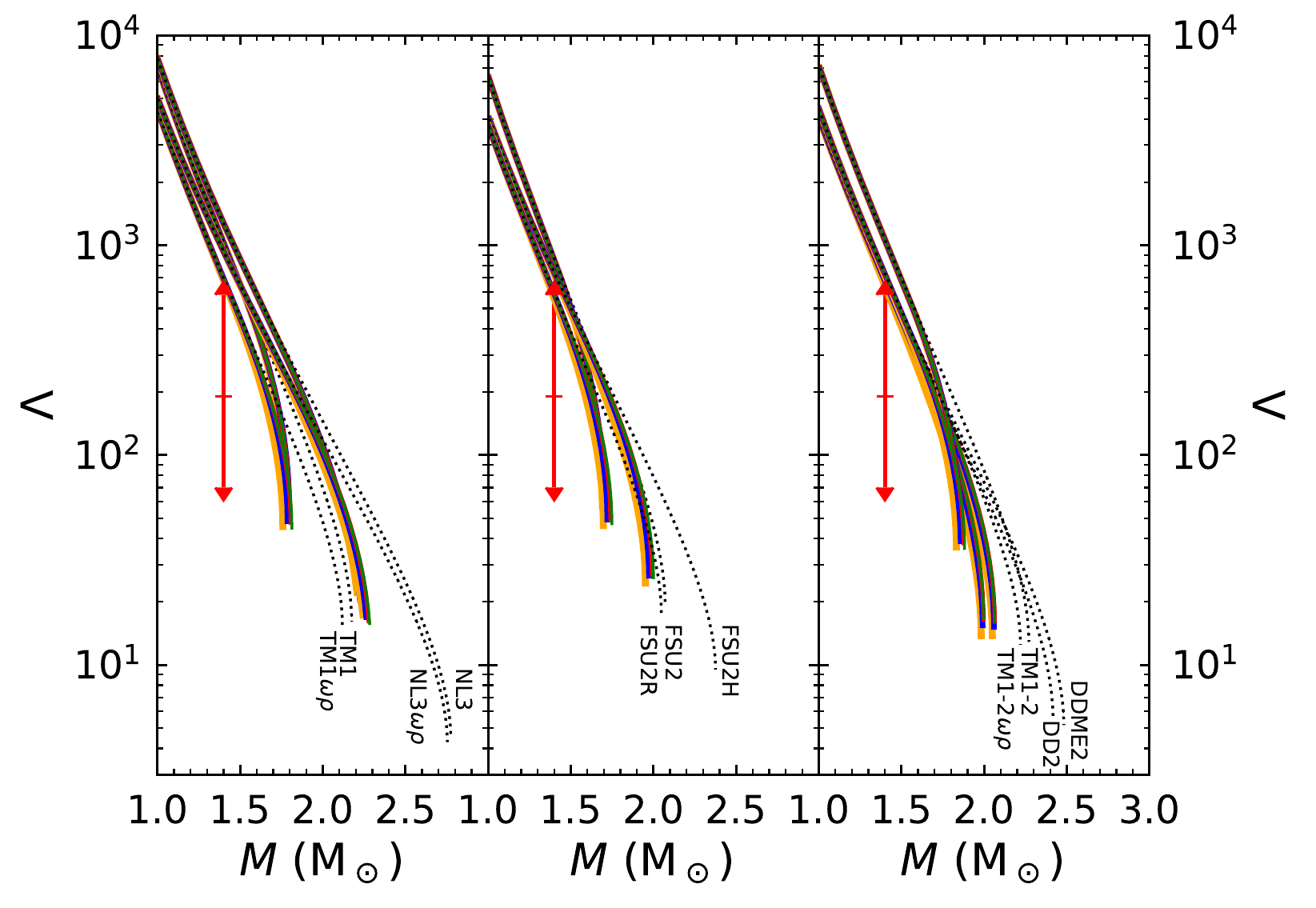}\\
\includegraphics[width=.95\linewidth]{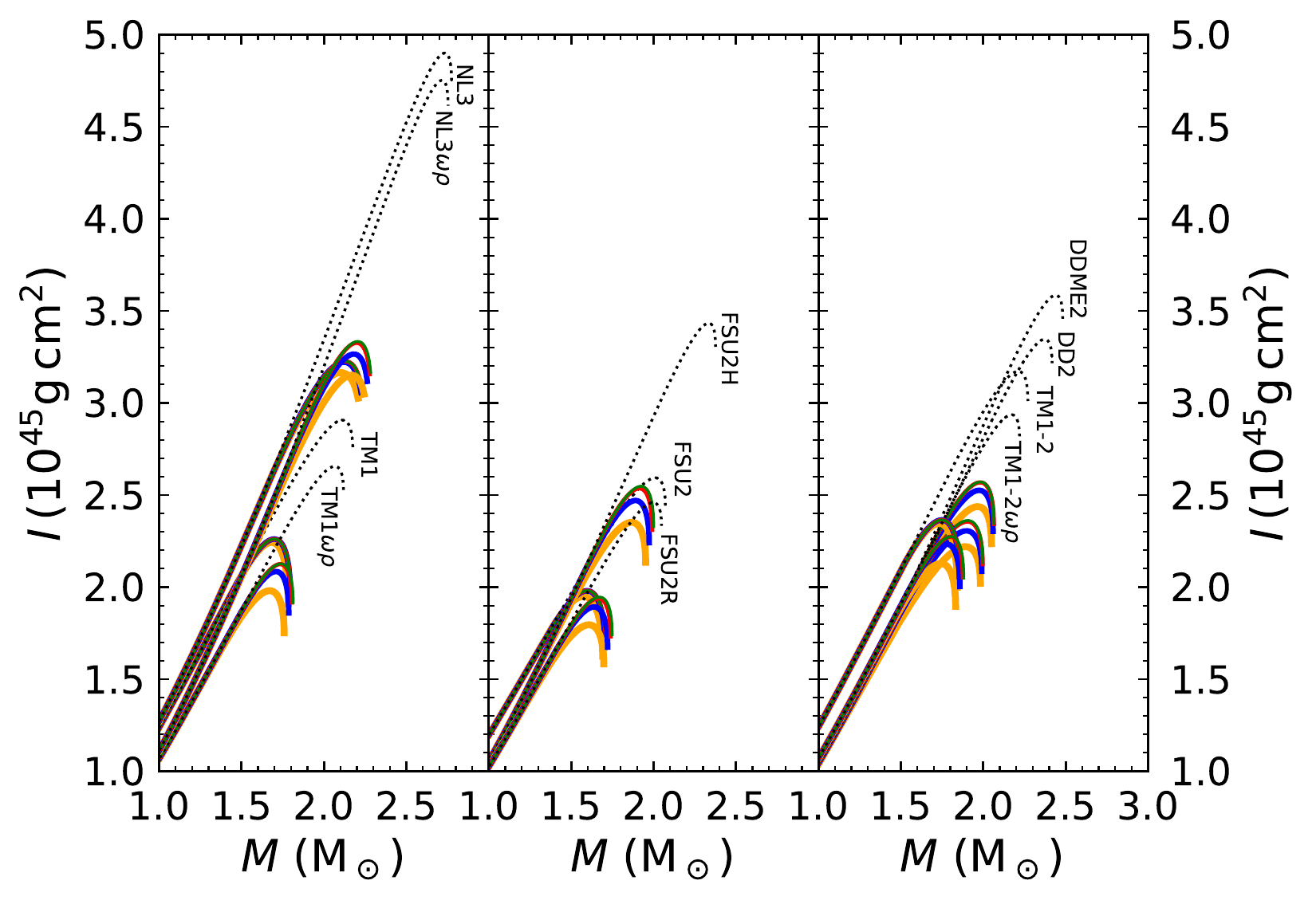}\\
\end{tabular}
\caption{Mass-radius diagrams (top panels)
  and tidal deformabilities (second middle panels) and
  moments of inertia (third bottom panels)
  as a function of the star mass expressed in units of solar masses,
  for hypernuclear stars with calibrated $\Lambda$ and $\Xi$ interactions.
  The different nucleonic effective interactions
  discussed in Sec.~\ref{ssec:model} are considered.
  Predictions corresponding to different values of $U^{(N)}_{\Sigma}$ are plotted
  with different colors: -10 (gold), 10 (blue), 30 (red) and 50 MeV (green) and line thicknesses: the thiner the  line
the more repulsive the potential.  Predictions corresponding to purely nucleonic stars are illustrated
  by dotted curves.
  The horizontal bands on $M-R$ diagrams correspond to the mass measurements of PSR J$1614-2230$ and J$0348+0432$ with a 1-$\sigma$ uncertainty. The colored rectangles correspond to the mass and radius constraints at the $1-\sigma$ level obtained for PSR J0030$+$0451 by two teams analyzing NICER x-ray data \cite{Riley19,Miller19}.
  For indication, the red vertical bars on $\Lambda-M$ correspond to limits obtained on the
tidal deformability of a $1.4M_\odot$ NS, $70<\Lambda_{1.4}<580$, as derived
from the observation of GW170817 by the LVC collaboration \cite{Abbott18}, using the  waveform model PhenomPNRT.}
\label{fig:star}
\end{figure}

\begin{figure}[th]
\begin{tabular}{cc}
  \includegraphics[width=.95\linewidth]{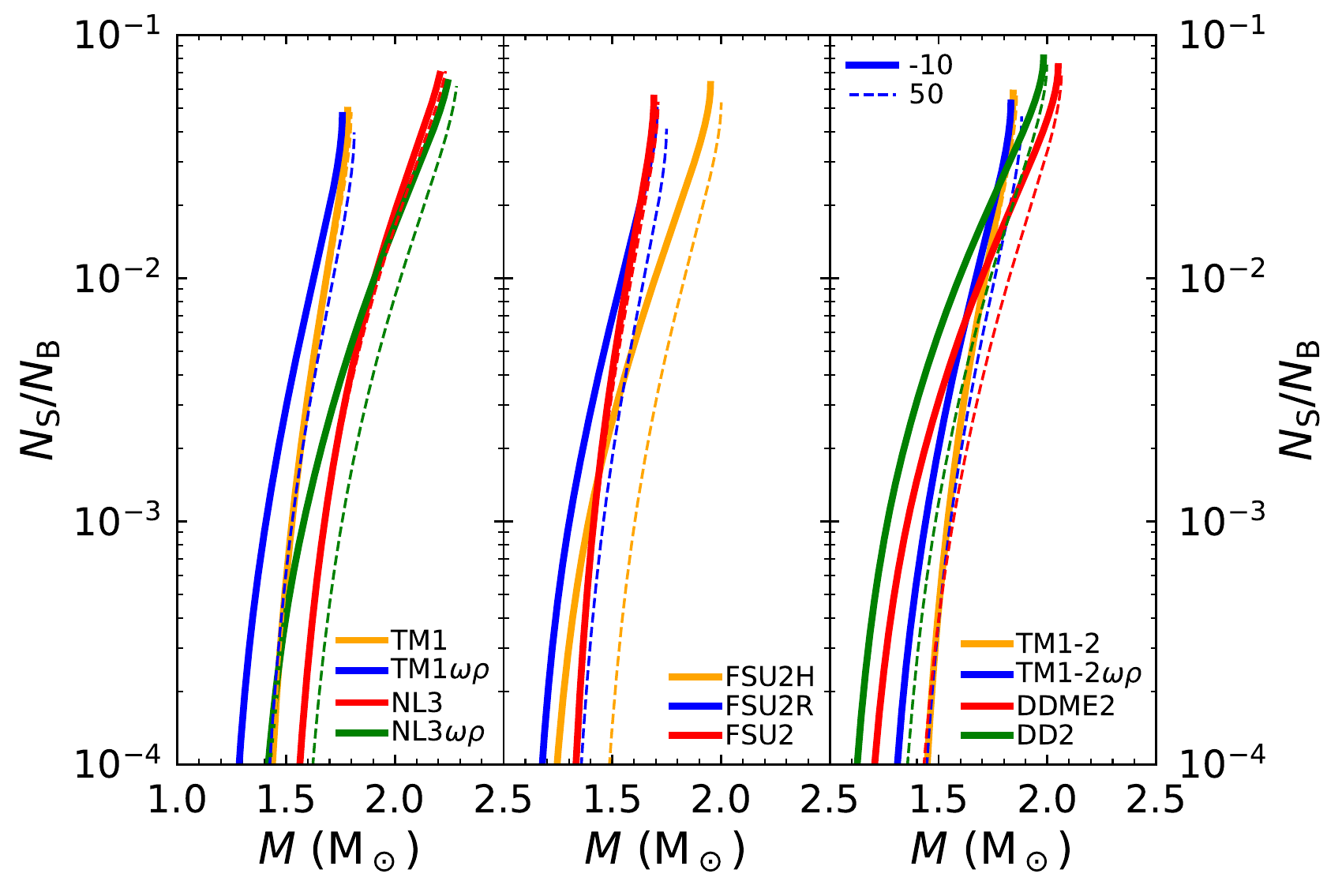}\\
  \end{tabular}
\caption{
  Strangeness fraction as a function of the star mass
  expressed in units of solar masses for the same models
  considered in Fig. \ref{fig:star}. The thick lines correspond to  $U^{(N)}_\Sigma=-10$ MeV and the thin ones to $+50$ MeV.}
\label{fig:strangeness}
\end{figure}

In the following we use unified EoS for neutron star matter for the 11
parametrizations in this work. We employ crust models computed
consistently with the core \cite{Fortin16,Providencia19} following the
approach presented in the first reference. EoS tables are available as supplemental material to this paper assuming that the $\Xi^-$ hyperon in $^{15}_\Xi{\rm C}$ of the Kiso event is in a $1p$ state, for each of the 11 RMF models and four values of the $U_\Sigma$ potential: $-10, 10, 30$ and 50 MeV.\\

For a spherical star in hydrostatic equilibrium we determine the mass-radius relation
by solving the Tolman-Oppenheimer-Volkoff \cite{TOV1,TOV2} equations. The radius of some NSs has been determined, most recently by two teams who modeled the pulsed x-ray emission of the millisecond pulsar PSR J0030+0451 \cite{Riley19,Miller19}. However uncertainties in the derived constraints and in the modeling of the source itself are still large and so far no strong constraint on the radius of NS has been  obtained (see e.g. \cite{Miller16,Haensel16}). However the determination of the radius of few NSs with a precision of a few percent is expected from the currently operating NICER mission \cite{NICER} and also from future x-ray observatories like Athena x-ray telescope \citep{Athena} and eXTP \cite{eXTP}.

The moments of inertia $I$ and tidal deformabilities $\Lambda$ are calculated
as following Refs. \cite{Hartle_ApJ_1967,Bejger05} and Ref. \cite{Hinderer2008}, respectively. Some constraints on the deformabilities of the two NS that composed the binary before they merge during the GW170817 event have been obtained thanks to the multimessenger observations (see Refs. \cite{Abbott19,Abbott18} for the latest results), and many more are expected in the near future from the current and future observational runs of the LVC collaborations. As far as the moment of inertia is concerned, no measurement has been obtained so far. However it could be measured in a binary of two radio pulsars, such as PSR J0737-3039. It could not be achieved for this system so far as the radio beam of one of the NSs cannot be observed anymore due to precession. With more observations with current radiotelescopes and future ones like the SKA \citep{SKA}, the number of known pulsars is expected to increase by orders of magnitude, including many thousands of millisecond pulsars, and among them possibly binary systems with two pulsars.

Figure \ref{fig:star} gives a general summary
of the properties of the NSs built upon the considered
models. Regarding observational constraints on the
  maximum mass we note that, thanks to radio observations of the Shapiro delay, the mass of MSP J$0740+6620$ has been determined to be $2.14^{+0.20}_{-0.18}M_\odot$ at a 2-$\sigma$ level (equivalently 95.4\% credibility interval) i.e. $1.96<M/M_\odot< 2.34$ \cite{Cromartie2019}.
  We consider that the uncertainty in this measurement is still
too large to use this mass as a strong constraint.  Let us recall
indeed that the mass of PSR J1614$-$2230 inferred from Shapiro delay
was initially determined to be $1.97 \pm0.04M_\odot$ ($1-\sigma$
level) \cite{Demorest10}. After more data were accumulated this
number went down to  $M = 1.908\pm0.016M_\odot$ (also $1-\sigma$
level) which is compatible with the previous reported mass at more
than 1 standard deviation. Consequently in this work we only
consider and plot the mass constraints from PSR J$1614-2230$ and
J$0348 + 0432$. We note that among all the models considered in this
paper, DD2, DDME2, FSU2H, NL3 and NL3$\omega\rho$ are compatible with
the mass constraint from these objects. We also show in the $M-R$ plot
the mass and radius constraints at the $1-\sigma$ level obtained for
PSR J0030+0451 by two teams analyzing NICER data
\cite{Riley19,Miller19}. The stiffest EoS NL3, which is already ruled
out because of its too large slope of the symmetry energy at saturation, is not compatible with these measurements, while TM1 and TM1-2 are marginally consistent.

In the middle panel of Fig.~\ref{fig:star} we plot for indication the limits imposed on $\Lambda_{1.4}$ taken from Ref. \cite{Abbott18}. These have been deduced from the effective $\tilde \Lambda$ obtained within the waveform model PhenomPNRT at a 90\% confidence level. None of the models we use satisfy this constraint. However, in Ref.  \cite{Abbott19} the authors show  the
dependence of the analysis of the GW170817  observation on the waveform model, and, in particular, the TaylorF2 model predicts effective tidal deformabilities larger by $\sim 100$ . New observations are needed to impose stricter constraints.

Figure \ref{fig:strangeness} represents strangeness fractions $N_S/N_B$
as a function of the star mass.
Baryonic and strangeness numbers entering the definition of the
strangeness fraction are defined as
\begin{eqnarray}
  N_B&=&4\pi\int dr \frac{n_i \, r^2}{\sqrt{1-2 m(r)/r}}, \\
  N_S&=&\frac{4\pi}{3}\int dr \frac{q_{Si} n_i \, r^2}{\sqrt{1-2 m(r)/r}},
  \nonumber
\end{eqnarray}

where $n_i$  and  $q_{Si}$ stand for particle number density and, respectively,  strangeness charge of particle $i$, and $m(r)$ denotes the gravitational
mass corresponding to the radial coordinate $r$. 
The dependence of NS properties and strangeness composition on the magnitude
of the $\Sigma$-N interaction potential is illustrated by using
different colors and thicknesses, the thinner the  line
the more repulsive the potential.
It comes out that the most important role is played by the
nucleonic sector. The reason is that nucleons represent the dominant component.
Quite remarkably, also the modifications brought by nucleation of
$\Sigma$-hyperons and the associated value of $U^{(N)}_{\Sigma}$ show strong
dependence on the nucleonic EoS. 
For instance, models with large values of the slope of the symmetry energy
({\it e.g.} TM1, TM1-2, NL3 and FSU2) show very little sensitivity of geometric, deformability
and chemical composition to the value of $U^{(N)}_\Sigma$.
At variance with them, models with moderate $L$ values lead to
smaller NS radii and masses, when attractive or less repulsive $\Sigma$N
potentials are assumed.
From Fig. \ref{fig:strangeness} one may see that the maximum strangeness fraction
reached in NS cores does not depend on $U^{(N)}_\Sigma$.
The strangeness related quantity which does depend on $U^{(N)}_\Sigma$ is
the density and, implicitly, the NS mass where strangeness sets in.
As is easy to anticipate, attractive or less repulsive potentials favor early
nucleation of $\Sigma$.

In Table \ref{tabii}, we have compiled, for each model and each value
of the $U^{(N)}_\Sigma$ potential the information covering several properties
of neutron stars: maximum mass and respective central baryonic number density,
onset density of the three hyperonic species and
threshold densities of nucleonic and various hyperonic dUrca
channels. Also given are the NS masses with
central densities equal to these values. 

\begin{figure}[th]
\begin{tabular}{cc}
  \includegraphics[width=1\linewidth]{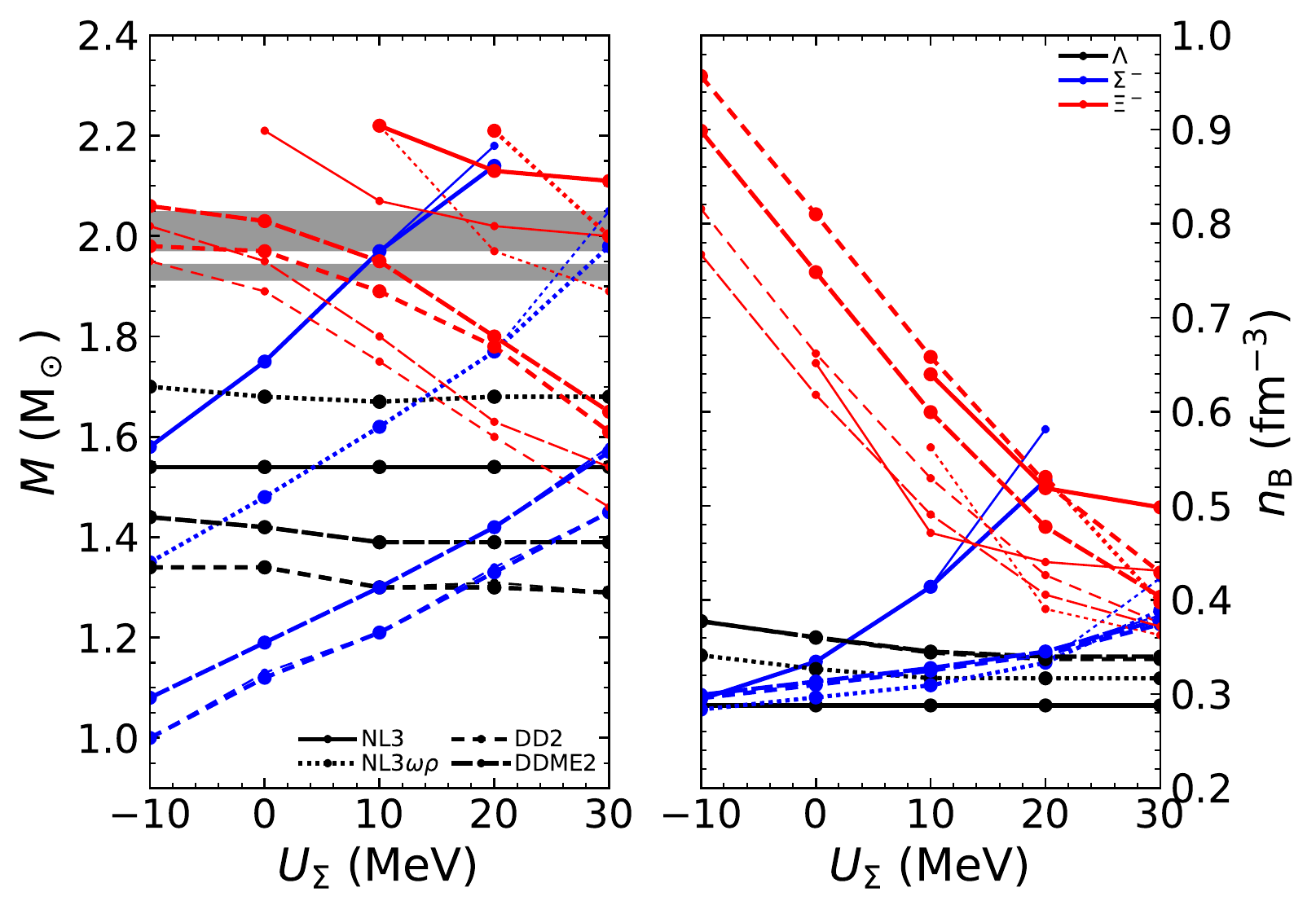}\\
  \includegraphics[width=1\linewidth]{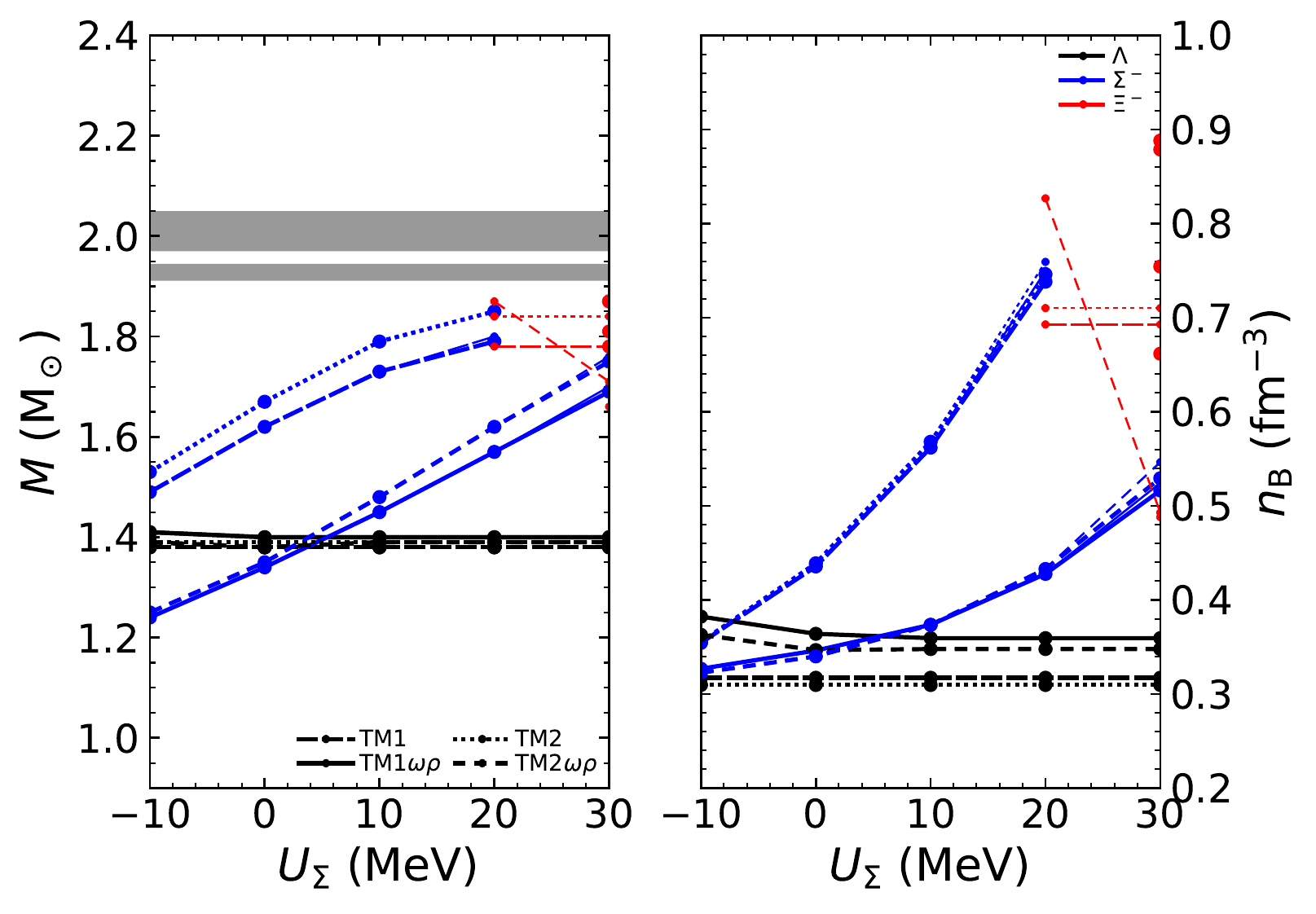}\\
  \includegraphics[width=1\linewidth]{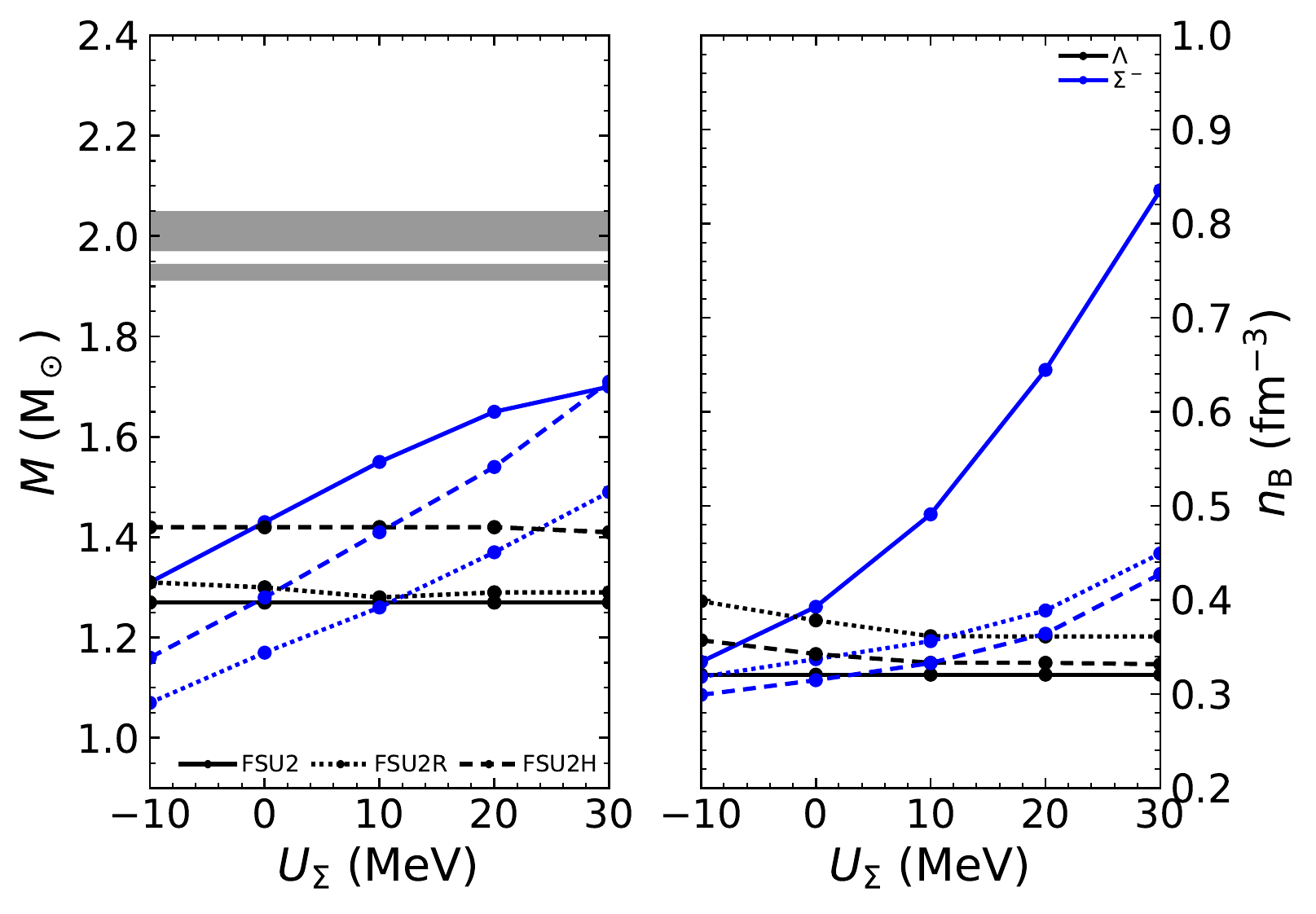}\\    
  \end{tabular}
\caption{$U^{(N)}_\Sigma$ dependence of the onset density (right panel)
  and corresponding NS mass  with this central density (left panel) of the 
hyperonic species that nucleate in NS cores.
  The same models as in Fig. \ref{xi14A} are considered. The horizontal gray strips on 
  the left panel correspond to the mass measurements of PSR J$1614-2230$ and J$0348+0432$ 
  with a 1-$\sigma$ uncertainty. }
\label{fig:Hyperons}
\end{figure}

\begin{figure}[th]
\begin{tabular}{cc}
  \includegraphics[width=1\linewidth]{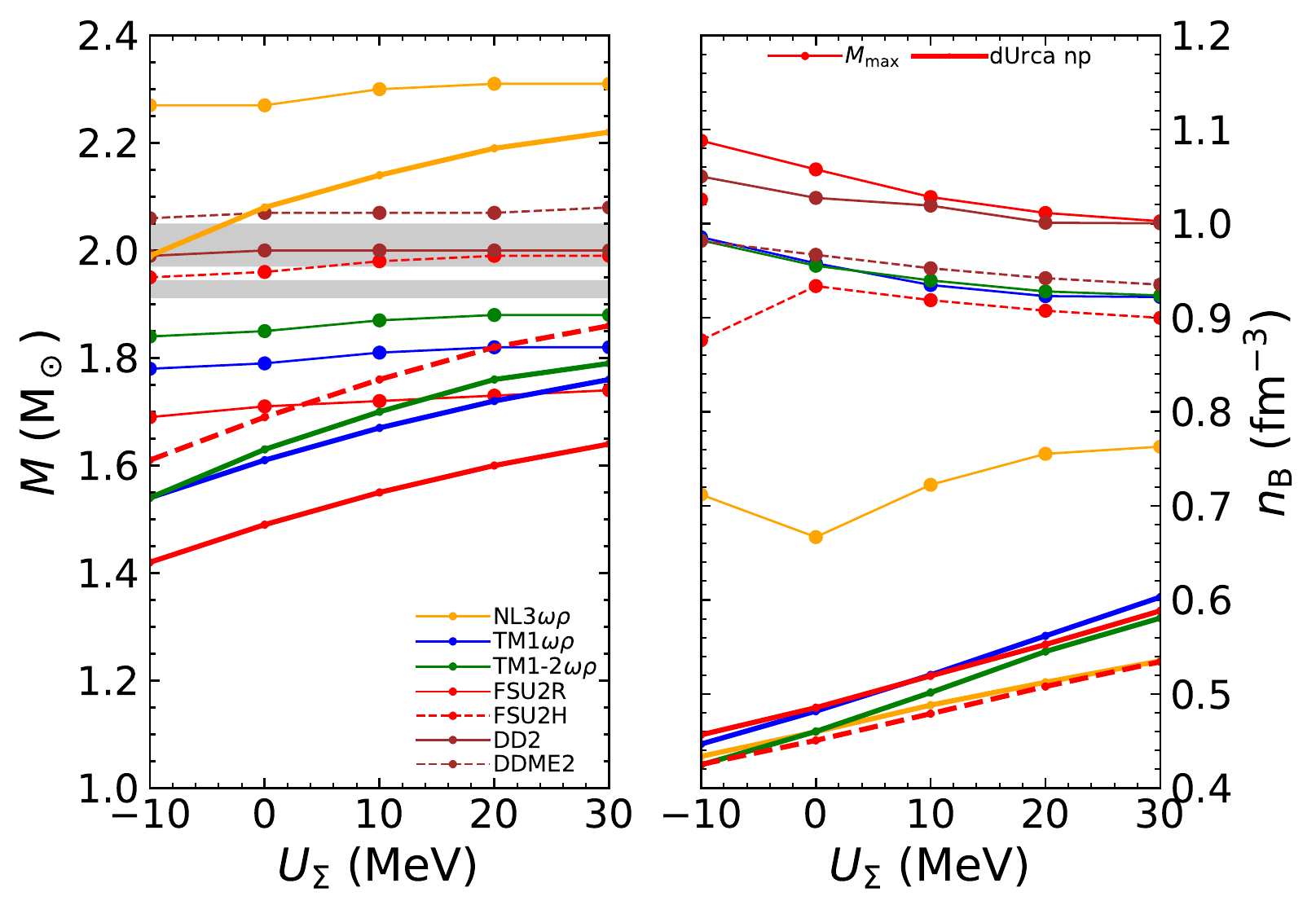}\\
  \end{tabular}
\caption{Left: dependence on $U^{(N)}_\Sigma$ of the NS maximum mass $M_{\rm max}$ (thin lines) and of the minimum NS mass which allows nucleonic dUrca to operate (thick lines).  The horizontal gray strips correspond to the mass measurements of PSR J$1614-2230$ and J$0348+0432$ with a 1-$\sigma$ uncertainty.  Right: dependence on $U^{(N)}_\Sigma$ of the central density at the maximum mass (thin dashed lines) and of the onset density of nucleonic dUrca (thick solid lines).}
\label{fig:DU}
\end{figure}

In the following, we discuss these results. 

\begin{itemize}

\item Some of the considered models, {\it e.g.} FSU2R, FSU2, TM1,
  TM1$\omega\rho$, TM1-2 and TM1-2$\omega\rho$, are not able to reach
  the 2$M_\odot$ lower bound of maximum NS mass, when hyperons are introduced.

\item Irrespective of the nucleonic EoS, the only hyperonic species that are
  present are $\Lambda$, $\Sigma^-$ and $\Xi^-$. The explanation relies
  on the attractive character of $\Lambda$N- and $\Xi$N interactions
  and dominance of negatively charged particles. Note that other models
  of hypernuclear compact stars allow also for $\Xi^0$ \cite{Fortin16}.
  
\item For most of the NL models studied here only two species of hyperons nucleate
  in the core, the $\Lambda$ and $\Sigma^-$ hyperons. However for repulsive enough
   $U^{(N)}_\Sigma$ potential the $\Xi^-$ nucleates either instead of the $\Sigma^-$ 
   for the TM1, TM1-2 and NL3 models or in addition to it. For DD models the three 
   hyperons $\Lambda$, $\Sigma^-$  and  $\Xi^-$ appear, the latter species the last 
   except for all models but one.

\item For models with a large $L$, {\it e.g.} TM1, TM1-2, NL3 and FSU2, the nucleonic dUrca
  is insensitive to the onset of hyperons. The reason is that it becomes
  active well before the onset of any hyperon species. We recall that, the nucleonic dUrca process
   corresponds to the neutron $\beta$-decay followed by the electron capture on the proton:
$n \rightarrow p + e^- + \bar{\nu}_e$  and $p+e^- \rightarrow n+\nu_e$,
which  operates when the Fermi momenta of involved baryons and charged
lepton verify the triangle inequality:
$p_{F,i}+p_{F,j} \geq p_{F,k}$ \cite{DU91}. This latter condition also applies to the hyperonic dUrca processes \cite{DUY92}.

\item With the exception of the above cited models, attractive
  $U^{(N)}_\Sigma$ potentials modify NS composition such that the nucleonic
  dUrca threshold is shifted to lower densities. 
  
\item NL and DD models provide different results in regards to the
  nucleonic dUrca process. More precisely, DD models either completely
  forbid this process or allow for it at densities beyond the central density
  of the maximum mass, which is equivalent with saying that it never operates.
  All NL models allow for nucleonic dUrca. Depending on the value of the symmetry
  energy it starts to operate at densities slightly above $n_0$ or several times $n_0$.

 \item DD models only allow for hyperonic dUrca \cite{DUY92}.
   The allowed processes are $\Lambda \to p+e+\tilde \nu_e$,
   $\Sigma^- \to \Lambda+e+\tilde \nu_e$
   and, for repulsive values of $U^{(N)}_\Sigma$,
   also $\Xi^- \to \Lambda+e+\tilde \nu_e$
   (not mentioned in Table \ref{tabii})
   For strongly repulsive $U^{(N)}_\Sigma$, $\Lambda \to p+e+\tilde \nu_e$
   sets in much before $\Sigma^- \to \Lambda+e+\tilde \nu_e$.
   For less repulsive $U^{(N)}_\Sigma$ the two processes have close
   density thresholds and, thus, compete.
   For $U^{(N)}_\Sigma \geq 20$~MeV, $\Xi^- \to \Lambda+e+\tilde \nu_e$
   sets in at densities of the order of $0.37-0.46~{\rm fm}^{-3}$, which
   corresponds to $1.46 M_{\odot} \leq M \leq 1.66 M_{\odot}$.
   
 \item in NL models with $L\sim 50-60$ MeV the hyperonic dUrca starts
   operating before the nucleonic dUrca. Repulsive $U^{(N)}_\Sigma$ values
   shift $\Lambda \to p+e+\tilde \nu_e $ to lower densities.
  
 \item $U_\Sigma$ defines the process that first operates:
   a less repulsive potential favors $\Sigma^- \to \Lambda+e+\tilde \nu_e$,
   which is 20 times more efficient than $\Lambda \to p+e+\tilde \nu_e$
   \cite{DUY92}.
   
 \item With the exception of NL3$\omega\rho$, under the assumption
   that $U^{(N)}_\Sigma$ is repulsive,
   the $\Lambda \to p+e+\tilde \nu_e$ process starts operating
   at $n \approx 2 n_0$, which corresponds to 
   $M/M_{\odot} \approx 1.3 - 1.4 M_\odot $. The relative stability
   of this threshold is attributable to the constraints imposed
   to nuclear matter around the $n_0$ and $\Lambda N$ potential.
\end{itemize}

Figures \ref{fig:Hyperons} and \ref{fig:DU} illustrate the dependence of some
quantities reported in Table \ref{tabii} on the nucleon effective interaction and
$U^{(N)}_\Sigma$ potential. These are the densities at which the
three
hyperonic species nucleate in NS matter and the corresponding NS masses
(Fig. \ref{fig:Hyperons}) and, respectively,
the density and corresponding NS masses where nucleonic dUrca becomes active
(Fig. \ref{fig:DU}).
Figure \ref{fig:Hyperons} shows that the onset density of $\Lambda$s depends little
on $U^{(N)}_\Sigma$ and the nucleonic EoS.
The explanation of the first feature is that, with the exception of attractive
$U^{(N)}_\Sigma$ values and FSU2H, $\Lambda$ onset before $\Sigma$.
The explanation of the second one is that, up to $n_{\Lambda}$,
the nucleonic EoS is relatively well constrained.
Despite small dispersion on $n_{\Lambda}$, the mimimum NS mass that accommodates $\Lambda$s
varies over $0.4M_{\odot}$. The size of this interval reflects the
integrated variation among the EoS, especially in the isovector channel, up to
$n_{\Lambda}$.
Nucleation of $\Sigma$ depends much on both $U^{(N)}_\Sigma$ and
nucleon-nucleon effective interactions. In terms of density the domain of variation
is $\approx 0.5$ fm$^{-3}$ wide, while in terms of NS masses it is $\approx 1.2M_{\odot}$.
Nucleation of $\Xi$, which appear only in some models, is inversely correlated with that
of $\Sigma^-$. Even larger uncertainties affect these latter quantities
and the explanation is obviously due to the increased uncertainties that affect
the EoSs as the density increases.
Figure \ref{fig:DU} illustrates the already discussed huge dispersion that
concerns the onset density and mass of the nucleonic dUrca process.
The variation of this quantity with the nucleon-nucleon effective interaction stems
from the isovector channel. The variation with $U^{(N)}_\Sigma$ reflects the way in which
negatively charged particles affect the whole composition of matter and, implicitly,
the relative abundances of neutrons, protons and electrons. 

Modification of chemical composition is expected to impact the thermal evolution
of isolated and accreting neutron stars. The extent to which one may
constrain the effective interactions from effective surface
temperature will be
considered in a future work.

\begin{figure}[th]
\includegraphics[width=.75\linewidth]{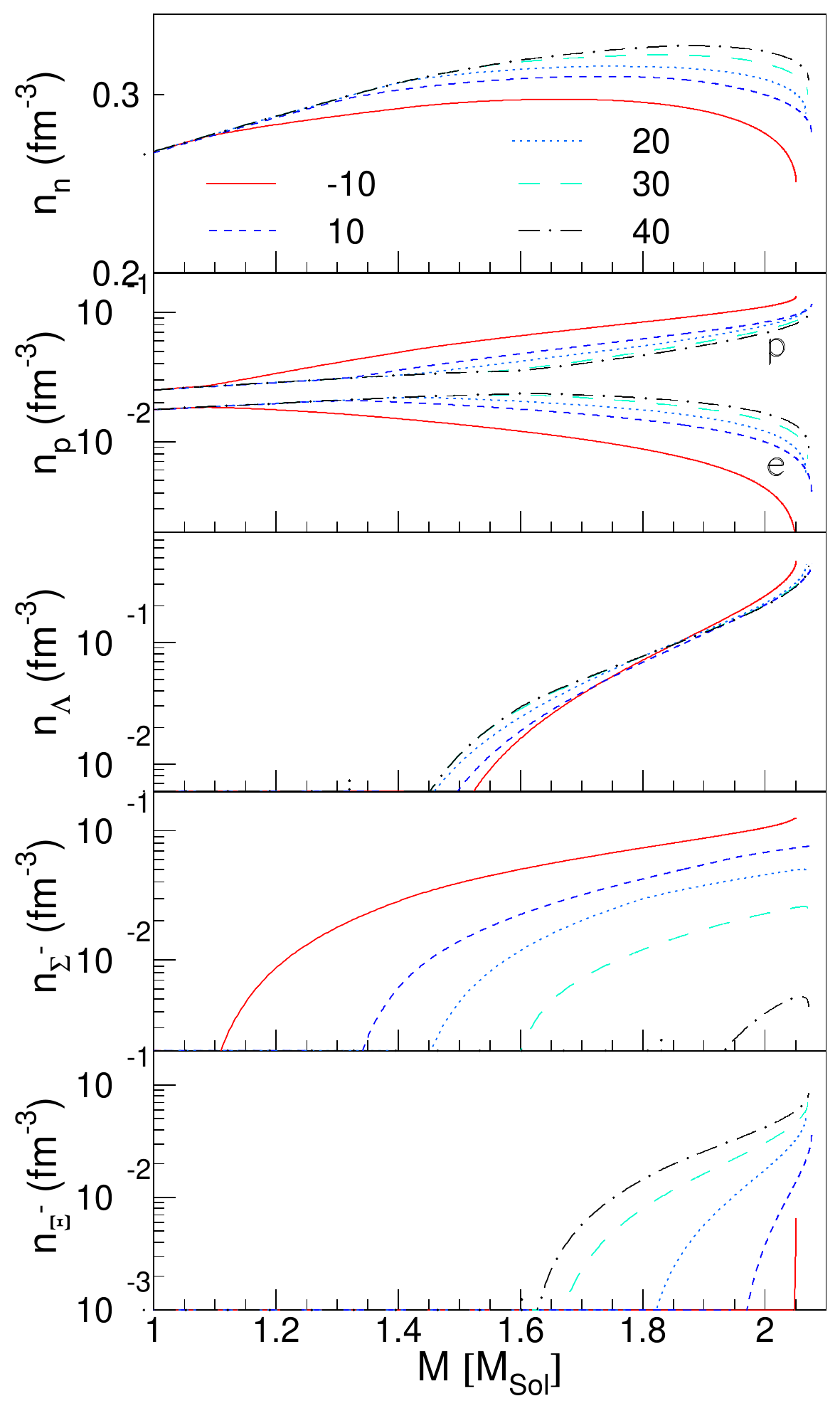}\\
\caption{For the DDME2 model, from top to bottom, partial densities of neutrons, protons, $\Lambda$s, $\Sigma$s and
  $\Xi$s in the center of the star as a function of the gravitational
  mass in units of solar masses. The different curves correspond to
  different values of the $U^{(N)}_\Sigma$, as mentioned in the key legend
  (in MeV).
  Also plotted in the second panel are the electron densities.  }
\label{fig:dens}
\end{figure}

With the aim of understanding how $U^{(N)}_\Sigma$ affects NS chemical composition
and, consequently, the dUrca threshold Fig.~\ref{fig:dens} illustrates,  for the DDME2 parametrization,
the individual central particle number densities $n_{\rm i}$
as a function of the gravitational mass (in units of solar masses)
for different values of $U^{(N)}_\Sigma$ between -10 and 40 MeV.
The considered species are: neutrons, protons, electrons,
$\Lambda$s, $\Sigma$s and $\Xi$s.  Attractive or less repulsive values
of $U^{(N)}_\Sigma$ favor the onset of $\Sigma^-$ at lower values of $n_B$.
By partially replacing the electrons, which compensate the positive electric charge
of protons, $\Sigma^-$s modify both neutron and proton densities, as the first two
panels confirm. Smaller values of $n_n$ together with larger values of $n_p$
act in the direction of allowing the nucleonic dUrca to operate at lower densities.
Chemical equilibrium with unconserved strangeness asks that $\Lambda$ chemical potential
is equal with the neutron one. As a result, $n_{\Lambda}$ qualitatively follows
the evolution of $n_n$. Quantitatively, the less abundant $\Lambda$s are less
affected than the more abundant neutrons.
Based on similar arguments one could expect that $n_{\Xi^-}$ follows $n_{\Sigma^-}$.
The bottom panel of Fig. \ref{fig:dens} shows the opposite; {\it i.e.}
the more attractive $U^{(N)}_\Sigma$ is the smaller the amount of $\Xi$-hyperons.
The effect is attributable to the net neutrality condition, where the role
of $\Xi$s is overtaken by $\Sigma$s.

\begin{figure}[t!]
\begin{tabular}{cc}
\includegraphics[width=1\linewidth]{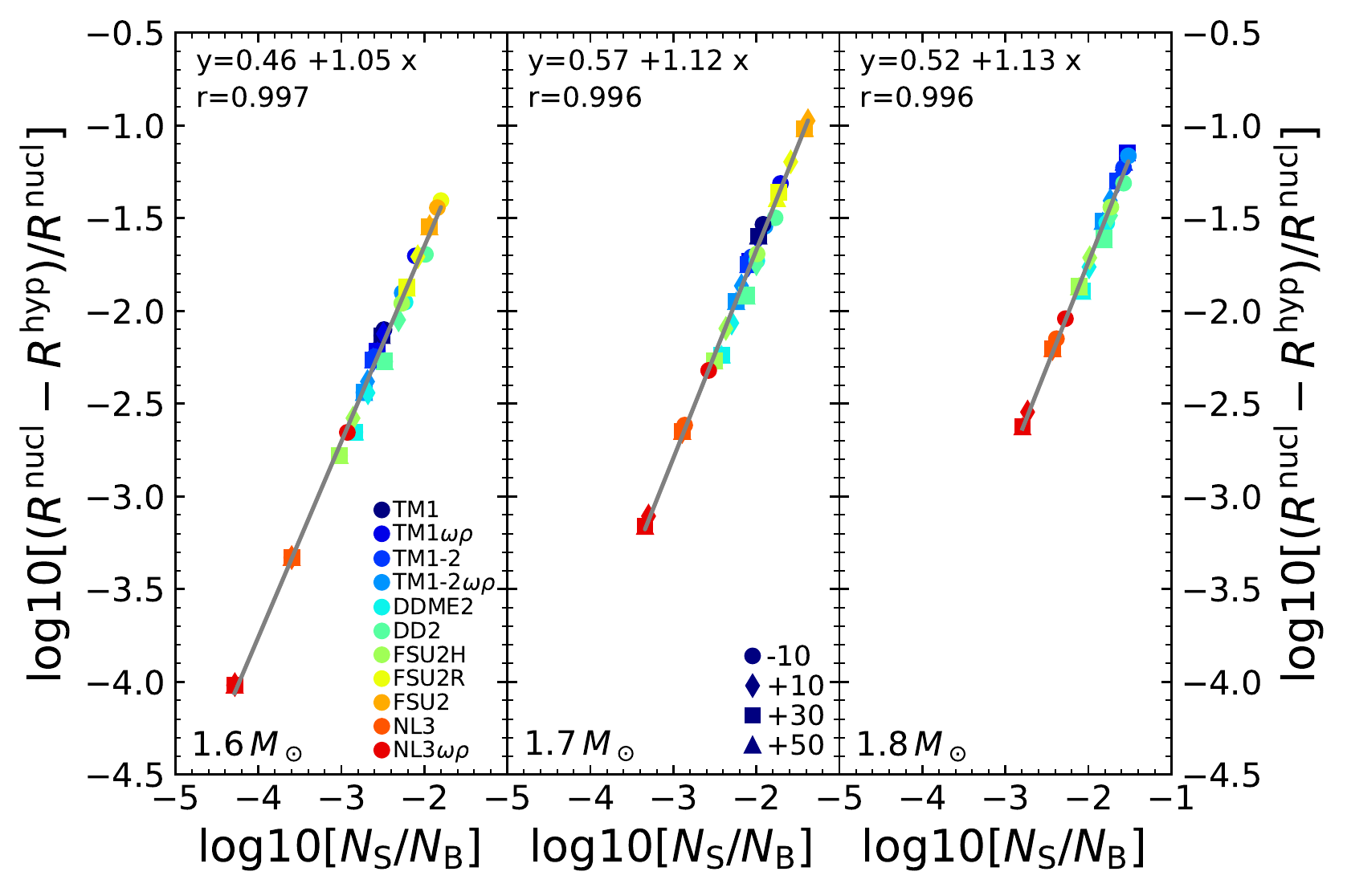}\\
\includegraphics[width=1\linewidth]{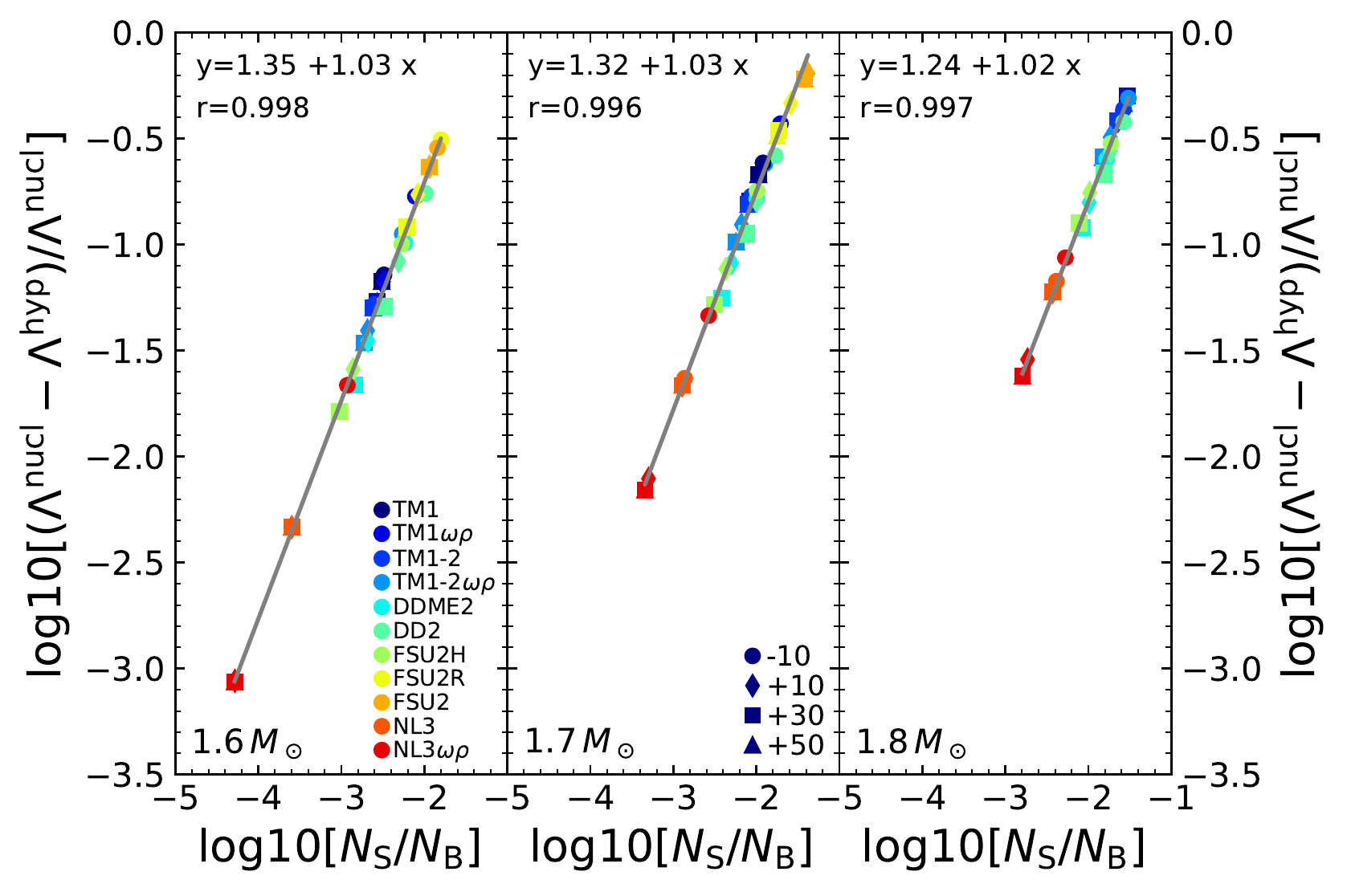}\\
\includegraphics[width=1\linewidth]{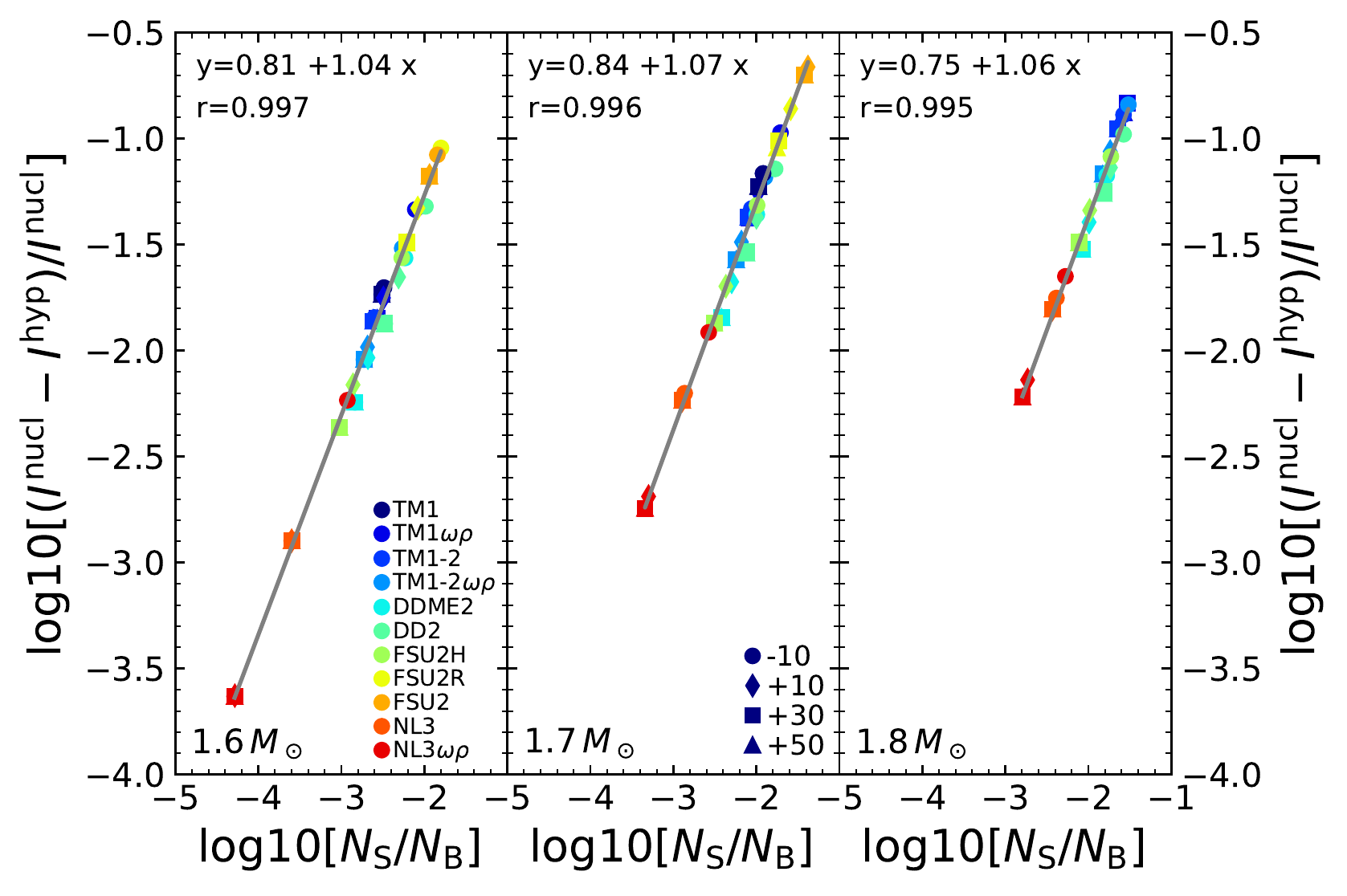}\\
\end{tabular}
\caption{
  Correlations between relative deviations of quantities that characterize
  hypernuclear stars from those that characterize purely nucleonic stars,
  and the strangeness fraction. 
  Considered are: the radius (top panels), the tidal deformability (middle panels)
  and the moment of inertia (bottom panels).
  The results correspond to various nucleonic EoS and $U^{(N)}_{\Sigma}$
  values, as mentioned in the key legend.
  Results are for $1.6M_{\odot}$, $1.7M_{\odot}$ and $1.8M_{\odot}$.
  The correlation coefficient $r$ and the parameters of the linear fit are mentioned on each panel.}
\label{fig:DLRI}
\end{figure}

The most important consequence of hyperon nucleation in the NS core
is the drastic reduction of the maximum mass. Other quantities,
like NS radius, tidal deformability and moment of inertia are also
affected though to a much lesser extent, as illustrated in
Fig. \ref{fig:star}. The question raised is which is the relation
between the strangeness density or fraction and the magnitude by which
different parameters that characterize hypernuclear stars deviate from
those that characterize purely nucleonic stars.
Figure \ref{fig:DLRI} shows, in log-log scale, the relative deviations
which affect the  radii (top panels), the  tidal deformabilities
(middle panels) and the moments of inertia (bottom panels) for NSs with
masses equal to $1.6 M_{\odot}$,  $1.7 M_{\odot}$ and  $1.8 M_{\odot}$.
The predictions corresponding to different nucleonic models are plotted
with different colors, while different symbols signal the different values of
$U^{(N)}_{\Sigma}$=-10, 10, 30 and 50 MeV.
It comes out that each of the three quantities is strongly correlated with
$N_S/N_B$.
The Pearson correlation factor $r$, indicated on each panel, is only slightly smaller than one indicating the strong correlations between the quantities of interest in each panel.
Moreover, the relation between $\log \left(|F_{\rm nucl}-F_{\rm hyp}|/F_{\rm nucl} \right)$ and
$\log \left( N_S/N_B \right)$, where $F=R, \Lambda, I$, is linear.
The parameters of the linear fit are mentioned on each panel.
In all cases, the slopes of the lines are slightly larger than 1.

These correlations can be understood by considering that
all the above quantities $F= R, \Lambda, I$ depend on $R$
to a given power $n$, ($n=1$, 5 and 2 for $R$, $\Lambda$ and $I$).
Assuming that $R$ gets modified by $\delta R$, 
$\delta F \sim n R_{\rm nucl}^{(n-1)} \delta R$.
In its turn, $\delta R$ depends linearly on $\delta M$,
\begin{equation}
\delta R=-\left( \frac{\partial R}{\partial M}\right)_{M_{\rm nucl}}
\delta M,
\nonumber
\end{equation}
where $\delta M$ stands for the reduction of NS mass due to the onset
of hyperons, $\delta M=M_{\rm nucl}-M_{\rm hyp}$.
Finally, for  small strangeness fractions, $\delta M \sim N_S/N_B$.

According to Fig. \ref{fig:DLRI}, the modifications entailed by
hyperons on the radius, tidal deformability and moment of inertia
of NS are of the order of 5\%, 30\% and, respectively,
10\% for 1.6 $M_\odot$.
Larger values, of the order of 10\%, 60\% and, respectively, 15\% are
obtained for more massive NS.

\begin{figure}[th]
\includegraphics[width=1\linewidth]{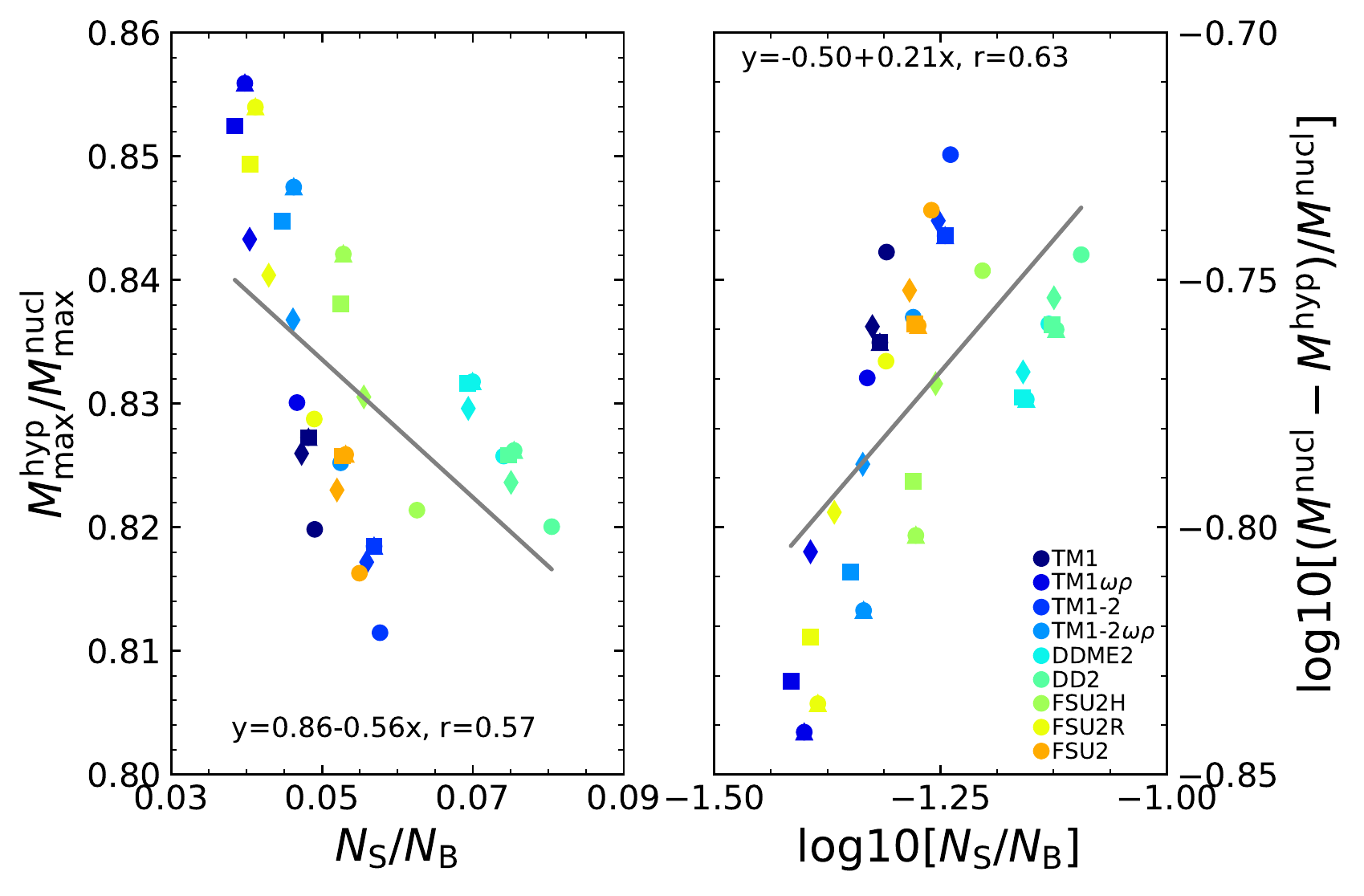}\\
\caption{Correlations between the strangeness fraction and
  the ratio (left panel) and, respectively, relative deviation (right panel)
  between maximum mass of hypernuclear stars and
  maximum mass of purely nucleonic stars.
  The same nucleonic EoS and values of $U^{(N)}_{\Sigma}$
  as in Fig. \ref{fig:DLRI} are considered.
  The correlation coefficients $r$
  and the linear fit parameters are mentioned on each
  panel.}
\label{fig:dmax}
\end{figure}

Finally, Fig. \ref{fig:dmax} plots the ratio and relative reduction
of the maximum mass of hypernuclear stars with respect to the
maximum mass of purely nucleonic stars.
Possible correlations can be judged upon by considering the variety of nucleonic
EoS introduced in Sec. \ref{ssec:model} and a range of $U^{(N)}_{\Sigma}$ values.
The first conclusion is that the inclusion of strangeness reduces the maximum
mass by $\approx 15\%-20\%$, out of which 5\% comes from insufficient knowledge
on the interaction between $\Sigma$-hyperons and nucleons.
This means that, in order to produce hypernuclear stars with masses
$\gtrsim 2M{_\odot}$, the nucleonic EoS should be stiff enough
to provide for purely nucleonic stars maximum masses larger than 2.35$-$2.5 $M_\odot$.
Contrary to what we have seen in Fig. \ref{fig:DLRI}, only a loose
correlation exists between $M^{\rm hyp}_{\rm max}/M^{\rm nucl}_{\rm max}$ and
$N_S/N_B$ and $\log(1-M^{\rm hyp}_{\rm max}/M^{\rm nucl}_{\rm max})$ and $\log(N_S/N_B)$. 
This result is easy to understand considering that no constraint is imposed
on the high density behavior of the dominant nucleonic component, as no data
exist so far in that region. As a consequence, properties of
stars with quite different masses are being compared.

\section{Conclusions}
\label{sec:concl}

Starting from a set of nucleonic RMF models that explore widely different behaviors
in the isoscalar and isovector channels and predict NS maximum masses
in excess to the astrophysical $2M_{\odot}$ constraint, we have studied the effect
of including hyperons on the properties of NS.
With the aim of constraining as much as possible the hyperon-nucleon interaction,
we employ $\sigma Y$ coupling constants calibrated on experimental data
on hypernuclei.
The method relies on comparison between the values of the binding energy
of nuclei with a variable number of nucleons and one or more hyperons, as obtained
by solving the Dirac equation, with corresponding experimental data.
More precisely, we employ the $g_{\sigma \Lambda}$ values determined in
Refs. \cite{Fortin17,Fortin18,Providencia19}, based on a vast collection
of data of single-$\Lambda$ hypernuclei,
and further determine $g_{\sigma \Xi}$ from the binding energy of $\Xi^-$ in
the hypernucleus $^{15}_{\Xi}{\rm C}$ \cite{khaustov00}.
Experimental uncertainties related to the final state of the daughter nucleus
$^{10}_{\Lambda}{\rm Be}$ are accounted for by considering the possibility that
$\Xi^-$ occupies the $1s$ or, alternatively, the $1p$ state of $^{14}{\rm N}$,
as previously done in Ref. \cite{Sun2016}.
Uncertainties related to the $\Sigma N$ interaction are dealt with
by allowing the $U^{(N)}_{\Sigma}$ potential to span a wide domain.

For all considered models and $U^{(N)}_{\Sigma}$ potentials,
we have investigated NS geometric, deformability and chemical properties.
Special attention was devoted to the density at which various hyperonic species
nucleate and their sensitivity to the nucleonic EoS and $\Sigma N$ interaction.
Dependence on these ingredients of the nucleonic and hyperonic dUrca thresholds
was discussed as well.
In regards to $\Sigma^-$ we noticed that 
a) in most NL models only two hyperon species are present: the $\Lambda$ and $\Sigma^-$. For the DD models the $\Xi^-$ also appears. 
b) The value of the $U^{(N)}_\Sigma$ determines the hyperonic dUrca process
that opens up first, with a less (more) repulsive potential favoring the more (less)
efficient $\Sigma\Lambda$ ($\Lambda p$) process \cite{DUY92}; 
(c) for repulsive values of $U^{(N)}_\Sigma$ potential, as customarily assumed in the
literature, the $\Lambda p$ dUrca process starts operating
at $n \approx 2 n_0$, which corresponds to $M/M_{\odot} \approx 1.3 - 1.4 M_\odot $. 
We have obtained very strong linear correlations between the strangeness fraction
in NS core and the relative deviation of the radius, tidal deformability and
moment of inertia of hypernuclear compact stars from values
characterizing purely nucleonic stars.
Quantitatively speaking, for NS with masses in excess of $1.6M_\odot$
hyperonic degrees of freedom are responsible for a reduction
of radii, tidal deformabilities and moments of inertia of the order of
$\sim$ 10\%,  60\% and, respectively, 15\%.
In regards to the maximum mass, the decrease is
of 15\% to 20\%, with a non-negligible role played by $U^{(N)}_\Sigma$.
\begin{table*}
\label{tabii}
\caption{Properties of NS built upon various relativistic density
  functional models, assuming different values of $U_\Sigma^{(N)}$ and
  $x_{s\Xi}$. $M_{\rm max}$ and $n_{\rm c}$ refer 
to the maximum NS mass and the corresponding central baryonic number
density. The next columns list the 
hyperonic species that
nucleate, the onset density and the associated NS mass with the same central baryonic density. The following
columns list the densities and mass thresholds above which the
nucleonic and some hyperonic dUrca processes operate. Below the name of each model we indicate between parentheses the maximum mass obtained for a purely nucleonic EoS in solar mass units. Particle number
densities are expressed in fm$^{-3}$. NS masses are expressed in
$M_\odot$.
$np$, $\Lambda p$, $\Sigma^- \Lambda$ and $\Sigma^- n$ are abbreviations for
the dUrca processes that involve the specified baryons. }
{\scriptsize
\hspace*{-1cm}
\vspace*{0cm}
\begin{tabular*}
    {\linewidth}{ @{\extracolsep{\fill}} l| c| c| c| c| lcc | lcc | lcc | cc | cc | cc | cc @{}}
\toprule
 \hline
  \hline
\multicolumn{5}{c|}{}&\multicolumn{9}{c|}{Y species} & \multicolumn{2}{c}{np} & \multicolumn{2}{c}{$\Lambda p$}
 & \multicolumn{2}{c}{$\Sigma^- \Lambda$} & \multicolumn{2}{c}{$\Sigma^- n$}   \\
 \hline
Model & $U_{\Sigma}^{(N)}$ & $x_{s \Xi}$ & $M_{\rm max}$ & $n_{\rm c}$&& $n_{Y}$&  $M_{Y}$&& $n_{Y}$& $M_{Y}$&& $n_{Y}$&  $M_{Y}$& $n_{DU}$&  $M_{DU}$&$n_{DU}$& $M_{DU}$&$n_{DU}$&  $M_{DU}$&$n_{DU}$&  $M_{DU}$\\
& (MeV)& & ($M_\odot$) & (fm$^{-3}$)& &  (fm$^{-3}$)& ($M_\odot$) & &(fm$^{-3}$)&($M_\odot$) & &  (fm$^{-3}$)& ($M_\odot$) & (fm$^{-3}$)& ($M_\odot$) &(fm$^{-3}$)& ($M_\odot$) &(fm$^{-3}$)& ($M_\odot$) &(fm$^{-3}$)& ($M_\odot$) \\
\hline
DD2  & -10  & 0.304  & 1.99  & 1.053  & $\Sigma^-$ &  0.295 &  1.00  & $\Lambda$ &  0.379 &  1.35  & $\Xi^-$ &  0.959 &  1.98       &        &            &  0.395 &  1.40   & 0.383 &  1.36      &     &    \\ 
(2.42) & & 0.320  & 1.99  & 1.050  & $\Sigma^-$ &  0.295 &  1.00  & $\Lambda$ &  0.379 &  1.35  & $\Xi^-$ &  0.819 &  1.95       &        &            &  0.395 &  1.40   & 0.383 &  1.36      &     &    \\ 
 & 10  & 0.304  & 2.00  & 1.019  & $\Sigma^-$ &  0.325 &  1.21  & $\Lambda$ &  0.345 &  1.31  & $\Xi^-$ &  0.662 &  1.89       &        &            &  0.346 &  1.31   & 0.347 &  1.32      &     &    \\ 
 & & 0.320  & 2.00  & 1.019  & $\Sigma^-$ &  0.325 &  1.21  & $\Lambda$ &  0.345 &  1.31  & $\Xi^-$ &  0.532 &  1.76       &        &            &  0.346 &  1.31   & 0.347 &  1.32      &     &    \\ 
 & 30  & 0.304  & 2.01  & 0.996  & $\Lambda$ &  0.337 &  1.29  & $\Sigma^-$ &  0.374 &  1.45  & $\Xi^-$ &  0.430 &  1.61       &        &            &  0.337 &  1.29   & 0.374 &  1.45      &     &    \\ 
 & & 0.320  & 2.00  & 1.000  & $\Lambda$ &  0.337 &  1.29  & $\Sigma^-$ &  0.374 &  1.45  & $\Xi^-$ &  0.375 &  1.46       &        &            &  0.337 &  1.29   & 0.374 &  1.45      &     &    \\ 
 DDME2  & -10  & 0.306  & 2.06  & 0.985  & $\Sigma^-$ &  0.299 &  1.08  & $\Lambda$ &  0.378 &  1.45  & $\Xi^-$ &  0.901 &  2.06       &        &            &  0.393 &  1.50   & 0.382 &  1.46      &     &    \\ 
(2.48) & & 0.321  & 2.06  & 0.981  & $\Sigma^-$ &  0.299 &  1.08  & $\Lambda$ &  0.378 &  1.45  & $\Xi^-$ &  0.771 &  2.03       &        &            &  0.393 &  1.50   & 0.382 &  1.46      &     &    \\ 
 & 10  & 0.306  & 2.07  & 0.952  & $\Sigma^-$ &  0.328 &  1.30  & $\Lambda$ &  0.346 &  1.40  & $\Xi^-$ &  0.601 &  1.95       &        &            &  0.346 &  1.40   & 0.349 &  1.41      &     &    \\ 
 & & 0.321  & 2.07  & 0.952  & $\Sigma^-$ &  0.328 &  1.30  & $\Lambda$ &  0.346 &  1.40  & $\Xi^-$ &  0.493 &  1.81       &        &            &  0.346 &  1.40   & 0.349 &  1.41      &     &    \\ 
 & 30  & 0.306  & 2.08  & 0.931  & $\Lambda$ &  0.340 &  1.39  & $\Sigma^-$ &  0.379 &  1.57  & $\Xi^-$ &  0.405 &  1.65       &        &            &  0.340 &  1.39   & 0.379 &  1.57      &     &    \\ 
 & & 0.321  & 2.08  & 0.935  & $\Lambda$ &  0.340 &  1.39  & $\Xi^-$ &  0.372 &  1.54  & $\Sigma^-$ &  0.386 &  1.59       &        &            &  0.340 &  1.39   & 0.386 &  1.59      &     &    \\ 
 FSU2R  & -10  & 0.296  & 1.69  & 1.088  & $\Sigma^-$ &  0.318 &  1.07  & $\Lambda$ &  0.402 &  1.32      &         &        &    & 0.457 &  1.42    &  0.412 &  1.34   & 0.402 &  1.32  & 0.732 &  1.64   \\ 
(2.05) & & 0.316  & 1.69  & 1.088  & $\Sigma^-$ &  0.318 &  1.07  & $\Lambda$ &  0.402 &  1.32      &         &        &    & 0.457 &  1.42    &  0.412 &  1.34   & 0.402 &  1.32  & 0.732 &  1.64   \\ 
 & 10  & 0.296  & 1.72  & 1.028  & $\Sigma^-$ &  0.356 &  1.26  & $\Lambda$ &  0.361 &  1.28      &         &        &    & 0.519 &  1.55    &  0.361 &  1.28   & 0.371 &  1.31      &     &    \\ 
 & & 0.316  & 1.72  & 1.028  & $\Sigma^-$ &  0.356 &  1.26  & $\Lambda$ &  0.361 &  1.28      &         &        &    & 0.519 &  1.55    &  0.361 &  1.28   & 0.371 &  1.31      &     &    \\ 
 & 30  & 0.296  & 1.74  & 1.002  & $\Lambda$ &  0.361 &  1.29  & $\Sigma^-$ &  0.449 &  1.49      &         &        &    & 0.588 &  1.64    &  0.361 &  1.29   & 0.449 &  1.49      &     &    \\ 
 & & 0.316  & 1.74  & 1.002  & $\Lambda$ &  0.361 &  1.29  & $\Sigma^-$ &  0.449 &  1.49      &         &        &    & 0.588 &  1.64    &  0.361 &  1.29   & 0.449 &  1.49      &     &    \\ 
 FSU2  & -10  & 0.296  & 1.69  & 1.026  & $\Lambda$ &  0.321 &  1.27  & $\Sigma^-$ &  0.334 &  1.31      &         &        &    & 0.216 &  0.77    &  0.321 &  1.27   & 0.348 &  1.34  & 0.539 &  1.58   \\ 
(2.07) & & 0.311  & 1.69  & 1.026  & $\Lambda$ &  0.321 &  1.27  & $\Sigma^-$ &  0.334 &  1.31      &         &        &    & 0.216 &  0.77    &  0.321 &  1.27   & 0.348 &  1.34  & 0.539 &  1.58   \\ 
 & 10  & 0.296  & 1.71  & 1.004  & $\Lambda$ &  0.321 &  1.27  & $\Sigma^-$ &  0.491 &  1.55      &         &        &    & 0.216 &  0.77    &  0.321 &  1.27   & 0.544 &  1.60  & 0.865 &  1.70   \\ 
 & & 0.311  & 1.71  & 1.004  & $\Lambda$ &  0.321 &  1.27  & $\Sigma^-$ &  0.491 &  1.55      &         &        &    & 0.216 &  0.77    &  0.321 &  1.27   & 0.544 &  1.60  & 0.865 &  1.70   \\ 
 & 30  & 0.296  & 1.71  & 1.011  & $\Lambda$ &  0.321 &  1.27  & $\Sigma^-$ &  0.835 &  1.70      &         &        &    & 0.216 &  0.77    &  0.321 &  1.27       &      &               &     &    \\ 
 & & 0.311  & 1.71  & 1.011  & $\Lambda$ &  0.321 &  1.27  & $\Sigma^-$ &  0.835 &  1.70      &         &        &    & 0.216 &  0.77    &  0.321 &  1.27       &      &               &     &    \\ 
 FSU2H  & -10  & 0.296  & 1.95  & 0.876  & $\Sigma^-$ &  0.299 &  1.16  & $\Lambda$ &  0.357 &  1.42      &         &        &    & 0.425 &  1.61    &  0.365 &  1.45   & 0.357 &  1.42  & 0.627 &  1.87   \\ 
(2.38) & & 0.310  & 1.95  & 0.876  & $\Sigma^-$ &  0.299 &  1.16  & $\Lambda$ &  0.357 &  1.42      &         &        &    & 0.425 &  1.61    &  0.365 &  1.45   & 0.357 &  1.42  & 0.627 &  1.87   \\ 
 & 10  & 0.296  & 1.98  & 0.919  & $\Lambda$ &  0.333 &  1.42  & $\Sigma^-$ &  0.333 &  1.41      &         &        &    & 0.479 &  1.76    &  0.333 &  1.41   & 0.341 &  1.45      &     &    \\ 
 & & 0.310  & 1.98  & 0.919  & $\Lambda$ &  0.333 &  1.42  & $\Sigma^-$ &  0.333 &  1.41      &         &        &    & 0.479 &  1.76    &  0.333 &  1.41   & 0.341 &  1.45      &     &    \\ 
 & 30  & 0.296  & 1.99  & 0.900  & $\Lambda$ &  0.332 &  1.41  & $\Sigma^-$ &  0.428 &  1.71      &         &        &    & 0.536 &  1.86    &  0.332 &  1.41   & 0.446 &  1.74      &     &    \\ 
 & & 0.310  & 1.99  & 0.900  & $\Lambda$ &  0.332 &  1.41  & $\Sigma^-$ &  0.428 &  1.71  & $\Xi^-$ &  0.4939 &  1.81   & 0.534 &  1.86    &  0.332 &  1.41   & 0.446 &  1.74      &     &    \\ 
 TM1  & -10  & 0.295  & 1.79  & 0.912  & $\Lambda$ &  0.317 &  1.38  & $\Sigma^-$ &  0.354 &  1.49      &         &        &    & 0.211 &  0.82    &  0.317 &  1.38   & 0.357 &  1.49  & 0.553 &  1.71   \\ 
(2.18) & & 0.310  & 1.79  & 0.912  & $\Lambda$ &  0.317 &  1.38  & $\Sigma^-$ &  0.354 &  1.49      &         &        &    & 0.211 &  0.82    &  0.317 &  1.38   & 0.357 &  1.49  & 0.553 &  1.71   \\ 
 & 10  & 0.295  & 1.81  & 0.904  & $\Lambda$ &  0.317 &  1.38  & $\Sigma^-$ &  0.563 &  1.73      &         &        &    & 0.211 &  0.82    &  0.317 &  1.38       &      &           & 0.882 &  1.80   \\ 
 & & 0.310  & 1.81  & 0.904  & $\Lambda$ &  0.317 &  1.38  & $\Sigma^-$ &  0.563 &  1.73      &         &        &    & 0.211 &  0.82    &  0.317 &  1.38       &      &           & 0.882 &  1.80   \\ 
 & 30  & 0.295  & 1.81  & 0.910  & $\Lambda$ &  0.317 &  1.38  & $\Xi^-$ &  0.889 &  1.81      &         &        &    & 0.211 &  0.82    &  0.317 &  1.38       &      &               &     &    \\ 
 & & 0.310  & 1.81  & 0.910  & $\Lambda$ &  0.317 &  1.38  & $\Xi^-$ &  0.693 &  1.78      &         &        &    & 0.211 &  0.82    &  0.317 &  1.38       &      &               &     &    \\ 
 TM1$\omega\rho$  & -10  & 0.295  & 1.78  & 0.985  & $\Sigma^-$ &  0.327 &  1.24  & $\Lambda$ &  0.382 &  1.41      &         &        &    & 0.447 &  1.54    &  0.389 &  1.43   & 0.382 &  1.41  & 0.699 &  1.74   \\ 
(2.12) & & 0.310  & 1.78  & 0.985  & $\Sigma^-$ &  0.327 &  1.24  & $\Lambda$ &  0.382 &  1.41      &         &        &    & 0.447 &  1.54    &  0.389 &  1.43   & 0.382 &  1.41  & 0.699 &  1.74   \\ 
 & 10  & 0.295  & 1.81  & 0.935  & $\Lambda$ &  0.359 &  1.40  & $\Sigma^-$ &  0.374 &  1.45      &         &        &    & 0.520 &  1.67    &  0.359 &  1.40   & 0.379 &  1.46      &     &    \\ 
 & & 0.310  & 1.81  & 0.935  & $\Lambda$ &  0.359 &  1.40  & $\Sigma^-$ &  0.374 &  1.45      &         &        &    & 0.520 &  1.67    &  0.359 &  1.40   & 0.379 &  1.46      &     &    \\ 
 & 30  & 0.295  & 1.82  & 0.918  & $\Lambda$ &  0.359 &  1.40  & $\Sigma^-$ &  0.516 &  1.69  & $\Xi^-$ &  0.698 &  1.80   & 0.607 &  1.76    &  0.359 &  1.40       &      &               &     &    \\ 
 & & 0.310  & 1.82  & 0.922  & $\Lambda$ &  0.359 &  1.40  & $\Xi^-$ &  0.488 &  1.66  & $\Sigma^-$ &  0.524 &  1.70   & 0.603 &  1.76    &  0.359 &  1.40       &      &               &     &    \\ 
 TM1-2  & -10  & 0.292  & 1.85  & 0.919  & $\Lambda$ &  0.310 &  1.39  & $\Sigma^-$ &  0.355 &  1.53      &         &        &    & 0.209 &  0.81    &  0.310 &  1.39   & 0.355 &  1.53  & 0.550 &  1.76   \\ 
(2.27) & & 0.309  & 1.85  & 0.919  & $\Lambda$ &  0.310 &  1.39  & $\Sigma^-$ &  0.355 &  1.53      &         &        &    & 0.209 &  0.81    &  0.310 &  1.39   & 0.355 &  1.53  & 0.550 &  1.76   \\ 
 & 10  & 0.292  & 1.86  & 0.910  & $\Lambda$ &  0.310 &  1.39  & $\Sigma^-$ &  0.568 &  1.79      &         &        &    & 0.209 &  0.81    &  0.310 &  1.39       &      &           & 0.870 &  1.86   \\ 
 & & 0.309  & 1.86  & 0.910  & $\Lambda$ &  0.310 &  1.39  & $\Sigma^-$ &  0.568 &  1.79      &         &        &    & 0.209 &  0.81    &  0.310 &  1.39       &      &           & 0.870 &  1.86   \\ 
 & 30  & 0.292  & 1.87  & 0.917  & $\Lambda$ &  0.310 &  1.39  & $\Xi^-$ &  0.888 &  1.87      &         &        &    & 0.209 &  0.81    &  0.310 &  1.39       &      &               &     &    \\ 
 & & 0.309  & 1.87  & 0.913  & $\Lambda$ &  0.310 &  1.39  & $\Xi^-$ &  0.710 &  1.84      &         &        &    & 0.209 &  0.81    &  0.310 &  1.39       &      &               &     &    \\ 
 TM1-2$\omega\rho$  & -10  & 0.292  & 1.84  & 0.982  & $\Sigma^-$ &  0.323 &  1.25  & $\Lambda$ &  0.366 &  1.40      &         &        &    & 0.425 &  1.54    &  0.368 &  1.41   & 0.366 &  1.40  & 0.680 &  1.79   \\ 
(2.22) & & 0.309  & 1.84  & 0.982  & $\Sigma^-$ &  0.323 &  1.25  & $\Lambda$ &  0.366 &  1.40      &         &        &    & 0.425 &  1.54    &  0.368 &  1.41   & 0.366 &  1.40  & 0.680 &  1.79   \\ 
 & 10  & 0.292  & 1.87  & 0.940  & $\Lambda$ &  0.348 &  1.39  & $\Sigma^-$ &  0.373 &  1.48      &         &        &    & 0.505 &  1.70    &  0.348 &  1.39   & 0.373 &  1.48      &     &    \\ 
 & & 0.309  & 1.87  & 0.940  & $\Lambda$ &  0.348 &  1.39  & $\Sigma^-$ &  0.373 &  1.48      &         &        &    & 0.505 &  1.70    &  0.348 &  1.39   & 0.373 &  1.48      &     &    \\ 
 & 30  & 0.292  & 1.88  & 0.923  & $\Lambda$ &  0.348 &  1.39  & $\Sigma^-$ &  0.529 &  1.75  & $\Xi^-$ &  0.787 &  1.87   & 0.585 &  1.80    &  0.348 &  1.39       &      &               &     &    \\ 
 & & 0.309  & 1.88  & 0.923  & $\Lambda$ &  0.348 &  1.39  & $\Xi^-$ &  0.493 &  1.71  & $\Sigma^-$ &  0.546 &  1.76   & 0.581 &  1.79    &  0.348 &  1.39       &      &               &     &    \\ 
 NL3  & -10  & 0.296  & 2.22  & 0.685  & $\Lambda$ &  0.288 &  1.54  & $\Sigma^-$ &  0.294 &  1.58      &         &        &    & 0.205 &  0.84    &  0.288 &  1.54   & 0.302 &  1.62  & 0.410 &  1.93   \\ 
(2.77) & & 0.310  & 2.22  & 0.685  & $\Lambda$ &  0.288 &  1.54  & $\Sigma^-$ &  0.294 &  1.58      &         &        &    & 0.205 &  0.84    &  0.288 &  1.54   & 0.302 &  1.62  & 0.410 &  1.93   \\ 
 & 10  & 0.296  & 2.23  & 0.684  & $\Lambda$ &  0.288 &  1.54  & $\Sigma^-$ &  0.414 &  1.97  & $\Xi^-$ &  0.640 &  2.22   & 0.205 &  0.84    &  0.288 &  1.54       &      &           & 0.597 &  2.19   \\ 
 & & 0.310  & 2.24  & 0.703  & $\Lambda$ &  0.288 &  1.54  & $\Sigma^-$ &  0.414 &  1.97  & $\Xi^-$ &  0.471 &  2.07   & 0.205 &  0.84    &  0.288 &  1.54       &      &           & 0.676 &  2.23   \\ 
 & 30  & 0.296  & 2.25  & 0.713  & $\Lambda$ &  0.288 &  1.54  & $\Xi^-$ &  0.498 &  2.11      &         &        &    & 0.205 &  0.84    &  0.288 &  1.54       &      &               &     &    \\ 
 & & 0.310  & 2.24  & 0.723  & $\Lambda$ &  0.288 &  1.54  & $\Xi^-$ &  0.442 &  2.02      &         &        &    & 0.205 &  0.84    &  0.288 &  1.54       &      &               &     &    \\ 
 NL3$\omega\rho$  & -10  & 0.296  & 2.27  & 0.712  & $\Sigma^-$ &  0.283 &  1.35  & $\Lambda$ &  0.341 &  1.70      &         &        &    & 0.434 &  1.99    &  0.349 &  1.73   & 0.341 &  1.70  & 0.572 &  2.20   \\ 
(2.75) & & 0.311  & 2.27  & 0.712  & $\Sigma^-$ &  0.283 &  1.35  & $\Lambda$ &  0.341 &  1.70      &         &        &    & 0.434 &  1.99    &  0.349 &  1.73   & 0.341 &  1.70  & 0.572 &  2.20   \\ 
 & 10  & 0.296  & 2.30  & 0.712  & $\Sigma^-$ &  0.309 &  1.62  & $\Lambda$ &  0.317 &  1.67      &         &        &    & 0.488 &  2.14    &  0.317 &  1.67   & 0.320 &  1.69      &     &    \\ 
 & & 0.311  & 2.30  & 0.722  & $\Sigma^-$ &  0.309 &  1.62  & $\Lambda$ &  0.317 &  1.67  & $\Xi^-$ &  0.562 &  2.22   & 0.488 &  2.14    &  0.317 &  1.67   & 0.320 &  1.69      &     &    \\ 
 & 30  & 0.296  & 2.31  & 0.742  & $\Lambda$ &  0.317 &  1.68  & $\Sigma^-$ &  0.389 &  1.98  & $\Xi^-$ &  0.397 &  2.00   & 0.532 &  2.22    &  0.317 &  1.68       &      &               &     &    \\ 
 & & 0.311  & 2.31  & 0.763  & $\Lambda$ &  0.317 &  1.68  & $\Xi^-$ &  0.363 &  1.89  & $\Sigma^-$ &  0.423 &  2.05   & 0.535 &  2.22    &  0.317 &  1.68       &      &               &     &    \\ 

 \hline
 \hline
\bottomrule
\end{tabular*}
}
\end{table*}
\newpage

{\bf Acknowledgments}:
This  work  was  supported  by  Funda\c{c}\~ao  para  a Ciência e Tecnologia,  Portugal,
under the Grants No. UID/FIS/04564/2019 and No. POCI-01-0145-FEDER-029912
with financial support from POCI, in its FEDER component,
and by the FCT/MCTES budget through national funds  (OE),
and by the Polish National Science This work was partially supported by Conselho Nacional de Desenvolvimento Científico e Tecnológico (CNPq) under Grant No. 6484/2016-1 (S. S. A.), and as a part of the project INCT-FNA (Instituto Nacional de Ciência e Tecnologia— Física Nuclear e Aplicações) No. 464898/2014-5 (S. S. A.)
Centre (NCN) under the Grant No. 2017/26/D/ST9/00591.
A. R. R. acknowledges the support provided by the European
COST Action “PHAROS” (Grant No. CA16214), through a STSM grant as well as the kind hospitality
of the Department of Physics, University of Coimbra. C.P. acknowledges
the support of THEIA networking activity of the Strong 2020 Project.

\bibliography{biblio.bib}

\begin{thebibliography}{92}
\expandafter\ifx\csname natexlab\endcsname\relax\def\natexlab#1{#1}\fi
\expandafter\ifx\csname bibnamefont\endcsname\relax
  \def\bibnamefont#1{#1}\fi
\expandafter\ifx\csname bibfnamefont\endcsname\relax
  \def\bibfnamefont#1{#1}\fi
\expandafter\ifx\csname citenamefont\endcsname\relax
  \def\citenamefont#1{#1}\fi
\expandafter\ifx\csname url\endcsname\relax
  \def\url#1{\texttt{#1}}\fi
\expandafter\ifx\csname urlprefix\endcsname\relax\def\urlprefix{URL }\fi
\providecommand{\bibinfo}[2]{#2}
\providecommand{\eprint}[2][]{\url{#2}}

\bibitem[{\citenamefont{Abbott et~al.}(2017{\natexlab{a}})}]{gw170817}
\bibinfo{author}{\bibfnamefont{B.~P.} \bibnamefont{Abbott}}
  \bibnamefont{et~al.} (\bibinfo{collaboration}{the Virgo, The LIGO
  Scientific}), \bibinfo{journal}{Phys. Rev. Lett.}
  \textbf{\bibinfo{volume}{119}}, \bibinfo{eid}{161101}
  (\bibinfo{year}{2017}{\natexlab{a}}).

\bibitem[{\citenamefont{Abbott et~al.}(2017{\natexlab{b}})}]{GBM2017}
\bibinfo{author}{\bibfnamefont{B.~P.} \bibnamefont{Abbott}}
  \bibnamefont{et~al.} (\bibinfo{collaboration}{LIGO Scientific, Virgo, Fermi
  GBM, INTEGRAL, IceCube, AstroSat Cadmium Zinc Telluride Imager Team, IPN,
  Insight-Hxmt, ANTARES, Swift, AGILE Team, 1M2H Team, Dark Energy Camera
  GW-EM, DES, DLT40, GRAWITA, Fermi-LAT, ATCA, ASKAP, Las Cumbres Observatory
  Group, OzGrav, DWF (Deeper Wider Faster Program), AST3, CAASTRO, VINROUGE,
  MASTER, J-GEM, GROWTH, JAGWAR, CaltechNRAO, TTU-NRAO, NuSTAR, Pan-STARRS,
  MAXI Team, TZAC Consortium, KU, Nordic Optical Telescope, ePESSTO, GROND,
  Texas Tech University, SALT Group, TOROS, BOOTES, MWA, CALET, IKI-GW
  Follow-up, H.E.S.S., LOFAR, LWA, HAWC, Pierre Auger, ALMA, Euro VLBI Team, Pi
  of Sky, Chandra Team at McGill University, DFN, ATLAS Telescopes, High Time
  Resolution Universe Survey, RIMAS, RATIR, SKA South Africa/MeerKAT}),
  \bibinfo{journal}{Astrophys. J.} \textbf{\bibinfo{volume}{848}},
  \bibinfo{pages}{L12} (\bibinfo{year}{2017}{\natexlab{b}}),
  \eprint{1710.05833}.

\bibitem[{\citenamefont{Glendenning}(2000)}]{Glend2000}
\bibinfo{author}{\bibfnamefont{N.~K.} \bibnamefont{Glendenning}},
  \emph{\bibinfo{title}{{Compact stars: Nuclear physics, particle physics, and
  general relativity}}} (\bibinfo{publisher}{Springer-Verlag New York},
  \bibinfo{year}{2000}), \bibinfo{edition}{2nd} ed.

\bibitem[{\citenamefont{Drago et~al.}(2014)\citenamefont{Drago, Lavagno,
  Pagliara, and Pigato}}]{Drago2014}
\bibinfo{author}{\bibfnamefont{A.}~\bibnamefont{Drago}},
  \bibinfo{author}{\bibfnamefont{A.}~\bibnamefont{Lavagno}},
  \bibinfo{author}{\bibfnamefont{G.}~\bibnamefont{Pagliara}}, \bibnamefont{and}
  \bibinfo{author}{\bibfnamefont{D.}~\bibnamefont{Pigato}},
  \bibinfo{journal}{Phys. Rev.} \textbf{\bibinfo{volume}{C90}},
  \bibinfo{pages}{065809} (\bibinfo{year}{2014}), \eprint{1407.2843}.

\bibitem[{\citenamefont{Cai et~al.}(2015)\citenamefont{Cai, Fattoyev, Li, and
  Newton}}]{Cai2015}
\bibinfo{author}{\bibfnamefont{B.-J.} \bibnamefont{Cai}},
  \bibinfo{author}{\bibfnamefont{F.~J.} \bibnamefont{Fattoyev}},
  \bibinfo{author}{\bibfnamefont{B.-A.} \bibnamefont{Li}}, \bibnamefont{and}
  \bibinfo{author}{\bibfnamefont{W.~G.} \bibnamefont{Newton}},
  \bibinfo{journal}{Phys. Rev.} \textbf{\bibinfo{volume}{C92}},
  \bibinfo{pages}{015802} (\bibinfo{year}{2015}), \eprint{1501.01680}.

\bibitem[{\citenamefont{Ribes et~al.}(2019)\citenamefont{Ribes, Ramos, Tolos,
  Gonzalez-Boquera, and Centelles}}]{Ribes2019}
\bibinfo{author}{\bibfnamefont{P.}~\bibnamefont{Ribes}},
  \bibinfo{author}{\bibfnamefont{A.}~\bibnamefont{Ramos}},
  \bibinfo{author}{\bibfnamefont{L.}~\bibnamefont{Tolos}},
  \bibinfo{author}{\bibfnamefont{C.}~\bibnamefont{Gonzalez-Boquera}},
  \bibnamefont{and}
  \bibinfo{author}{\bibfnamefont{M.}~\bibnamefont{Centelles}},
  \bibinfo{journal}{Astrophys. J.} \textbf{\bibinfo{volume}{883}},
  \bibinfo{pages}{168} (\bibinfo{year}{2019}), \eprint{1907.08583}.

\bibitem[{\citenamefont{Li and Sedrakian}(2019)}]{Li:2019}
\bibinfo{author}{\bibfnamefont{J.~J.} \bibnamefont{Li}} \bibnamefont{and}
  \bibinfo{author}{\bibfnamefont{A.}~\bibnamefont{Sedrakian}},
  \bibinfo{journal}{Astrophys. J.} \textbf{\bibinfo{volume}{874}},
  \bibinfo{pages}{L22} (\bibinfo{year}{2019}), \eprint{1904.02006}.

\bibitem[{\citenamefont{Fortin et~al.}(2017)\citenamefont{Fortin, Avancini,
  Providência, and Vidaña}}]{Fortin17}
\bibinfo{author}{\bibfnamefont{M.}~\bibnamefont{Fortin}},
  \bibinfo{author}{\bibfnamefont{S.~S.} \bibnamefont{Avancini}},
  \bibinfo{author}{\bibfnamefont{C.}~\bibnamefont{Providência}},
  \bibnamefont{and} \bibinfo{author}{\bibfnamefont{I.}~\bibnamefont{Vidaña}},
  \bibinfo{journal}{Phys. Rev. C} \textbf{\bibinfo{volume}{95}},
  \bibinfo{eid}{065803} (\bibinfo{year}{2017}).

\bibitem[{\citenamefont{Fortin et~al.}(2018)\citenamefont{Fortin, Oertel, and
  Providência}}]{Fortin18}
\bibinfo{author}{\bibfnamefont{M.}~\bibnamefont{Fortin}},
  \bibinfo{author}{\bibfnamefont{M.}~\bibnamefont{Oertel}}, \bibnamefont{and}
  \bibinfo{author}{\bibfnamefont{C.}~\bibnamefont{Providência}},
  \bibinfo{journal}{Publ. Astron. Soc. Austral.} \textbf{\bibinfo{volume}{35}},
  \bibinfo{pages}{44} (\bibinfo{year}{2018}), \eprint{1711.09427}.

\bibitem[{\citenamefont{Providência et~al.}(2019)\citenamefont{Providência,
  Fortin, Pais, and Rabhi}}]{Providencia19}
\bibinfo{author}{\bibfnamefont{C.}~\bibnamefont{Providência}},
  \bibinfo{author}{\bibfnamefont{M.}~\bibnamefont{Fortin}},
  \bibinfo{author}{\bibfnamefont{H.}~\bibnamefont{Pais}}, \bibnamefont{and}
  \bibinfo{author}{\bibfnamefont{A.}~\bibnamefont{Rabhi}},
  \bibinfo{journal}{Frontiers in Astronomy and Space Sciences}
  \textbf{\bibinfo{volume}{6}}, \bibinfo{pages}{13} (\bibinfo{year}{2019}),
  ISSN \bibinfo{issn}{2296-987X},
  \urlprefix\url{https://www.frontiersin.org/article/10.3389/fspas.2019.00013}.

\bibitem[{\citenamefont{{Nakazawa} et~al.}(2015)\citenamefont{{Nakazawa},
  {Endo}, {Fukunaga}, {Hoshino}, {Hwang}, {Imai}, {Ito}, {Itonaga}, {Kand a},
  {Kawasaki} et~al.}}]{kiso}
\bibinfo{author}{\bibfnamefont{K.}~\bibnamefont{{Nakazawa}}},
  \bibinfo{author}{\bibfnamefont{Y.}~\bibnamefont{{Endo}}},
  \bibinfo{author}{\bibfnamefont{S.}~\bibnamefont{{Fukunaga}}},
  \bibinfo{author}{\bibfnamefont{K.}~\bibnamefont{{Hoshino}}},
  \bibinfo{author}{\bibfnamefont{S.~H.} \bibnamefont{{Hwang}}},
  \bibinfo{author}{\bibfnamefont{K.}~\bibnamefont{{Imai}}},
  \bibinfo{author}{\bibfnamefont{H.}~\bibnamefont{{Ito}}},
  \bibinfo{author}{\bibfnamefont{K.}~\bibnamefont{{Itonaga}}},
  \bibinfo{author}{\bibfnamefont{T.}~\bibnamefont{{Kand a}}},
  \bibinfo{author}{\bibfnamefont{M.}~\bibnamefont{{Kawasaki}}},
  \bibnamefont{et~al.}, \bibinfo{journal}{Progress of Theoretical and
  Experimental Physics} \textbf{\bibinfo{volume}{2015}}, \bibinfo{eid}{033D02}
  (\bibinfo{year}{2015}).

\bibitem[{\citenamefont{Gal}(2010)}]{Gal2010}
\bibinfo{author}{\bibfnamefont{A.}~\bibnamefont{Gal}}, \bibinfo{journal}{Prog.
  Theor. Phys. Suppl.} \textbf{\bibinfo{volume}{186}}, \bibinfo{pages}{270}
  (\bibinfo{year}{2010}), \eprint{1008.3510}.

\bibitem[{\citenamefont{Gal et~al.}(2016)\citenamefont{Gal, Hungerford, and
  Millener}}]{Gal2016}
\bibinfo{author}{\bibfnamefont{A.}~\bibnamefont{Gal}},
  \bibinfo{author}{\bibfnamefont{E.~V.} \bibnamefont{Hungerford}},
  \bibnamefont{and} \bibinfo{author}{\bibfnamefont{D.~J.}
  \bibnamefont{Millener}}, \bibinfo{journal}{Rev. Mod. Phys.}
  \textbf{\bibinfo{volume}{88}}, \bibinfo{pages}{035004}
  (\bibinfo{year}{2016}),
  \urlprefix\url{https://link.aps.org/doi/10.1103/RevModPhys.88.035004}.

\bibitem[{\citenamefont{Sugimura et~al.}(2014)}]{Sugimura2014}
\bibinfo{author}{\bibfnamefont{H.}~\bibnamefont{Sugimura}} \bibnamefont{et~al.}
  (\bibinfo{collaboration}{J-PARC E10}), \bibinfo{journal}{Phys. Lett.}
  \textbf{\bibinfo{volume}{B729}}, \bibinfo{pages}{39} (\bibinfo{year}{2014}),
  \eprint{1310.6104}.

\bibitem[{\citenamefont{Honda et~al.}(2017)\citenamefont{Honda, Agnello, Ahn,
  Ajimura, Akazawa, Amano, Aoki, Bhang, Chiga, Endo et~al.}}]{Honda2017}
\bibinfo{author}{\bibfnamefont{R.}~\bibnamefont{Honda}},
  \bibinfo{author}{\bibfnamefont{M.}~\bibnamefont{Agnello}},
  \bibinfo{author}{\bibfnamefont{J.~K.} \bibnamefont{Ahn}},
  \bibinfo{author}{\bibfnamefont{S.}~\bibnamefont{Ajimura}},
  \bibinfo{author}{\bibfnamefont{Y.}~\bibnamefont{Akazawa}},
  \bibinfo{author}{\bibfnamefont{N.}~\bibnamefont{Amano}},
  \bibinfo{author}{\bibfnamefont{K.}~\bibnamefont{Aoki}},
  \bibinfo{author}{\bibfnamefont{H.~C.} \bibnamefont{Bhang}},
  \bibinfo{author}{\bibfnamefont{N.}~\bibnamefont{Chiga}},
  \bibinfo{author}{\bibfnamefont{M.}~\bibnamefont{Endo}}, \bibnamefont{et~al.}
  (\bibinfo{collaboration}{J-PARC E10 Collaboration}), \bibinfo{journal}{Phys.
  Rev. C} \textbf{\bibinfo{volume}{96}}, \bibinfo{pages}{014005}
  (\bibinfo{year}{2017}),
  \urlprefix\url{https://link.aps.org/doi/10.1103/PhysRevC.96.014005}.

\bibitem[{\citenamefont{Harada et~al.}(2018)\citenamefont{Harada, Honda, and
  Hirabayashi}}]{Harada2018}
\bibinfo{author}{\bibfnamefont{T.}~\bibnamefont{Harada}},
  \bibinfo{author}{\bibfnamefont{R.}~\bibnamefont{Honda}}, \bibnamefont{and}
  \bibinfo{author}{\bibfnamefont{Y.}~\bibnamefont{Hirabayashi}},
  \bibinfo{journal}{Phys. Rev. C} \textbf{\bibinfo{volume}{97}},
  \bibinfo{pages}{024601} (\bibinfo{year}{2018}),
  \urlprefix\url{https://link.aps.org/doi/10.1103/PhysRevC.97.024601}.

\bibitem[{\citenamefont{Haidenbauer and Meißner}(2015)}]{Haidenbauer2015}
\bibinfo{author}{\bibfnamefont{J.}~\bibnamefont{Haidenbauer}} \bibnamefont{and}
  \bibinfo{author}{\bibfnamefont{U.-G.} \bibnamefont{Meißner}},
  \bibinfo{journal}{Nuclear Physics A} \textbf{\bibinfo{volume}{936}},
  \bibinfo{pages}{29 } (\bibinfo{year}{2015}), ISSN \bibinfo{issn}{0375-9474},
  \urlprefix\url{http://www.sciencedirect.com/science/article/pii/S0375947415000160}.

\bibitem[{\citenamefont{Demorest et~al.}(2010)\citenamefont{Demorest, Pennucci,
  Ransom, Roberts, and Hessels}}]{Demorest10}
\bibinfo{author}{\bibfnamefont{P.}~\bibnamefont{Demorest}},
  \bibinfo{author}{\bibfnamefont{T.}~\bibnamefont{Pennucci}},
  \bibinfo{author}{\bibfnamefont{S.}~\bibnamefont{Ransom}},
  \bibinfo{author}{\bibfnamefont{M.}~\bibnamefont{Roberts}}, \bibnamefont{and}
  \bibinfo{author}{\bibfnamefont{J.}~\bibnamefont{Hessels}},
  \bibinfo{journal}{Nature} \textbf{\bibinfo{volume}{467}},
  \bibinfo{pages}{1081} (\bibinfo{year}{2010}), \eprint{1010.5788}.

\bibitem[{\citenamefont{Arzoumanian et~al.}(2018)}]{j1614a}
\bibinfo{author}{\bibfnamefont{Z.}~\bibnamefont{Arzoumanian}}
  \bibnamefont{et~al.} (\bibinfo{collaboration}{NANOGrav}),
  \bibinfo{journal}{Astrophys. J. Suppl.} \textbf{\bibinfo{volume}{235}},
  \bibinfo{pages}{37} (\bibinfo{year}{2018}), \eprint{1801.01837}.

\bibitem[{\citenamefont{Vidana et~al.}(2011)\citenamefont{Vidana, Logoteta,
  Providencia, Polls, and Bombaci}}]{Vidana10}
\bibinfo{author}{\bibfnamefont{I.}~\bibnamefont{Vidana}},
  \bibinfo{author}{\bibfnamefont{D.}~\bibnamefont{Logoteta}},
  \bibinfo{author}{\bibfnamefont{C.}~\bibnamefont{Providencia}},
  \bibinfo{author}{\bibfnamefont{A.}~\bibnamefont{Polls}}, \bibnamefont{and}
  \bibinfo{author}{\bibfnamefont{I.}~\bibnamefont{Bombaci}},
  \bibinfo{journal}{EPL} \textbf{\bibinfo{volume}{94}}, \bibinfo{pages}{11002}
  (\bibinfo{year}{2011}), \eprint{1006.5660}.

\bibitem[{\citenamefont{Chatterjee and Vidaña}(2016)}]{Chatterjee15}
\bibinfo{author}{\bibfnamefont{D.}~\bibnamefont{Chatterjee}} \bibnamefont{and}
  \bibinfo{author}{\bibfnamefont{I.}~\bibnamefont{Vidaña}},
  \bibinfo{journal}{Eur. Phys. J.} \textbf{\bibinfo{volume}{A52}},
  \bibinfo{pages}{29} (\bibinfo{year}{2016}), \eprint{1510.06306}.

\bibitem[{\citenamefont{{Bednarek, I.} et~al.}(2012)\citenamefont{{Bednarek,
  I.}, {Haensel, P.}, {Zdunik, J. L.}, {Bejger, M.}, and {Ma\'{}nka,
  R.}}}]{Bednarek2012}
\bibinfo{author}{\bibnamefont{{Bednarek, I.}}},
  \bibinfo{author}{\bibnamefont{{Haensel, P.}}},
  \bibinfo{author}{\bibnamefont{{Zdunik, J. L.}}},
  \bibinfo{author}{\bibnamefont{{Bejger, M.}}}, \bibnamefont{and}
  \bibinfo{author}{\bibnamefont{{Ma\'{}nka, R.}}}, \bibinfo{journal}{A\&A}
  \textbf{\bibinfo{volume}{543}}, \bibinfo{pages}{A157} (\bibinfo{year}{2012}),
  \urlprefix\url{https://doi.org/10.1051/0004-6361/201118560}.

\bibitem[{\citenamefont{Weissenborn et~al.}(2012)\citenamefont{Weissenborn,
  Chatterjee, and Schaffner-Bielich}}]{Weissenborn12}
\bibinfo{author}{\bibfnamefont{S.}~\bibnamefont{Weissenborn}},
  \bibinfo{author}{\bibfnamefont{D.}~\bibnamefont{Chatterjee}},
  \bibnamefont{and}
  \bibinfo{author}{\bibfnamefont{J.}~\bibnamefont{Schaffner-Bielich}},
  \bibinfo{journal}{Phys. Rev.} \textbf{\bibinfo{volume}{C85}},
  \bibinfo{pages}{065802} (\bibinfo{year}{2012}), \bibinfo{note}{[Erratum:
  Phys. Rev.C90,no.1,019904(2014)]}, \eprint{1112.0234}.

\bibitem[{\citenamefont{Weissenborn et~al.}(2013)\citenamefont{Weissenborn,
  Chatterjee, and Schaffner-Bielich}}]{Weissenborn13}
\bibinfo{author}{\bibfnamefont{S.}~\bibnamefont{Weissenborn}},
  \bibinfo{author}{\bibfnamefont{D.}~\bibnamefont{Chatterjee}},
  \bibnamefont{and}
  \bibinfo{author}{\bibfnamefont{J.}~\bibnamefont{Schaffner-Bielich}},
  \bibinfo{journal}{Nucl. Phys.} \textbf{\bibinfo{volume}{A914}},
  \bibinfo{pages}{421} (\bibinfo{year}{2013}).

\bibitem[{\citenamefont{{Colucci} and {Sedrakian}}(2013)}]{colucci_13}
\bibinfo{author}{\bibfnamefont{G.}~\bibnamefont{{Colucci}}} \bibnamefont{and}
  \bibinfo{author}{\bibfnamefont{A.}~\bibnamefont{{Sedrakian}}},
  \bibinfo{journal}{Phys. Rev. C} \textbf{\bibinfo{volume}{87}},
  \bibinfo{eid}{055806} (\bibinfo{year}{2013}).

\bibitem[{\citenamefont{{van Dalen} et~al.}(2014)\citenamefont{{van Dalen},
  {Colucci}, and {Sedrakian}}}]{vandalen_14}
\bibinfo{author}{\bibfnamefont{E.~N.~E.} \bibnamefont{{van Dalen}}},
  \bibinfo{author}{\bibfnamefont{G.}~\bibnamefont{{Colucci}}},
  \bibnamefont{and}
  \bibinfo{author}{\bibfnamefont{A.}~\bibnamefont{{Sedrakian}}},
  \bibinfo{journal}{Phys. Lett. B} \textbf{\bibinfo{volume}{734}},
  \bibinfo{pages}{383} (\bibinfo{year}{2014}).

\bibitem[{\citenamefont{Fortin et~al.}(2016)\citenamefont{Fortin, Providencia,
  Raduta, Gulminelli, Zdunik, Haensel, and Bejger}}]{Fortin16}
\bibinfo{author}{\bibfnamefont{M.}~\bibnamefont{Fortin}},
  \bibinfo{author}{\bibfnamefont{C.}~\bibnamefont{Providencia}},
  \bibinfo{author}{\bibfnamefont{A.~R.} \bibnamefont{Raduta}},
  \bibinfo{author}{\bibfnamefont{F.}~\bibnamefont{Gulminelli}},
  \bibinfo{author}{\bibfnamefont{J.~L.} \bibnamefont{Zdunik}},
  \bibinfo{author}{\bibfnamefont{P.}~\bibnamefont{Haensel}}, \bibnamefont{and}
  \bibinfo{author}{\bibfnamefont{M.}~\bibnamefont{Bejger}},
  \bibinfo{journal}{Phys. Rev.} \textbf{\bibinfo{volume}{C94}},
  \bibinfo{pages}{035804} (\bibinfo{year}{2016}), \eprint{1604.01944}.

\bibitem[{\citenamefont{Sun et~al.}(2019)\citenamefont{Sun, Zhang, Zhang, and
  Xia}}]{Sun19}
\bibinfo{author}{\bibfnamefont{T.-T.} \bibnamefont{Sun}},
  \bibinfo{author}{\bibfnamefont{S.-S.} \bibnamefont{Zhang}},
  \bibinfo{author}{\bibfnamefont{Q.-L.} \bibnamefont{Zhang}}, \bibnamefont{and}
  \bibinfo{author}{\bibfnamefont{C.-J.} \bibnamefont{Xia}},
  \bibinfo{journal}{Phys. Rev. D} \textbf{\bibinfo{volume}{99}},
  \bibinfo{pages}{023004} (\bibinfo{year}{2019}),
  \urlprefix\url{https://link.aps.org/doi/10.1103/PhysRevD.99.023004}.

\bibitem[{\citenamefont{Antoniadis et~al.}(2013)}]{Antoniadis13}
\bibinfo{author}{\bibfnamefont{J.}~\bibnamefont{Antoniadis}}
  \bibnamefont{et~al.}, \bibinfo{journal}{Science}
  \textbf{\bibinfo{volume}{340}}, \bibinfo{pages}{6131} (\bibinfo{year}{2013}),
  \eprint{1304.6875}.

\bibitem[{\citenamefont{{Cromartie} et~al.}(2019)\citenamefont{{Cromartie},
  {Fonseca}, {Ransom}, {Demorest}, {Arzoumanian}, {Blumer}, {Brook}, {DeCesar},
  {Dolch}, {Ellis} et~al.}}]{Cromartie2019}
\bibinfo{author}{\bibfnamefont{H.~T.} \bibnamefont{{Cromartie}}},
  \bibinfo{author}{\bibfnamefont{E.}~\bibnamefont{{Fonseca}}},
  \bibinfo{author}{\bibfnamefont{S.~M.} \bibnamefont{{Ransom}}},
  \bibinfo{author}{\bibfnamefont{P.~B.} \bibnamefont{{Demorest}}},
  \bibinfo{author}{\bibfnamefont{Z.}~\bibnamefont{{Arzoumanian}}},
  \bibinfo{author}{\bibfnamefont{H.}~\bibnamefont{{Blumer}}},
  \bibinfo{author}{\bibfnamefont{P.~R.} \bibnamefont{{Brook}}},
  \bibinfo{author}{\bibfnamefont{M.~E.} \bibnamefont{{DeCesar}}},
  \bibinfo{author}{\bibfnamefont{T.}~\bibnamefont{{Dolch}}},
  \bibinfo{author}{\bibfnamefont{J.~A.} \bibnamefont{{Ellis}}},
  \bibnamefont{et~al.}, \bibinfo{journal}{Nature Astronomy} p.
  \bibinfo{pages}{439} (\bibinfo{year}{2019}), \eprint{1904.06759}.

\bibitem[{\citenamefont{Alford et~al.}(2007)\citenamefont{Alford, Blaschke,
  Drago, Klahn, Pagliara, and Schaffner-Bielich}}]{Alford06}
\bibinfo{author}{\bibfnamefont{M.}~\bibnamefont{Alford}},
  \bibinfo{author}{\bibfnamefont{D.}~\bibnamefont{Blaschke}},
  \bibinfo{author}{\bibfnamefont{A.}~\bibnamefont{Drago}},
  \bibinfo{author}{\bibfnamefont{T.}~\bibnamefont{Klahn}},
  \bibinfo{author}{\bibfnamefont{G.}~\bibnamefont{Pagliara}}, \bibnamefont{and}
  \bibinfo{author}{\bibfnamefont{J.}~\bibnamefont{Schaffner-Bielich}},
  \bibinfo{journal}{Nature} \textbf{\bibinfo{volume}{445}}, \bibinfo{pages}{E7}
  (\bibinfo{year}{2007}), \eprint{astro-ph/0606524}.

\bibitem[{\citenamefont{Weissenborn et~al.}(2011)\citenamefont{Weissenborn,
  Sagert, Pagliara, Hempel, and Schaffner-Bielich}}]{Weissenborn11}
\bibinfo{author}{\bibfnamefont{S.}~\bibnamefont{Weissenborn}},
  \bibinfo{author}{\bibfnamefont{I.}~\bibnamefont{Sagert}},
  \bibinfo{author}{\bibfnamefont{G.}~\bibnamefont{Pagliara}},
  \bibinfo{author}{\bibfnamefont{M.}~\bibnamefont{Hempel}}, \bibnamefont{and}
  \bibinfo{author}{\bibfnamefont{J.}~\bibnamefont{Schaffner-Bielich}},
  \bibinfo{journal}{The Astrophysical Journal Letters}
  \textbf{\bibinfo{volume}{740}}, \bibinfo{pages}{L14} (\bibinfo{year}{2011}),
  \urlprefix\url{http://stacks.iop.org/2041-8205/740/i=1/a=L14}.

\bibitem[{\citenamefont{Bonanno and Sedrakian}(2012)}]{Bonanno2012}
\bibinfo{author}{\bibfnamefont{L.}~\bibnamefont{Bonanno}} \bibnamefont{and}
  \bibinfo{author}{\bibfnamefont{A.}~\bibnamefont{Sedrakian}},
  \bibinfo{journal}{Astron. Astrophys.} \textbf{\bibinfo{volume}{539}},
  \bibinfo{pages}{A16} (\bibinfo{year}{2012}), \eprint{1108.0559}.

\bibitem[{\citenamefont{Masuda et~al.}(2013)\citenamefont{Masuda, Hatsuda, and
  Takatsuka}}]{Masuda2013}
\bibinfo{author}{\bibfnamefont{K.}~\bibnamefont{Masuda}},
  \bibinfo{author}{\bibfnamefont{T.}~\bibnamefont{Hatsuda}}, \bibnamefont{and}
  \bibinfo{author}{\bibfnamefont{T.}~\bibnamefont{Takatsuka}},
  \bibinfo{journal}{Astrophys. J.} \textbf{\bibinfo{volume}{764}},
  \bibinfo{pages}{12} (\bibinfo{year}{2013}), \eprint{1205.3621}.

\bibitem[{\citenamefont{Alford et~al.}(2013)\citenamefont{Alford, Han, and
  Prakash}}]{Alford2013}
\bibinfo{author}{\bibfnamefont{M.~G.} \bibnamefont{Alford}},
  \bibinfo{author}{\bibfnamefont{S.}~\bibnamefont{Han}}, \bibnamefont{and}
  \bibinfo{author}{\bibfnamefont{M.}~\bibnamefont{Prakash}},
  \bibinfo{journal}{Phys. Rev.} \textbf{\bibinfo{volume}{D88}},
  \bibinfo{pages}{083013} (\bibinfo{year}{2013}), \eprint{1302.4732}.

\bibitem[{\citenamefont{Klähn et~al.}(2013)\citenamefont{Klähn, Łastowiecki,
  and Blaschke}}]{Klahn2013}
\bibinfo{author}{\bibfnamefont{T.}~\bibnamefont{Klähn}},
  \bibinfo{author}{\bibfnamefont{R.}~\bibnamefont{Łastowiecki}},
  \bibnamefont{and} \bibinfo{author}{\bibfnamefont{D.~B.}
  \bibnamefont{Blaschke}}, \bibinfo{journal}{Phys. Rev.}
  \textbf{\bibinfo{volume}{D88}}, \bibinfo{pages}{085001}
  (\bibinfo{year}{2013}), \eprint{1307.6996}.

\bibitem[{\citenamefont{Zdunik and Haensel}(2013)}]{Zdunik2013}
\bibinfo{author}{\bibfnamefont{J.~L.} \bibnamefont{Zdunik}} \bibnamefont{and}
  \bibinfo{author}{\bibfnamefont{P.}~\bibnamefont{Haensel}},
  \bibinfo{journal}{Astron. Astrophys.} \textbf{\bibinfo{volume}{551}},
  \bibinfo{pages}{A61} (\bibinfo{year}{2013}), \eprint{1211.1231}.

\bibitem[{\citenamefont{Logoteta et~al.}(2013)\citenamefont{Logoteta,
  Providência, and Vidaña}}]{logoteta2013}
\bibinfo{author}{\bibfnamefont{D.}~\bibnamefont{Logoteta}},
  \bibinfo{author}{\bibfnamefont{C.}~\bibnamefont{Providência}},
  \bibnamefont{and} \bibinfo{author}{\bibfnamefont{I.}~\bibnamefont{Vidaña}},
  \bibinfo{journal}{Phys. Rev.} \textbf{\bibinfo{volume}{C88}},
  \bibinfo{pages}{055802} (\bibinfo{year}{2013}), \eprint{1311.0618}.

\bibitem[{\citenamefont{Drago et~al.}(2016)\citenamefont{Drago, Lavagno,
  Pagliara, and Pigato}}]{Drago2016}
\bibinfo{author}{\bibfnamefont{A.}~\bibnamefont{Drago}},
  \bibinfo{author}{\bibfnamefont{A.}~\bibnamefont{Lavagno}},
  \bibinfo{author}{\bibfnamefont{G.}~\bibnamefont{Pagliara}}, \bibnamefont{and}
  \bibinfo{author}{\bibfnamefont{D.}~\bibnamefont{Pigato}},
  \bibinfo{journal}{Eur. Phys. J.} \textbf{\bibinfo{volume}{A52}},
  \bibinfo{pages}{40} (\bibinfo{year}{2016}), \eprint{1509.02131}.

\bibitem[{\citenamefont{Pereira et~al.}(2016)\citenamefont{Pereira, Costa, and
  Providência}}]{Pereira2016}
\bibinfo{author}{\bibfnamefont{R.~C.} \bibnamefont{Pereira}},
  \bibinfo{author}{\bibfnamefont{P.}~\bibnamefont{Costa}}, \bibnamefont{and}
  \bibinfo{author}{\bibfnamefont{C.}~\bibnamefont{Providência}},
  \bibinfo{journal}{Phys. Rev.} \textbf{\bibinfo{volume}{D94}},
  \bibinfo{pages}{094001} (\bibinfo{year}{2016}), \eprint{1610.06435}.

\bibitem[{\citenamefont{Fukushima and Kojo}(2016)}]{Fukushima2016}
\bibinfo{author}{\bibfnamefont{K.}~\bibnamefont{Fukushima}} \bibnamefont{and}
  \bibinfo{author}{\bibfnamefont{T.}~\bibnamefont{Kojo}},
  \bibinfo{journal}{Astrophys. J.} \textbf{\bibinfo{volume}{817}},
  \bibinfo{pages}{180} (\bibinfo{year}{2016}), \eprint{1509.00356}.

\bibitem[{\citenamefont{Alford and Sedrakian}(2017)}]{Alford2017}
\bibinfo{author}{\bibfnamefont{M.~G.} \bibnamefont{Alford}} \bibnamefont{and}
  \bibinfo{author}{\bibfnamefont{A.}~\bibnamefont{Sedrakian}},
  \bibinfo{journal}{Phys. Rev. Lett.} \textbf{\bibinfo{volume}{119}},
  \bibinfo{pages}{161104} (\bibinfo{year}{2017}), \eprint{1706.01592}.

\bibitem[{\citenamefont{Dutra et~al.}(2014)\citenamefont{Dutra, Lourenço,
  Avancini, Carlson, Delfino et~al.}}]{Dutra2014}
\bibinfo{author}{\bibfnamefont{M.}~\bibnamefont{Dutra}},
  \bibinfo{author}{\bibfnamefont{O.}~\bibnamefont{Lourenço}},
  \bibinfo{author}{\bibfnamefont{S.}~\bibnamefont{Avancini}},
  \bibinfo{author}{\bibfnamefont{B.}~\bibnamefont{Carlson}},
  \bibinfo{author}{\bibfnamefont{A.}~\bibnamefont{Delfino}},
  \bibnamefont{et~al.}, \bibinfo{journal}{Phys.Rev.}
  \textbf{\bibinfo{volume}{C90}}, \bibinfo{pages}{055203}
  (\bibinfo{year}{2014}).

\bibitem[{\citenamefont{Chen and Piekarewicz}(2014)}]{Chen2014}
\bibinfo{author}{\bibfnamefont{W.-C.} \bibnamefont{Chen}} \bibnamefont{and}
  \bibinfo{author}{\bibfnamefont{J.}~\bibnamefont{Piekarewicz}},
  \bibinfo{journal}{Phys. Rev.} \textbf{\bibinfo{volume}{C90}},
  \bibinfo{pages}{044305} (\bibinfo{year}{2014}), \eprint{1408.4159}.

\bibitem[{\citenamefont{Tolos et~al.}(2017)\citenamefont{Tolos, Centelles, and
  Ramos}}]{Tolos17}
\bibinfo{author}{\bibfnamefont{L.}~\bibnamefont{Tolos}},
  \bibinfo{author}{\bibfnamefont{M.}~\bibnamefont{Centelles}},
  \bibnamefont{and} \bibinfo{author}{\bibfnamefont{A.}~\bibnamefont{Ramos}},
  \bibinfo{journal}{Publ. Astron. Soc. Austral.} \textbf{\bibinfo{volume}{34}},
  \bibinfo{pages}{e065} (\bibinfo{year}{2017}), \eprint{1708.08681}.

\bibitem[{\citenamefont{Negreiros et~al.}(2018)\citenamefont{Negreiros, Tolos,
  Centelles, Ramos, and Dexheimer}}]{Negreiros18}
\bibinfo{author}{\bibfnamefont{R.}~\bibnamefont{Negreiros}},
  \bibinfo{author}{\bibfnamefont{L.}~\bibnamefont{Tolos}},
  \bibinfo{author}{\bibfnamefont{M.}~\bibnamefont{Centelles}},
  \bibinfo{author}{\bibfnamefont{A.}~\bibnamefont{Ramos}}, \bibnamefont{and}
  \bibinfo{author}{\bibfnamefont{V.}~\bibnamefont{Dexheimer}},
  \bibinfo{journal}{Astrophys. J.} \textbf{\bibinfo{volume}{863}},
  \bibinfo{pages}{104} (\bibinfo{year}{2018}), \eprint{1804.00334}.

\bibitem[{\citenamefont{Lalazissis et~al.}(1997)\citenamefont{Lalazissis,
  Konig, and Ring}}]{nl3}
\bibinfo{author}{\bibfnamefont{G.~A.} \bibnamefont{Lalazissis}},
  \bibinfo{author}{\bibfnamefont{J.}~\bibnamefont{Konig}}, \bibnamefont{and}
  \bibinfo{author}{\bibfnamefont{P.}~\bibnamefont{Ring}},
  \bibinfo{journal}{Phys. Rev.} \textbf{\bibinfo{volume}{C55}},
  \bibinfo{pages}{540} (\bibinfo{year}{1997}), \eprint{nucl-th/9607039}.

\bibitem[{\citenamefont{Pais and Providência}(2016)}]{Pais16}
\bibinfo{author}{\bibfnamefont{H.}~\bibnamefont{Pais}} \bibnamefont{and}
  \bibinfo{author}{\bibfnamefont{C.}~\bibnamefont{Providência}},
  \bibinfo{journal}{Phys. Rev.} \textbf{\bibinfo{volume}{C94}},
  \bibinfo{pages}{015808} (\bibinfo{year}{2016}), \eprint{1607.05899}.

\bibitem[{\citenamefont{Horowitz and Piekarewicz}(2001)}]{Horowitz01}
\bibinfo{author}{\bibfnamefont{C.~J.} \bibnamefont{Horowitz}} \bibnamefont{and}
  \bibinfo{author}{\bibfnamefont{J.}~\bibnamefont{Piekarewicz}},
  \bibinfo{journal}{Phys. Rev. Lett.} \textbf{\bibinfo{volume}{86}},
  \bibinfo{pages}{5647} (\bibinfo{year}{2001}), \eprint{astro-ph/0010227}.

\bibitem[{\citenamefont{Sugahara and Toki}(1994{\natexlab{a}})}]{tm1}
\bibinfo{author}{\bibfnamefont{Y.}~\bibnamefont{Sugahara}} \bibnamefont{and}
  \bibinfo{author}{\bibfnamefont{H.}~\bibnamefont{Toki}},
  \bibinfo{journal}{Nucl. Phys.} \textbf{\bibinfo{volume}{A579}},
  \bibinfo{pages}{557} (\bibinfo{year}{1994}{\natexlab{a}}).

\bibitem[{\citenamefont{Providencia and Rabhi}(2013)}]{Providencia13}
\bibinfo{author}{\bibfnamefont{C.}~\bibnamefont{Providencia}} \bibnamefont{and}
  \bibinfo{author}{\bibfnamefont{A.}~\bibnamefont{Rabhi}},
  \bibinfo{journal}{Phys. Rev.} \textbf{\bibinfo{volume}{C87}},
  \bibinfo{pages}{055801} (\bibinfo{year}{2013}), \eprint{1212.5911}.

\bibitem[{\citenamefont{Bao and Shen}(2014)}]{Bao2014}
\bibinfo{author}{\bibfnamefont{S.~S.} \bibnamefont{Bao}} \bibnamefont{and}
  \bibinfo{author}{\bibfnamefont{H.}~\bibnamefont{Shen}},
  \bibinfo{journal}{Phys. Rev.} \textbf{\bibinfo{volume}{C89}},
  \bibinfo{pages}{045807} (\bibinfo{year}{2014}), \eprint{1405.3837}.

\bibitem[{\citenamefont{Typel et~al.}(2010)\citenamefont{Typel, R{\"o}pke,
  Kl{\"a}hn, Blaschke, and Wolter}}]{typel10}
\bibinfo{author}{\bibfnamefont{S.}~\bibnamefont{Typel}},
  \bibinfo{author}{\bibfnamefont{G.}~\bibnamefont{R{\"o}pke}},
  \bibinfo{author}{\bibfnamefont{T.}~\bibnamefont{Kl{\"a}hn}},
  \bibinfo{author}{\bibfnamefont{D.}~\bibnamefont{Blaschke}}, \bibnamefont{and}
  \bibinfo{author}{\bibfnamefont{H.}~\bibnamefont{Wolter}},
  \bibinfo{journal}{Phys.Rev.} \textbf{\bibinfo{volume}{C81}},
  \bibinfo{pages}{015803} (\bibinfo{year}{2010}).

\bibitem[{\citenamefont{Lalazissis et~al.}(2005)\citenamefont{Lalazissis,
  Nik\ifmmode \check{s}\else \v{s}\fi{}i\ifmmode~\acute{c}\else \'{c}\fi{},
  Vretenar, and Ring}}]{ddme2}
\bibinfo{author}{\bibfnamefont{G.~A.} \bibnamefont{Lalazissis}},
  \bibinfo{author}{\bibfnamefont{T.}~\bibnamefont{Nik\ifmmode \check{s}\else
  \v{s}\fi{}i\ifmmode~\acute{c}\else \'{c}\fi{}}},
  \bibinfo{author}{\bibfnamefont{D.}~\bibnamefont{Vretenar}}, \bibnamefont{and}
  \bibinfo{author}{\bibfnamefont{P.}~\bibnamefont{Ring}},
  \bibinfo{journal}{Phys. Rev. C} \textbf{\bibinfo{volume}{71}},
  \bibinfo{pages}{024312} (\bibinfo{year}{2005}),
  \urlprefix\url{https://link.aps.org/doi/10.1103/PhysRevC.71.024312}.

\bibitem[{\citenamefont{{Shlomo} et~al.}(2006)\citenamefont{{Shlomo},
  {Kolomietz}, and {Colo}}}]{Shlomo_EPJA_2006}
\bibinfo{author}{\bibfnamefont{S.}~\bibnamefont{{Shlomo}}},
  \bibinfo{author}{\bibfnamefont{V.~M.} \bibnamefont{{Kolomietz}}},
  \bibnamefont{and} \bibinfo{author}{\bibfnamefont{G.}~\bibnamefont{{Colo}}},
  \bibinfo{journal}{European Physical Journal A} \textbf{\bibinfo{volume}{30}},
  \bibinfo{pages}{23} (\bibinfo{year}{2006}).

\bibitem[{\citenamefont{De et~al.}(2015)\citenamefont{De, Samaddar, and
  Agrawal}}]{De_PRC_2015}
\bibinfo{author}{\bibfnamefont{J.~N.} \bibnamefont{De}},
  \bibinfo{author}{\bibfnamefont{S.~K.} \bibnamefont{Samaddar}},
  \bibnamefont{and} \bibinfo{author}{\bibfnamefont{B.~K.}
  \bibnamefont{Agrawal}}, \bibinfo{journal}{Phys. Rev.}
  \textbf{\bibinfo{volume}{C92}}, \bibinfo{pages}{014304}
  (\bibinfo{year}{2015}), \eprint{1506.06461}.

\bibitem[{\citenamefont{Lattimer and Lim}(2013)}]{Lattimer13}
\bibinfo{author}{\bibfnamefont{J.~M.} \bibnamefont{Lattimer}} \bibnamefont{and}
  \bibinfo{author}{\bibfnamefont{Y.}~\bibnamefont{Lim}},
  \bibinfo{journal}{Astrophys. J.} \textbf{\bibinfo{volume}{771}},
  \bibinfo{pages}{51} (\bibinfo{year}{2013}), \eprint{1203.4286}.

\bibitem[{\citenamefont{Oertel et~al.}(2017)\citenamefont{Oertel, Hempel,
  Kl\"ahn, and Typel}}]{Oertel17}
\bibinfo{author}{\bibfnamefont{M.}~\bibnamefont{Oertel}},
  \bibinfo{author}{\bibfnamefont{M.}~\bibnamefont{Hempel}},
  \bibinfo{author}{\bibfnamefont{T.}~\bibnamefont{Kl\"ahn}}, \bibnamefont{and}
  \bibinfo{author}{\bibfnamefont{S.}~\bibnamefont{Typel}},
  \bibinfo{journal}{Rev. Mod. Phys.} \textbf{\bibinfo{volume}{89}},
  \bibinfo{pages}{015007} (\bibinfo{year}{2017}),
  \urlprefix\url{https://link.aps.org/doi/10.1103/RevModPhys.89.015007}.

\bibitem[{\citenamefont{Abrahamyan et~al.}(2012)\citenamefont{Abrahamyan,
  Ahmed, Albataineh, Aniol, Armstrong, Armstrong, Averett, Babineau, Barbieri,
  Bellini et~al.}}]{prex}
\bibinfo{author}{\bibfnamefont{S.}~\bibnamefont{Abrahamyan}},
  \bibinfo{author}{\bibfnamefont{Z.}~\bibnamefont{Ahmed}},
  \bibinfo{author}{\bibfnamefont{H.}~\bibnamefont{Albataineh}},
  \bibinfo{author}{\bibfnamefont{K.}~\bibnamefont{Aniol}},
  \bibinfo{author}{\bibfnamefont{D.~S.} \bibnamefont{Armstrong}},
  \bibinfo{author}{\bibfnamefont{W.}~\bibnamefont{Armstrong}},
  \bibinfo{author}{\bibfnamefont{T.}~\bibnamefont{Averett}},
  \bibinfo{author}{\bibfnamefont{B.}~\bibnamefont{Babineau}},
  \bibinfo{author}{\bibfnamefont{A.}~\bibnamefont{Barbieri}},
  \bibinfo{author}{\bibfnamefont{V.}~\bibnamefont{Bellini}},
  \bibnamefont{et~al.} (\bibinfo{collaboration}{PREX Collaboration}),
  \bibinfo{journal}{Phys. Rev. Lett.} \textbf{\bibinfo{volume}{108}},
  \bibinfo{pages}{112502} (\bibinfo{year}{2012}),
  \urlprefix\url{https://link.aps.org/doi/10.1103/PhysRevLett.108.112502}.

\bibitem[{\citenamefont{{Cozma}}(2018)}]{Cozma2018}
\bibinfo{author}{\bibfnamefont{M.~D.} \bibnamefont{{Cozma}}},
  \bibinfo{journal}{European Physical Journal A} \textbf{\bibinfo{volume}{54}},
  \bibinfo{eid}{40} (\bibinfo{year}{2018}), \eprint{1706.01300}.

\bibitem[{\citenamefont{Shen et~al.}(2006)\citenamefont{Shen, Yang, and
  Toki}}]{Shen06}
\bibinfo{author}{\bibfnamefont{H.}~\bibnamefont{Shen}},
  \bibinfo{author}{\bibfnamefont{F.}~\bibnamefont{Yang}}, \bibnamefont{and}
  \bibinfo{author}{\bibfnamefont{H.}~\bibnamefont{Toki}},
  \bibinfo{journal}{Prog. Theor. Phys.} \textbf{\bibinfo{volume}{115}},
  \bibinfo{pages}{325} (\bibinfo{year}{2006}), \eprint{nucl-th/0602046}.

\bibitem[{\citenamefont{Avancini et~al.}(2007)\citenamefont{Avancini,
  Marinelli, Menezes, de~Moraes, and Providencia}}]{Avancini07}
\bibinfo{author}{\bibfnamefont{S.~S.} \bibnamefont{Avancini}},
  \bibinfo{author}{\bibfnamefont{J.~R.} \bibnamefont{Marinelli}},
  \bibinfo{author}{\bibfnamefont{D.~P.} \bibnamefont{Menezes}},
  \bibinfo{author}{\bibfnamefont{M.~M.~W.} \bibnamefont{de~Moraes}},
  \bibnamefont{and}
  \bibinfo{author}{\bibfnamefont{C.}~\bibnamefont{Providencia}},
  \bibinfo{journal}{Phys. Rev.} \textbf{\bibinfo{volume}{C75}},
  \bibinfo{pages}{055805} (\bibinfo{year}{2007}), \eprint{0704.0407}.

\bibitem[{\citenamefont{Sun et~al.}(2016)\citenamefont{Sun, Hiyama, Sagawa,
  Schulze, and Meng}}]{Sun2016}
\bibinfo{author}{\bibfnamefont{T.~T.} \bibnamefont{Sun}},
  \bibinfo{author}{\bibfnamefont{E.}~\bibnamefont{Hiyama}},
  \bibinfo{author}{\bibfnamefont{H.}~\bibnamefont{Sagawa}},
  \bibinfo{author}{\bibfnamefont{H.~J.} \bibnamefont{Schulze}},
  \bibnamefont{and} \bibinfo{author}{\bibfnamefont{J.}~\bibnamefont{Meng}},
  \bibinfo{journal}{Phys. Rev.} \textbf{\bibinfo{volume}{C94}},
  \bibinfo{pages}{064319} (\bibinfo{year}{2016}), \eprint{1611.03661}.

\bibitem[{\citenamefont{Mares and Zofka}(1989)}]{Mares1989}
\bibinfo{author}{\bibfnamefont{J.}~\bibnamefont{Mares}} \bibnamefont{and}
  \bibinfo{author}{\bibfnamefont{J.}~\bibnamefont{Zofka}}, \bibinfo{journal}{Z.
  Phys.} \textbf{\bibinfo{volume}{A333}}, \bibinfo{pages}{209}
  (\bibinfo{year}{1989}).

\bibitem[{\citenamefont{Jennings}(1990)}]{Jennings1990}
\bibinfo{author}{\bibfnamefont{B.~K.} \bibnamefont{Jennings}},
  \bibinfo{journal}{Phys. Lett.} \textbf{\bibinfo{volume}{B246}},
  \bibinfo{pages}{325} (\bibinfo{year}{1990}), \bibinfo{note}{[,325(1990)]}.

\bibitem[{\citenamefont{Sugahara and Toki}(1994{\natexlab{b}})}]{Sugahara1994}
\bibinfo{author}{\bibfnamefont{Y.}~\bibnamefont{Sugahara}} \bibnamefont{and}
  \bibinfo{author}{\bibfnamefont{H.}~\bibnamefont{Toki}},
  \bibinfo{journal}{Prog. Theor. Phys.} \textbf{\bibinfo{volume}{92}},
  \bibinfo{pages}{803} (\bibinfo{year}{1994}{\natexlab{b}}).

\bibitem[{\citenamefont{Mare\ifmmode~\check{s}\else \v{s}\fi{} and
  Jennings}(1994)}]{Mares1994}
\bibinfo{author}{\bibfnamefont{J.}~\bibnamefont{Mare\ifmmode~\check{s}\else
  \v{s}\fi{}}} \bibnamefont{and} \bibinfo{author}{\bibfnamefont{B.~K.}
  \bibnamefont{Jennings}}, \bibinfo{journal}{Phys. Rev. C}
  \textbf{\bibinfo{volume}{49}}, \bibinfo{pages}{2472} (\bibinfo{year}{1994}),
  \urlprefix\url{https://link.aps.org/doi/10.1103/PhysRevC.49.2472}.

\bibitem[{\citenamefont{Khaustov et~al.}(2000)}]{khaustov00}
\bibinfo{author}{\bibfnamefont{P.}~\bibnamefont{Khaustov}} \bibnamefont{et~al.}
  (\bibinfo{collaboration}{AGS E885}), \bibinfo{journal}{Phys. Rev.}
  \textbf{\bibinfo{volume}{C61}}, \bibinfo{pages}{054603}
  (\bibinfo{year}{2000}), \eprint{nucl-ex/9912007}.

\bibitem[{\citenamefont{Yoshida et~al.}(2019)}]{Yoshida2019}
\bibinfo{author}{\bibfnamefont{J.}~\bibnamefont{Yoshida}} \bibnamefont{et~al.},
  \bibinfo{journal}{JPS Conf. Proc.} \textbf{\bibinfo{volume}{26}},
  \bibinfo{pages}{023006} (\bibinfo{year}{2019}).

\bibitem[{\citenamefont{Hiyama et~al.}(2008)\citenamefont{Hiyama, Yamamoto,
  Motoba, Rijken, and Kamimura}}]{Hiyama2008}
\bibinfo{author}{\bibfnamefont{E.}~\bibnamefont{Hiyama}},
  \bibinfo{author}{\bibfnamefont{Y.}~\bibnamefont{Yamamoto}},
  \bibinfo{author}{\bibfnamefont{T.}~\bibnamefont{Motoba}},
  \bibinfo{author}{\bibfnamefont{T.~A.} \bibnamefont{Rijken}},
  \bibnamefont{and} \bibinfo{author}{\bibfnamefont{M.}~\bibnamefont{Kamimura}},
  \bibinfo{journal}{Phys. Rev. C} \textbf{\bibinfo{volume}{78}},
  \bibinfo{pages}{054316} (\bibinfo{year}{2008}),
  \urlprefix\url{https://link.aps.org/doi/10.1103/PhysRevC.78.054316}.

\bibitem[{\citenamefont{Matsumiya et~al.}(2011)\citenamefont{Matsumiya,
  Tsubakihara, Kimura, Dote, and Ohnishi}}]{Matsumya2011}
\bibinfo{author}{\bibfnamefont{H.}~\bibnamefont{Matsumiya}},
  \bibinfo{author}{\bibfnamefont{K.}~\bibnamefont{Tsubakihara}},
  \bibinfo{author}{\bibfnamefont{M.}~\bibnamefont{Kimura}},
  \bibinfo{author}{\bibfnamefont{A.}~\bibnamefont{Dote}}, \bibnamefont{and}
  \bibinfo{author}{\bibfnamefont{A.}~\bibnamefont{Ohnishi}},
  \bibinfo{journal}{Phys. Rev. C} \textbf{\bibinfo{volume}{83}},
  \bibinfo{pages}{024312} (\bibinfo{year}{2011}),
  \urlprefix\url{https://link.aps.org/doi/10.1103/PhysRevC.83.024312}.

\bibitem[{\citenamefont{Hiyama and Yamamoto}(2012)}]{Hiyama_2012}
\bibinfo{author}{\bibfnamefont{E.}~\bibnamefont{Hiyama}} \bibnamefont{and}
  \bibinfo{author}{\bibfnamefont{Y.}~\bibnamefont{Yamamoto}},
  \bibinfo{journal}{Prog. Theor. Phys.} \textbf{\bibinfo{volume}{128}},
  \bibinfo{pages}{105} (\bibinfo{year}{2012}), \eprint{1205.6551}.

\bibitem[{\citenamefont{Millener}(2012)}]{Millener_2012}
\bibinfo{author}{\bibfnamefont{D.}~\bibnamefont{Millener}},
  \bibinfo{journal}{Nuclear Physics A} \textbf{\bibinfo{volume}{881}},
  \bibinfo{pages}{298 } (\bibinfo{year}{2012}), ISSN \bibinfo{issn}{0375-9474},
  \bibinfo{note}{progress in Strangeness Nuclear Physics},
  \urlprefix\url{http://www.sciencedirect.com/science/article/pii/S0375947412000504}.

\bibitem[{\citenamefont{Gogami et~al.}(2016)\citenamefont{Gogami, Chen, Kawama,
  Achenbach, Ahmidouch, Albayrak, Androic, Asaturyan, Asaturyan, Ates
  et~al.}}]{Gogami_PRC_2016}
\bibinfo{author}{\bibfnamefont{T.}~\bibnamefont{Gogami}},
  \bibinfo{author}{\bibfnamefont{C.}~\bibnamefont{Chen}},
  \bibinfo{author}{\bibfnamefont{D.}~\bibnamefont{Kawama}},
  \bibinfo{author}{\bibfnamefont{P.}~\bibnamefont{Achenbach}},
  \bibinfo{author}{\bibfnamefont{A.}~\bibnamefont{Ahmidouch}},
  \bibinfo{author}{\bibfnamefont{I.}~\bibnamefont{Albayrak}},
  \bibinfo{author}{\bibfnamefont{D.}~\bibnamefont{Androic}},
  \bibinfo{author}{\bibfnamefont{A.}~\bibnamefont{Asaturyan}},
  \bibinfo{author}{\bibfnamefont{R.}~\bibnamefont{Asaturyan}},
  \bibinfo{author}{\bibfnamefont{O.}~\bibnamefont{Ates}}, \bibnamefont{et~al.}
  (\bibinfo{collaboration}{HKS(JLab E05-115) Collaboration}),
  \bibinfo{journal}{Phys. Rev. C} \textbf{\bibinfo{volume}{93}},
  \bibinfo{pages}{034314} (\bibinfo{year}{2016}),
  \urlprefix\url{https://link.aps.org/doi/10.1103/PhysRevC.93.034314}.

\bibitem[{\citenamefont{Glendenning and Moszkowski}(1991)}]{gm1991}
\bibinfo{author}{\bibfnamefont{N.~K.} \bibnamefont{Glendenning}}
  \bibnamefont{and} \bibinfo{author}{\bibfnamefont{S.~A.}
  \bibnamefont{Moszkowski}}, \bibinfo{journal}{Phys. Rev. Lett.}
  \textbf{\bibinfo{volume}{67}}, \bibinfo{pages}{2414} (\bibinfo{year}{1991}).

\bibitem[{\citenamefont{Riley et~al.}(2019)\citenamefont{Riley, Watts,
  Bogdanov, Ray, Ludlam, Guillot, Arzoumanian, Baker, Bilous, Chakrabarty
  et~al.}}]{Riley19}
\bibinfo{author}{\bibfnamefont{T.~E.} \bibnamefont{Riley}},
  \bibinfo{author}{\bibfnamefont{A.~L.} \bibnamefont{Watts}},
  \bibinfo{author}{\bibfnamefont{S.}~\bibnamefont{Bogdanov}},
  \bibinfo{author}{\bibfnamefont{P.~S.} \bibnamefont{Ray}},
  \bibinfo{author}{\bibfnamefont{R.~M.} \bibnamefont{Ludlam}},
  \bibinfo{author}{\bibfnamefont{S.}~\bibnamefont{Guillot}},
  \bibinfo{author}{\bibfnamefont{Z.}~\bibnamefont{Arzoumanian}},
  \bibinfo{author}{\bibfnamefont{C.~L.} \bibnamefont{Baker}},
  \bibinfo{author}{\bibfnamefont{A.~V.} \bibnamefont{Bilous}},
  \bibinfo{author}{\bibfnamefont{D.}~\bibnamefont{Chakrabarty}},
  \bibnamefont{et~al.}, \bibinfo{journal}{The Astrophysical Journal}
  \textbf{\bibinfo{volume}{887}}, \bibinfo{pages}{L21} (\bibinfo{year}{2019}),
  \urlprefix\url{https://doi.org/10.3847%2F2041-8213%2Fab481c}.

\bibitem[{\citenamefont{Miller et~al.}(2019)\citenamefont{Miller, Lamb,
  Dittmann, Bogdanov, Arzoumanian, Gendreau, Guillot, Harding, Ho, Lattimer
  et~al.}}]{Miller19}
\bibinfo{author}{\bibfnamefont{M.~C.} \bibnamefont{Miller}},
  \bibinfo{author}{\bibfnamefont{F.~K.} \bibnamefont{Lamb}},
  \bibinfo{author}{\bibfnamefont{A.~J.} \bibnamefont{Dittmann}},
  \bibinfo{author}{\bibfnamefont{S.}~\bibnamefont{Bogdanov}},
  \bibinfo{author}{\bibfnamefont{Z.}~\bibnamefont{Arzoumanian}},
  \bibinfo{author}{\bibfnamefont{K.~C.} \bibnamefont{Gendreau}},
  \bibinfo{author}{\bibfnamefont{S.}~\bibnamefont{Guillot}},
  \bibinfo{author}{\bibfnamefont{A.~K.} \bibnamefont{Harding}},
  \bibinfo{author}{\bibfnamefont{W.~C.~G.} \bibnamefont{Ho}},
  \bibinfo{author}{\bibfnamefont{J.~M.} \bibnamefont{Lattimer}},
  \bibnamefont{et~al.}, \bibinfo{journal}{The Astrophysical Journal}
  \textbf{\bibinfo{volume}{887}}, \bibinfo{pages}{L24} (\bibinfo{year}{2019}),
  \urlprefix\url{https://doi.org/10.3847%2F2041-8213%2Fab50c5}.

\bibitem[{\citenamefont{{Abbott} et~al.}(2018)\citenamefont{{Abbott}, {Abbott},
  {Abbott}, {Acernese}, {Ackley}, {Adams}, {Adams}, {Addesso}, {Adhikari},
  {Adya} et~al.}}]{Abbott18}
\bibinfo{author}{\bibfnamefont{B.~P.} \bibnamefont{{Abbott}}},
  \bibinfo{author}{\bibfnamefont{R.}~\bibnamefont{{Abbott}}},
  \bibinfo{author}{\bibfnamefont{T.~D.} \bibnamefont{{Abbott}}},
  \bibinfo{author}{\bibfnamefont{F.}~\bibnamefont{{Acernese}}},
  \bibinfo{author}{\bibfnamefont{K.}~\bibnamefont{{Ackley}}},
  \bibinfo{author}{\bibfnamefont{C.}~\bibnamefont{{Adams}}},
  \bibinfo{author}{\bibfnamefont{T.}~\bibnamefont{{Adams}}},
  \bibinfo{author}{\bibfnamefont{P.}~\bibnamefont{{Addesso}}},
  \bibinfo{author}{\bibfnamefont{R.~X.} \bibnamefont{{Adhikari}}},
  \bibinfo{author}{\bibfnamefont{V.~B.} \bibnamefont{{Adya}}},
  \bibnamefont{et~al.}, \bibinfo{journal}{\prl} \textbf{\bibinfo{volume}{121}},
  \bibinfo{eid}{161101} (\bibinfo{year}{2018}), \eprint{1805.11581}.

\bibitem[{\citenamefont{Tolman}(1939)}]{TOV1}
\bibinfo{author}{\bibfnamefont{R.~C.} \bibnamefont{Tolman}},
  \bibinfo{journal}{Phys. Rev.} \textbf{\bibinfo{volume}{55}},
  \bibinfo{pages}{364} (\bibinfo{year}{1939}).

\bibitem[{\citenamefont{Oppenheimer and Volkoff}(1939)}]{TOV2}
\bibinfo{author}{\bibfnamefont{J.~R.} \bibnamefont{Oppenheimer}}
  \bibnamefont{and} \bibinfo{author}{\bibfnamefont{G.~M.}
  \bibnamefont{Volkoff}}, \bibinfo{journal}{Phys. Rev.}
  \textbf{\bibinfo{volume}{55}}, \bibinfo{pages}{374} (\bibinfo{year}{1939}).

\bibitem[{\citenamefont{{Miller} and {Lamb}}(2016)}]{Miller16}
\bibinfo{author}{\bibfnamefont{M.~C.} \bibnamefont{{Miller}}} \bibnamefont{and}
  \bibinfo{author}{\bibfnamefont{F.~K.} \bibnamefont{{Lamb}}},
  \bibinfo{journal}{European Physical Journal A} \textbf{\bibinfo{volume}{52}},
  \bibinfo{eid}{63} (\bibinfo{year}{2016}), \eprint{1604.03894}.

\bibitem[{\citenamefont{{Haensel} et~al.}(2016)\citenamefont{{Haensel},
  {Bejger}, {Fortin}, and {Zdunik}}}]{Haensel16}
\bibinfo{author}{\bibfnamefont{P.}~\bibnamefont{{Haensel}}},
  \bibinfo{author}{\bibfnamefont{M.}~\bibnamefont{{Bejger}}},
  \bibinfo{author}{\bibfnamefont{M.}~\bibnamefont{{Fortin}}}, \bibnamefont{and}
  \bibinfo{author}{\bibfnamefont{L.}~\bibnamefont{{Zdunik}}},
  \bibinfo{journal}{European Physical Journal A} \textbf{\bibinfo{volume}{52}},
  \bibinfo{eid}{59} (\bibinfo{year}{2016}), \eprint{1601.05368}.

\bibitem[{\citenamefont{{Arzoumanian} et~al.}(2014)\citenamefont{{Arzoumanian},
  {Gendreau}, {Baker}, {Cazeau}, {Hestnes}, {Kellogg}, {Kenyon}, {Kozon},
  {Liu}, {Manthripragada} et~al.}}]{NICER}
\bibinfo{author}{\bibfnamefont{Z.}~\bibnamefont{{Arzoumanian}}},
  \bibinfo{author}{\bibfnamefont{K.~C.} \bibnamefont{{Gendreau}}},
  \bibinfo{author}{\bibfnamefont{C.~L.} \bibnamefont{{Baker}}},
  \bibinfo{author}{\bibfnamefont{T.}~\bibnamefont{{Cazeau}}},
  \bibinfo{author}{\bibfnamefont{P.}~\bibnamefont{{Hestnes}}},
  \bibinfo{author}{\bibfnamefont{J.~W.} \bibnamefont{{Kellogg}}},
  \bibinfo{author}{\bibfnamefont{S.~J.} \bibnamefont{{Kenyon}}},
  \bibinfo{author}{\bibfnamefont{R.~P.} \bibnamefont{{Kozon}}},
  \bibinfo{author}{\bibfnamefont{K.~C.} \bibnamefont{{Liu}}},
  \bibinfo{author}{\bibfnamefont{S.~S.} \bibnamefont{{Manthripragada}}},
  \bibnamefont{et~al.}, \emph{\bibinfo{title}{{The neutron star interior
  composition explorer (NICER): mission definition}}} (\bibinfo{year}{2014}),
  vol. \bibinfo{volume}{9144} of \emph{\bibinfo{series}{Society of
  Photo-Optical Instrumentation Engineers (SPIE) Conference Series}}, p.
  \bibinfo{pages}{914420}.

\bibitem[{\citenamefont{{Motch} et~al.}(2013)\citenamefont{{Motch}, {Wilms},
  {Barret}, {Becker}, {Bogdanov}, {Boirin}, {Corbel}, {Cackett}, {Campana}, {de
  Martino} et~al.}}]{Athena}
\bibinfo{author}{\bibfnamefont{C.}~\bibnamefont{{Motch}}},
  \bibinfo{author}{\bibfnamefont{J.}~\bibnamefont{{Wilms}}},
  \bibinfo{author}{\bibfnamefont{D.}~\bibnamefont{{Barret}}},
  \bibinfo{author}{\bibfnamefont{W.}~\bibnamefont{{Becker}}},
  \bibinfo{author}{\bibfnamefont{S.}~\bibnamefont{{Bogdanov}}},
  \bibinfo{author}{\bibfnamefont{L.}~\bibnamefont{{Boirin}}},
  \bibinfo{author}{\bibfnamefont{S.}~\bibnamefont{{Corbel}}},
  \bibinfo{author}{\bibfnamefont{E.}~\bibnamefont{{Cackett}}},
  \bibinfo{author}{\bibfnamefont{S.}~\bibnamefont{{Campana}}},
  \bibinfo{author}{\bibfnamefont{D.}~\bibnamefont{{de Martino}}},
  \bibnamefont{et~al.}, \bibinfo{journal}{arXiv e-prints}
  \bibinfo{eid}{arXiv:1306.2334} (\bibinfo{year}{2013}), \eprint{1306.2334}.

\bibitem[{\citenamefont{{Watts} et~al.}(2019)\citenamefont{{Watts}, {Yu},
  {Poutanen}, {Zhang}, {Bhattacharyya}, {Bogdanov}, {Ji}, {Patruno}, {Riley},
  {Bakala} et~al.}}]{eXTP}
\bibinfo{author}{\bibfnamefont{A.~L.} \bibnamefont{{Watts}}},
  \bibinfo{author}{\bibfnamefont{W.}~\bibnamefont{{Yu}}},
  \bibinfo{author}{\bibfnamefont{J.}~\bibnamefont{{Poutanen}}},
  \bibinfo{author}{\bibfnamefont{S.}~\bibnamefont{{Zhang}}},
  \bibinfo{author}{\bibfnamefont{S.}~\bibnamefont{{Bhattacharyya}}},
  \bibinfo{author}{\bibfnamefont{S.}~\bibnamefont{{Bogdanov}}},
  \bibinfo{author}{\bibfnamefont{L.}~\bibnamefont{{Ji}}},
  \bibinfo{author}{\bibfnamefont{A.}~\bibnamefont{{Patruno}}},
  \bibinfo{author}{\bibfnamefont{T.~E.} \bibnamefont{{Riley}}},
  \bibinfo{author}{\bibfnamefont{P.}~\bibnamefont{{Bakala}}},
  \bibnamefont{et~al.}, \bibinfo{journal}{Science China Physics, Mechanics, and
  Astronomy} \textbf{\bibinfo{volume}{62}}, \bibinfo{eid}{29503}
  (\bibinfo{year}{2019}), \eprint{1812.04021}.

\bibitem[{\citenamefont{{Hartle}}(1967)}]{Hartle_ApJ_1967}
\bibinfo{author}{\bibfnamefont{J.~B.} \bibnamefont{{Hartle}}},
  \bibinfo{journal}{The Astrophysical Journal} \textbf{\bibinfo{volume}{150}},
  \bibinfo{pages}{1005} (\bibinfo{year}{1967}).

\bibitem[{\citenamefont{{Bejger} et~al.}(2005)\citenamefont{{Bejger}, {Bulik},
  and {Haensel}}}]{Bejger05}
\bibinfo{author}{\bibfnamefont{M.}~\bibnamefont{{Bejger}}},
  \bibinfo{author}{\bibfnamefont{T.}~\bibnamefont{{Bulik}}}, \bibnamefont{and}
  \bibinfo{author}{\bibfnamefont{P.}~\bibnamefont{{Haensel}}},
  \bibinfo{journal}{Monthly Notices of the Royal Astronomical Society}
  \textbf{\bibinfo{volume}{364}}, \bibinfo{pages}{635} (\bibinfo{year}{2005}),
  \eprint{astro-ph/0508105}.

\bibitem[{\citenamefont{Hinderer}(2008)}]{Hinderer2008}
\bibinfo{author}{\bibfnamefont{T.}~\bibnamefont{Hinderer}},
  \bibinfo{journal}{Astrophys. J.} \textbf{\bibinfo{volume}{677}},
  \bibinfo{pages}{1216} (\bibinfo{year}{2008}), \eprint{0711.2420}.

\bibitem[{\citenamefont{{Abbott} et~al.}(2019)\citenamefont{{Abbott}, {Abbott},
  {Abbott}, {Acernese}, {Ackley}, {Adams}, {Adams}, {Addesso}, {Adhikari},
  {Adya} et~al.}}]{Abbott19}
\bibinfo{author}{\bibfnamefont{B.~P.} \bibnamefont{{Abbott}}},
  \bibinfo{author}{\bibfnamefont{R.}~\bibnamefont{{Abbott}}},
  \bibinfo{author}{\bibfnamefont{T.~D.} \bibnamefont{{Abbott}}},
  \bibinfo{author}{\bibfnamefont{F.}~\bibnamefont{{Acernese}}},
  \bibinfo{author}{\bibfnamefont{K.}~\bibnamefont{{Ackley}}},
  \bibinfo{author}{\bibfnamefont{C.}~\bibnamefont{{Adams}}},
  \bibinfo{author}{\bibfnamefont{T.}~\bibnamefont{{Adams}}},
  \bibinfo{author}{\bibfnamefont{P.}~\bibnamefont{{Addesso}}},
  \bibinfo{author}{\bibfnamefont{R.~X.} \bibnamefont{{Adhikari}}},
  \bibinfo{author}{\bibfnamefont{V.~B.} \bibnamefont{{Adya}}},
  \bibnamefont{et~al.}, \bibinfo{journal}{Physical Review X}
  \textbf{\bibinfo{volume}{9}}, \bibinfo{eid}{011001} (\bibinfo{year}{2019}),
  \eprint{1805.11579}.

\bibitem[{\citenamefont{Watts et~al.}(2015)}]{SKA}
\bibinfo{author}{\bibfnamefont{A.}~\bibnamefont{Watts}} \bibnamefont{et~al.},
  \bibinfo{journal}{PoS} \textbf{\bibinfo{volume}{AASKA14}},
  \bibinfo{pages}{043} (\bibinfo{year}{2015}), \eprint{1501.00042}.

\bibitem[{\citenamefont{{Lattimer} et~al.}(1991)\citenamefont{{Lattimer},
  {Pethick}, {Prakash}, and {Haensel}}}]{DU91}
\bibinfo{author}{\bibfnamefont{J.~M.} \bibnamefont{{Lattimer}}},
  \bibinfo{author}{\bibfnamefont{C.~J.} \bibnamefont{{Pethick}}},
  \bibinfo{author}{\bibfnamefont{M.}~\bibnamefont{{Prakash}}},
  \bibnamefont{and}
  \bibinfo{author}{\bibfnamefont{P.}~\bibnamefont{{Haensel}}},
  \bibinfo{journal}{Physical Review Letters} \textbf{\bibinfo{volume}{66}},
  \bibinfo{pages}{2701} (\bibinfo{year}{1991}).

\bibitem[{\citenamefont{{Prakash} et~al.}(1992)\citenamefont{{Prakash},
  {Prakash}, {Lattimer}, and {Pethick}}}]{DUY92}
\bibinfo{author}{\bibfnamefont{M.}~\bibnamefont{{Prakash}}},
  \bibinfo{author}{\bibfnamefont{M.}~\bibnamefont{{Prakash}}},
  \bibinfo{author}{\bibfnamefont{J.~M.} \bibnamefont{{Lattimer}}},
  \bibnamefont{and} \bibinfo{author}{\bibfnamefont{C.~J.}
  \bibnamefont{{Pethick}}}, \bibinfo{journal}{The Astrophysical Journal
  Letters} \textbf{\bibinfo{volume}{390}}, \bibinfo{pages}{L77}
  (\bibinfo{year}{1992}).

\end{thebibliography}

\end{document}